\documentclass[12 pt]{article}
\def\la{\mathrel{\mathpalette\fun <}}
\def\ga{\mathrel{\mathpalette\fun >}}
\def\fun#1#2{\lower3.6pt\vbox{\baselineskip0pt\lineskip.9pt
  \ialign{$\mathsurround=0pt#1\hfil##\hfil$\crcr#2\crcr\sim\crcr}}}

\begin{document}
\input psfig.sty

\title{Force-Free Models of Magnetically Linked Star--Disk Systems}
\author{Dmitri A. Uzdensky, Arieh K{\"o}nigl\thanks{Also at the Enrico 
Fermi Institute, University of Chicago.}, and Christof Litwin}
\date{}
\maketitle

\vspace{-20pt}
\centerline{\it Department of Astronomy and Astrophysics, 
University of Chicago} 
\centerline{\it 5640 S. Ellis Ave., Chicago, IL 60637}

\vspace{10pt}
\centerline{\tt uzdensky@oddjob.uchicago.edu \
arieh@jets.uchicago.edu\ litwin@aleph.uchicago.edu}

\vspace{10pt}
\centerline{November 14, 2000}

\begin{abstract}
We consider disk accretion onto a magnetic star under the
assumption that the stellar magnetic field permeates the disk
and that the magnetosphere that lies between the disk and the
star is force free. Using simplified axisymmetric models (both
semianalytic and numerical), we study the time evolution of the
magnetic field configuration induced by the relative rotation 
between the disk and the star. We show that, if both the star 
and the magnetosphere can be approximated as being perfectly 
conducting, then the behavior of the twisted field
lines depends on the magnitude of the surface conductivity
of the disk. For any given relative azimuthal speed $\Delta
v_\phi$ between the disk and the star (measured at the disk
surface), there is a maximum surface conductivity $\Sigma_{\rm
max} \sim c^2/|\Delta v_\phi|$ such that, if the actual surface conductivity
is smaller than $\Sigma_{\rm max}$, then a steady-state field
configuration can be established, whereas for larger values no
steady state is possible and the field lines inflate and open up
when a critical twist angle (which for an initially dipolar
field is $\sim 2\ {\rm rad}$) is attained. We argue that thin
astrophysical disks are likely to have surface conductivities
that exceed the local $\Sigma_{\rm max}$ except in regions where
$|\delta v_\phi|$ is particularly small (such as the
immediate vicinity of the corotation radius in a Keplerian disk).

We find that, if the disk conductivity is high enough for
the twist angle to continue growing until the field lines 
open, then, except under rather special circumstances, the 
radial magnetic field at the disk surface will grow to a 
level for which the radial migration of the field lines in 
the disk cannot be ignored. A similar conclusion was reached 
by Bardou \& Heyvaerts (1996). We demonstrate, however, that 
the radial diffusion in the disk is much slower than the 
field-line expansion in the magnetosphere, which suggests
that, contrary to the claim by Bardou \& Heyvaerts, the
twisting process does {\em not} result in a rapid expulsion 
of the field lines from the disk.

We calculate the magnetospheric density redistribution
brought about by the field-line expansion and show that
inertial effects in the magnetosphere will slow the expansion 
(and invalidate the assumption of a force-free field) before 
the critical twist angle is reached. We argue, however, that 
these effects merely slow down the opening process but do not 
inhibit it altogether. We find that the field opening drastically 
reduces the density near the apex of the expanding field lines 
(which typically elongate in a direction making an angle $\ga 
60^{\circ}$ to the rotation axis) while also creating a pronounced 
density enhancement near the rotation axis. The former effect is 
conducive to the triggering of microinstabilities (e.g., ion-acoustic) 
in the $j_\phi$ current layer that forms along the direction of 
elongation, whereas the latter is evidently related to a mechanism 
for the formation of an axially outflowing condensation that was 
previously identified in axisymmetric numerical simulations of 
such systems.

We examine the possibility that the expanding field lines
reconnect across the current layer before they open up, so that
the twisting leads to a quasi-periodic process of field-line
expansion and reconnection and not just to a one-time opening event. We
tentatively conclude that hyperresistivity associated with
tearing-mode turbulence could lead to efficient
reconnection, but we speculate that even faster reconnection
might be brought about by 3-D kinking of the twisted field lines
that gave rise to thin current sheets. This question
could be further pursued through MHD stability
calculations and 3-D numerical simulations.

\end{abstract}

{\it Subject headings:} accretion, accretion disks --- MHD --- 
stars: formation --- stars: magnetic fields --- stars: winds, 
outflows --- stars: pre--main-sequence

\section{Introduction}
\label{sec-intro}

The process of disk accretion onto a magnetic star is thought
to have important consequences for the spin evolution and
observational characteristics of neutron stars, white dwarfs, 
and young stellar objects (YSOs). In particular, if the stellar 
magnetic field lines penetrate the disk, then they may transmit 
torques between the disk and the star. Such an interaction has 
been invoked to explain the spin-up and spin-down episodes observed 
in X-ray pulsars (e.g., Ghosh \& Lamb 1978, 1979a, 1979b, hereafter 
GL; Wang 1987; Lovelace, Romanova, \& Bisnovatyi-Kogan 1995, hereafter 
LRBK; Yi, Wheeler, \& Vishniac 1997) as well as the observed rotation-period
distribution in YSOs (e.g., K\"onigl 1991; Edwards et al. 1993; Bouvier 
et al. 1993; Collier Cameron \& Campbell 1993; Yi 1994, 1995; Ghosh 1995; 
Collier Cameron, Campbell, \& Quaintrell 1995; Herbst et al. 2000). 
Furthermore, if the field is strong enough, then it may truncate the 
disk before it reaches the stellar surface and channel the inflowing
matter to high stellar latitudes, where the (by now nearly free-falling)
gas is stopped and thermalized in accretion shocks. This is the
accepted explanation for X-ray pulsars (accreting magnetic
neutron stars; e.g., Lamb 1989) and DM Herculis stars (accreting magnetic white
dwarfs; e.g., Patterson 1994), and it has also been proposed as
the origin of the optical/UV ``hot spots'' in classical T Tauri
stars (interpreted as accreting magnetic YSOs; e.g., Bertout,
Basri, \& Bouvier 1988; K\"onigl 1991; Edwards et al. 1994;
Hartmann, Hewett, \& Calvet 1994; Lamzin 1995; Bertout et
al. 1996; Johns-Krull \& Basri 1997; Johns-Krull \& Hatzes 1997;
Martin 1997; Muzerolle, Hartmann, \& Calvet 1998).

However, because of the complexity and diversity of the physical
processes that are involved, the theory of the interaction between 
a magnetic star and a surrounding accretion disk remains incomplete.
One key question is whether a steady-state description, which has been 
adopted in many of the models constructed to date, is appropriate. In 
a Keplerian disk around a star of mass $M$ that rotates with angular 
velocity $\Omega_*$, the gas interior to the {\em corotation radius} 
$r_{\rm co} = (GM/\Omega_*^2)^{1/3}$ rotates faster than the star, 
whereas the matter at larger radii rotates more slowly. If the star, 
the disk, and the magnetosphere that occupies the space between them 
can be treated as perfect conductors, then stellar magnetic field 
lines that thread the disk at any radius other than $r_{\rm co}$ will
undergo secular twisting. One possibility for establishing a steady 
state is for the twisting of the field lines to be countered by magnetic 
diffusivity in the disk (e.g., GL). However, in many cases a realistic 
estimate of the disk diffusivity leads to values that are much too small 
to justify a steady-state description of a system with a dipolar field 
configuration except in the immediate vicinity of $r_{\rm co}$ (see 
\S~\ref{subsec-resistive-radial} below).

The behavior of twisted magnetic field lines that are anchored
in a well-conducting medium has originally been considered in
the context of the solar corona by means of
semianalytic (e.g., Aly 1985; Low 1986; Wolfson 1994; Aly 1995)
and numerical (e.g., Miki\'c \& Linker 1994; Amari et
al. 1996a,b) techniques. However, direct insight into the field
evolution in rotating accretion disks has by now been gained also from
explicit investigations of magnetically linked star--disk systems, in which
both semianalytic [e.g., van Ballegooijen
1994 (hereafter VB); Lynden-Bell \& Boily 1994; LRBK;
Bardou \& Heyvaerts 1996 (hereafter BH)] and numerical
(e.g., Hayashi, Shibata, \& Matsumoto 1996; Goodson, Winglee, \&
B\"ohm 1997; Miller \& Stone 1997; Goodson, B\"ohm, \& Winglee
1999; Goodson \& Winglee 1999) approaches have again been employed. 
These studies have indicated that the applied twist leads to a
strong expansion of the field lines away from the star. Initially, the
evolution is quasi-static, with the azimuthal field component
building up while the poloidal field structure
changes relatively little, but past a certain point the
expansion accelerates rapidly and in a finite time
(corresponding to a total twist $\sim \pi$) approaches a
singular state characterized by the opening of at least a 
fraction of the field lines.

The evolution of the magnetic field lines after the singular
state is reached is another important question
on which there has been no unanimity in the
literature. According to one school of thought (e.g., LRBK), the
opening effectively severs the magnetic link between
the disk and the star once and for all. The opposite view is
that the star--disk link is reestablished through rapid
field-line reconnection in the magnetosphere, and that the
field then continues to undergo successive episodes of
twisting, expansion, and reconnection (e.g., Aly \& Kuijpers 1990, 
Goodson et at 1997, 1999). The latter picture is consistent with 
the results of numerical simulations of twisted coronal arcades, in 
which the addition of finite resistivity was found to result in 
reconnection across the current sheet that forms in that case 
(e.g., Miki\'c \& Linker 1994; 
Amari et al. 1996a). It has, furthermore, been argued that, in a real 
magnetosphere, the current concentration will itself facilitate 
reconnection through the initiation of plasma microinstabilities 
that give rise to an anomalous resistivity (e.g., Lamb, Hamilton, \& Miller
1993). To resolve this question in a quantitative manner, it is necessary 
to calculate the evolution of the magnetospheric current and particle 
densities as the field lines are being twisted in order to determine 
whether an instability is likely to develop before the field expands
to such a degree that it becomes effectively open.

When the expanding, twisted field lines approach the singular state, 
they typically become sharply bent at the disk surface, giving rise 
to a strong radial stress that tends to push the field outward through 
the disk for reasonable values of the disk diffusivity. Previous 
treatments of the problem have either ignored this issue altogether 
(e.g., LRBK), concluded that it would lead to the effective expulsion 
of the field lines from the disk (BH), or else considered the intermediate 
case in which the radial field excursions average to zero (VB). 
The resolution of this issue depends in part on whether reconnection 
in the magnetosphere terminates the field-line expansion phase before 
the radial stress at the disk surface becomes very large, and it is 
thus directly related to the field-opening question.

Even if the field configuration is not steady, the duration
of the successive expansion/reconnection cycles is sufficiently
short [of order the dynamical (rotation) time], that one could
consider averaging the relevant quantities over the cycle period
(cf. VB) and using them in modeling effectively steady-state
star--disk systems (e.g., GL; Zylstra 1988; Daumerie 1996).
However, the time-dependent nature of the magnetic interaction
could be crucial to the origin of the observed accretion and 
outflow properties of disk-fed magnetic stars (e.g., Hartmann 
1997; Goodson \& Winglee 1999), and it may also help resolve 
some of the difficulties identified in steady-state magnetic 
accretion models (e.g., Safier 1998).

In this paper we address various aspects of the basic issues listed 
above using a simple, largely semianalytic, approach. Our modeling 
framework, based on evolving a series of self-similar, force-free 
equilibrium magnetospheric field configurations, is described in 
\S~\ref{sec-model}. In \S~\ref{sec-resistive} we clarify the condition 
for steady-state evolution in a disk with a given distribution of 
magnetic flux and diffusivity, and we then consider also the radial 
migration of flux across the disk. In \S~\ref{sec-magnetosphere} we 
examine the opening of twisted magnetic field lines using a uniformly 
rotating, self-similar disk model, and we investigate the role of 
reconnection and inertial effects in the magnetosphere above the disk 
in limiting this process. In \S~\ref{sec-keplerian} we generalize our 
analysis of the opening of magnetic field lines by using a non--self-similar 
model that incorporates a differentially rotating disk. In \S~\ref
{sec-discussion} we discuss our results and relate them to recent 
numerical simulations. We summarize in \S~\ref{sec-summary}.

\section{Perfectly Conducting, Uniformly Rotating Disk Model}
\label{sec-model}

In this section we describe a semianalytic model of a 
force-free magnetic field above a perfectly conducting 
thin disk. This model, which was first developed in 1994 
by VB\footnote
{The self-similar model constructed independently, and at 
about the same time, by Lynden-Bell \& Boily (1994), is 
mathematically identical to that of VB. This ``spherical'' 
model has been adapted to extended star--disk systems by 
BH, who, however, postulated different boundary conditions 
at the disk surface than VB.}, 
is distinguished by its relative mathematical simplicity, 
and we use it to illustrate the relevant ideas and as a 
framework for our quantitative analysis.

\subsection{Self-Similar Configurations}
\label{subsec-van-Ballegooijen}

Following VB, we consider a uniformly rotating disk that is magnetically 
linked to a central, point-like star. This may be an adequate representation
of the outer parts of a Keplerian disk, at radii $r \gg r_{\rm 
co}$, where the beat frequency~$\Delta\Omega$ [the difference between 
the rotation rate~$\Omega_{\rm d}(r)$ of the disk and the rotation 
rate~$\Omega_*$ of the star] is almost independent of $r$ (and is
approximately equal to~$-\Omega_*$). We work in spherical coordinates 
($r$,$\theta$,$\phi$) and assume that the system is axisymmetric.

Since the disk is assumed to be infinitely conducting, the 
distribution of the poloidal magnetic flux $\Psi$ on its surface
is fixed and does not change as the star rotates relative
to the disk: $\Psi_{\rm d}(r,t)=\Psi_{\rm d}(r)$. This distribution serves as a
boundary condition at the disk surface. 

We are looking for a {\em self-similar} solution,
and, therefore, neither the boundary conditions nor the
solution itself can possess a characteristic scale in~$r$.
Therefore, $\Psi_{\rm d}(r)$ must be a power-law function,
$$ \Psi_{\rm d}(r) = {C\over n} r^{-n},                     \eqno{(2.1)} $$
where the coefficient $C$ and the power exponent~$n$ are 
positive real constants.

Our goal is to find a time-dependent solution for 
the magnetic field in the infinitely conducting magnetosphere 
above the disk. Assuming that the plasma density is 
low enough for the Alfv{\'e}n speed~$v_{\rm A}$ in the magnetosphere
to be much greater than both the relevant Keplerian rotation 
speed~$v_{\rm K}$ and the sound speed~$c_{\rm s}$, the field configuration 
above the disk at any given time is given by a force-free equilibrium:
$$  \nabla \times {\bf B} = \alpha\ {\bf B},                   \eqno{(2.2)} $$
where $\alpha({\bf r},t)$ is a scalar function, constant along
each field line. Since the boundary conditions of our 2-D problem
[given by $\Psi_{\rm d}(r)$] do not change with time, the only way time enters
the problem is through the function $\alpha$.

Thus, the time-dependent solution is described by a sequence
of equilibria parametrized by $t$, or, equivalently, by the
twist angle 
$$\Delta\Phi(t)=(\Omega_{\rm d}-\Omega_{*})t \equiv \Delta\Omega \, t \, .
\eqno{(2.3)} $$

Following VB, the self-similar magnetic field can be written as 
$$ {\bf B} \equiv [B_r, B_{\theta}, B_{\phi}] = 
{C\over r^{n+2}} \left[f(\theta),{{g(\theta)}\over{\sin{\theta}}},
h(\theta)\right]\, ,                                             \eqno{(2.4)}$$
corresponding to
$$ \Psi(r,\theta) = {C\over{nr^n}} g(\theta)\, ,                             $$
where the functions $f(\theta)$, $g(\theta)$, and $h(\theta)$ are
also functions of time. Thus, the assumptions of axisymmetry and
self-similarity enable the problem of finding 3-D force-free equilibria to be
reduced to a 1D problem of determining the above three functions. 

The boundary condition (2.1) implies $g(\pi/2)=1$, whereas the condition 
$\nabla \cdot {\bf B}=0$ gives
$$ f(\theta) = {1\over{n\sin{\theta}}}{{dg}\over{d\theta}}\, ,  \eqno{(2.5)} $$
and the condition of self-similarity requires that $\alpha$ scale
as $1/r$, i.e.,
$$ \alpha({\bf r},t)={{a(\theta,\Delta\Phi(t))}\over r}\, .     \eqno{(2.6)} $$

By integrating $dr/d\theta=r B_r/B_{\theta}$, the shape of the 
field line is
$$ r(\theta, \Psi)=r_0(\Psi) [g(\theta)]^{1/n},                \eqno{(2.7)} $$
where $r_0(\Psi)$ is the position of the footpoint of the
field line, labeled by the flux function~$\Psi$, on the disk surface.

Then, the condition that $\alpha$ is constant along the magnetic field
immediately gives
$$ a(\theta,\Delta\Phi) = a_0(\Delta\Phi)[g(\theta)]^{1/n}\, . \eqno{(2.8)} $$ 
[Note that the function $g(\theta)$ also depends on time, and therefore
on~$\Delta\Phi$.]

Next, the $\theta$~component of equation~(2.2) enables $h(\theta)$ to be 
expressed in terms of~$g(\theta)$:
$$ h(\theta)={a_0\over{(n+1)\sin{\theta}}}[g(\theta)]^{1+1/n}\, . 
\eqno{(2.9)} $$

Finally, the $\phi$~component of equation~(2.2) gives the following 
second-order nonlinear differential equation for~$g(\theta)$:
$$ \sin{\theta}{d\over{d\theta}}\left({1\over{\sin{\theta}}}
{{dg}\over{d\theta}}\right) +n(n+1)g(\theta)+
{n\over{n+1}} a_0^2 [g(\theta)]^{1+2/n} = 0\, .            \eqno{(2.10)} $$
The boundary conditions are $g(0)=0$ and $g(\pi/2)=1$.

We note that equation (2.10) can also be obtained directly from
the Grad--Shafranov equation for force-free magnetostatic 
equilibria:
$$ {{\partial^2 \Psi}\over{\partial r^2}}+
{\sin{\theta}\over{r^2}} {\partial\over{\partial\theta}}
\left( {1\over{\sin{\theta}}} {{\partial\Psi}\over{\partial\theta}}\right)=
-F(\Psi) F'(\Psi)\, ,                                         \eqno{(2.11)} $$
where the function~$F(\Psi)$ in the nonlinear term on the right-hand 
side represents the contribution of the toroidal field component
and is related to $\alpha(\Psi,t)$ via 
$$ \alpha = F'(\Psi)\, .                                      \eqno{(2.12)} $$ 
This can be verified by making the self-similar ansatz~(2.4) and using 
equations~(2.6)--(2.8).

Equation (2.10) contains the single parameter $a_0$, so one gets a
one-parameter family of solutions that describe a sequence of equilibria.
The dependence of $g(\theta)$ on time is implicitly parametrized
by~$a_0(t)$. In order to find how $a_0$ changes with time, one
needs to calculate the azimuthal twist angle $\Delta\Phi$ as
a function of $a_0$, and then use the relation~(2.3). 
The twist angle $\Delta\Phi$ can be obtained by integrating 
the equation $\sin{\theta}d\phi/d\theta=B_{\phi}/B_{\theta}$ 
along the field, yielding
$$ \Delta\Phi=\phi(\pi/2) = {a_0\over{n+1}} \int_0^{\pi/2}
[g(\theta)]^{1/n} {{d\theta}\over{\sin{\theta}}}\, .          \eqno{(2.13)} $$

\subsection{Magnetic Field Evolution}
\label{subsec-sequence}

For any given $a_0$, one can solve equation~(2.10) by direct 
integration and then use the obtained solution in equation (2.13) to derive
the corresponding value of the twist angle. This is the procedure followed
by VB. We have, however, found it advantageous to reformulate the problem so
as to make the twist angle $\Delta\Phi$, rather than~$a_0$, the control
parameter. In this way the time dependence of the sequence of equilibria 
becomes more transparent: for a given~$t$, $\Delta\Phi$ is given by 
equation~(2.3), and then the problem can be solved using~$\Delta\Phi$ as the  
input parameter. This approach also has the merit of expediting the solution
process.

We have accomplished this goal by replacing the dependent variable
$g(\theta)$ by the function~$\phi(\theta)$, the twist angle of a 
given field line as a function of~$\theta$:
$$ \phi(\theta)= {a_0\over{n+1}} \int_0^{\theta}
[g(\theta)]^{1/n} {{d\theta}\over{\sin{\theta}}}\, .          \eqno{(2.14)} $$

The idea here is to express the function $g(\theta)$ through
$\phi(\theta)$ using equation~(2.14), and then substitute it 
into equation~(2.10), thus obtaining a differential equation 
for~$\phi(\theta)$. Then the parameter $\Delta\Phi$ comes in 
through the boundary condition $\phi(\pi/2)=\Delta\Phi$, and, 
as we show, $a_0$ completely drops out.

We have:
$$ \phi'(\theta) = {a_0\over{n+1}} [g(\theta)]^{1/n}
{1\over{\sin{\theta}}}\, ,                                 \eqno{(2.15)} $$
and after differentiating two more times one gets
$$ \phi'''(\theta) = {{\phi''^2}\over{\phi'}}+\phi'{{g''}\over{ng}}+
{{\phi'}\over{\sin^2{\theta}}}-
n\phi' \left[ {{g'}\over{ng}} \right]^2\, .                         $$

Using equation~(2.10) to express $g''$ in terms of $g'$ and $g$, one 
finally gets a third order differential equation for $\phi(\theta)$,
which does not contain $a_0$:
$$ \phi'''=                                                              $$
$$ {{\phi''^2}\over{\phi'}}(1-n) + {{\phi'}\over{\sin^2{\theta}}}
(2\cos^2{\theta}-n)-\phi''(2n-1)/\tan{\theta}-(n+1)\phi'^3\sin^2{\theta}\, .  
                                                             \eqno{(2.16)} $$

The three boundary conditions are:\\
1) $\phi(0)=0$;\\
2) $\phi'(0)=0$ --- this can be used only if $n<2$:
indeed, near $\theta=0$, $g(\theta) \sim \theta^2$, 
and so $\phi'(\theta) \sim g^{1/n}/\sin{\theta}\sim 
\theta^{(2/n)-1} \rightarrow 0$ as $\theta \rightarrow 0$ 
if $n<2$;\\
3) $\phi(\pi/2)=\Delta\Phi$ --- prescribed value.

After the solution $\phi(\theta)$ is found, one can use
the boundary condition $g(\pi/2)=1$ to determine~$a_0$:
$$ a_0 (\Delta\Phi) = (n+1) \phi'(\pi/2)\, .                  \eqno{(2.17)} $$

\begin{figure} [tbp]
\centerline {\psfig{file=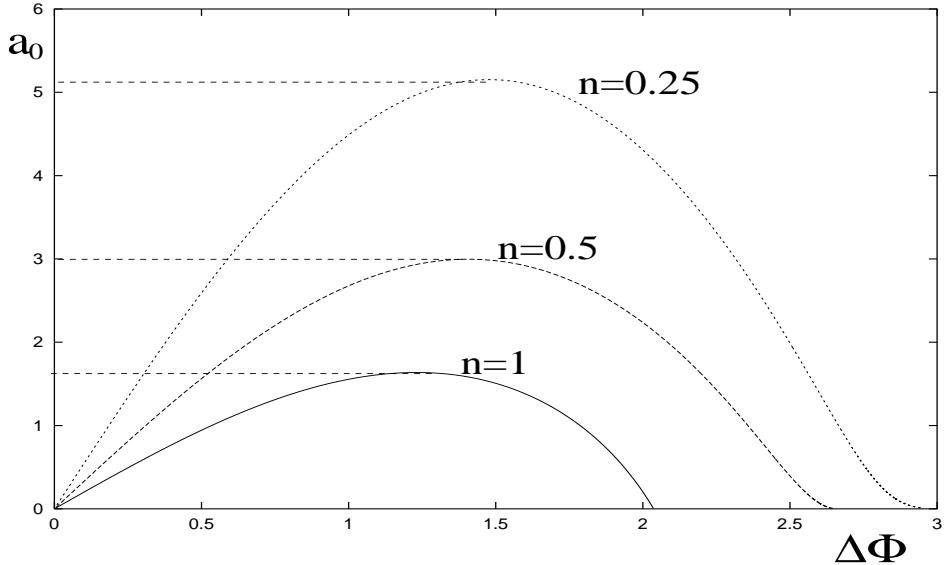,height=3 in,width=5 in}}
\caption{The dependence of $a_0$ on the twist angle $\Delta\Phi$.}
\label{fig-a0-of-phi}
\end{figure}

In Figure~\ref{fig-a0-of-phi} we plot $a_0(\Delta\Phi)$ for several 
values of the parameter~$n$. As has been noted by VB, the dependence 
of $a_0$ on $\Delta\Phi$ is nonmonotonic, and for any value of $a_0$ 
between 0 and $a_{0,\rm max}=a_0(\Delta\Phi_{\rm max})$ there exist 
two different solutions.\footnote
{Actually, there is an infinite series of bands of values
of $\Delta\Phi$ where solutions exist, separated by bands 
of forbidden values. Any two solutions with values of 
$\Delta\Phi$ in a given band belong to the same topological
class [e.g., the function $g(\theta)$ has the same number of 
nodes, with the first band, $0<\Delta\Phi<\Delta\Phi_{\rm c}$, 
having zero nodes], whereas two solutions in different bands 
are topologically distinct and cannot be smoothly transformed 
into each other.}

A purely poloidal field ($\Delta\Phi=0$) is potential ($a_0=0$).
As the field-line twist ($\Delta\Phi$) grows, $a_0$ increases and 
reaches a maximum at some finite twist angle $\Delta\Phi_{\rm max}(n)$. 
As $\Delta \Phi$ is increased even further, $a_0$ decreases and eventually
vanishes at a certain critical twist angle $\Delta\Phi_{\rm c}(n)$ 
[$>\Delta\Phi_{\rm max}(n)$].  We refer to the part of the curve that 
corresponds to $0< \Delta\Phi < \Delta\Phi_{\rm max}$ as the ascending branch, 
and to the portion corresponding to $\Delta\Phi_{\rm max} < \Delta\Phi < 
\Delta\Phi_{\rm c}$ as the descending branch. Both $\Delta\Phi_{\rm max}$ 
and $\Delta\Phi_{\rm c}$ are in general of the order of one radian and 
depend only on~$n$ (see Table~\ref{table-1}; $\Delta\Phi_0$ is the twist 
angle for which $f(\pi/2)=0$, see \S~\ref{subsec-resistive-radial}).

\begin{table}[tbp]
\caption{PARAMETERS OF THE SELF-SIMILAR MODEL}
\vskip 10 pt
\begin{tabular}{|@{\hspace{8mm}}c@{\hspace{8mm}}|
*{5}{@{\hspace{6mm}}c@{\hspace{6mm}}}|}
\hline \hline
$n$	& $\Delta\Phi_0$ & $\Delta\Phi_{\rm max}$ & $a_{0,{\rm max}}$ & 
$h_{\rm d, max}$ & $\Delta\Phi_{\rm c}$ \\ 
\hline
 1.0   	&     0.00   &   1.23	&     1.64	& 0.82	&  2.036 \\  
 0.5	&     1.05   &   1.41	&     3.00	& 2.00	&  2.645 \\
 0.25   &     1.30   &   1.48	&     5.15	& 4.12	&  2.944 \\
\hline \hline
\end{tabular}
\label{table-1}
\end{table}

The nonexistence of a solution for $a_0 > a_{0,\rm max}$ is 
in agreement with the result obtained by Aly (1984) for his 
Boundary Value Problem I, in which one prescribes values of 
the normal magnetic field component~$B_n$ and $\alpha$ on 
the boundary of an infinite domain. We also draw attention 
to the change of curvature (from convex to concave) exhibited 
by the solution (for $n<1$) near the critical point. The 
physical origin of this behavior is clarified in Appendix~A.

The behavior of the toroidal field component at the surface of 
the disk is given by the function $h_d \equiv h(\pi/2,\Delta\Phi)$. 
According to equation~(2.9), $h_d$ is equal to $a_0/(n+1)$, so it 
also first increases with~$\Delta\Phi$, reaches a maximum value 
$h_{\rm d, max} = a_{0,\rm max}/(n+1)$ at $\Delta\Phi = \Delta 
\Phi_{\rm max}(n)$ (see Table 1), and then decreases to zero at 
$\Delta\Phi=\Delta\Phi_{\rm c}$. The fact that twisting up the field 
lines does not result in an arbitrary increase of the azimuthal 
field component at the disk surface is central to our analysis of 
the evolution of resistive disks (\S~\ref{subsec-resistive-toroidal}). 
This behavior is emblematic of twisted flux tubes in general. As noted 
in \S~\ref{sec-intro}, such tubes initially exhibit a nearly linear 
growth of $B_\phi$ with $\Delta\Phi$, but after $\Delta\Phi$ exceeds 
$\sim \Delta\Phi_{\rm max}$ the built-up magnetic stress causes the 
field to expand rapidly (in the meridional plane), with the field-line 
twist traveling out to the region of weakest field near the apex of the 
flux tube. The propagation of the twist can be understood in terms of 
torque balance along the tube (e.g., Parker 1979). The evolution of 
the poloidal field components with increasing twist angle is illustrated 
in Figure 2. It is seen that, for $\Delta\Phi > \Delta\Phi_{\rm max}$, 
the flux tube expands and becomes elongated, with the direction of 
elongation defining the apex angle~$\theta_{\rm ap}$.\footnote
{Formally, the apex angle~$\theta_{\rm ap}$ is defined as the 
value of $\theta$ where $g(\theta)$ has a maximum. Since $g(\theta)$ 
and $r(\theta)$ are related via equation~(2.7), this angle also
corresponds to the point on the field line that is most distant
from the star.}

\begin{figure} [tbp]
\centerline {\psfig{file=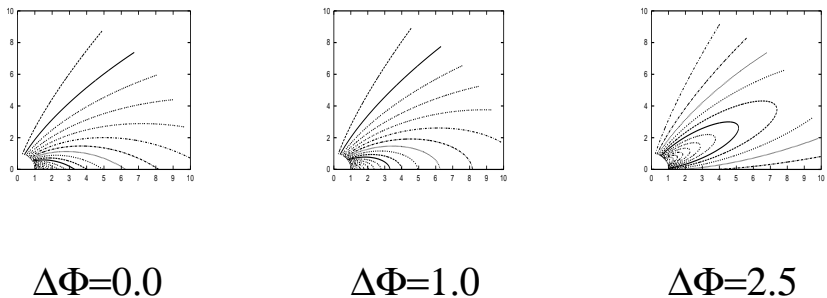,height=2 in,width=5 in}}
\caption{Meridional projection of the magnetic field lines for 
three values of the twist angle $\Delta\Phi$ in the case $n=0.5$,
plotted in the (arbitrarily chosen) interval $r\in [1,10]$.}
\label{fig-contour}
\end{figure}

\subsection{Finite-Time Singularity}
\label{subsec-singularity}

As the critical twist angle~$\Delta\Phi_{\rm c}$ is approached 
(corresponding to $a_0 \rightarrow 0$), the solution of equations~(2.10) 
and~(2.16) blows up, with the field lines expanding to infinity and thus 
opening up (see Fig.~\ref{fig-contour}). In this limit, the radial field 
component at the disk surface diverges even as the surface azimuthal field 
tends to zero. No equilibrium solution is found for twist angles 
$>\Delta\Phi_{\rm c}$. This behavior is generic to twisted flux 
tubes in 3-D and is characterized as a ``finite time singularity'' 
(e.g., Aly 1995): the magnetic field reaches a singular state after 
being twisted for a finite time (or, equivalently, by a finite angle).

To analyze the asymptotic ($a_0 \rightarrow 0$) properties of the function 
$g(\theta)$, we note that $a_0$ can be rescaled out of equation~(2.10)
by the substitution:
$$ G(\theta) = g(\theta) a_0^n\, .                            \eqno{(2.18)} $$
The equation for $G(\theta)$ then becomes 
$$ \sin{\theta}{d\over{d\theta}}\left({1\over{\sin{\theta}}}
{{dG}\over{d\theta}}\right) +n(n+1)G(\theta)+
{n\over{n+1}} [G(\theta)]^{1+2/n} = 0\, .                     \eqno{(2.19)} $$
The boundary conditions are: $G(0)=0$ and $G(\pi/2)=a_0^n$.
Thus, the parameter $a_0$ has been transferred from the 
equation to a boundary condition. But now the transition to the limit $a_0
\rightarrow 0$ can be easily made, since the solution of equation~(2.19)
does not blow up as the boundary condition at $\theta=\pi/2$ approaches zero. 
We designate the solution of equation~(2.19) with the boundary conditions 
$G_0(0)=G_0(\pi/2)=0$ as $G_0(\theta)$. This function depends only on $n$, and
remains finite in the entire interval $(0,\pi/2)$. The behavior of the original
function $g(\theta)$ as the critical twist angle is approached is then 
obtained from
$$ g(\theta, a_0,n) \rightarrow  G_0(\theta,n) a_0^{-n}
\qquad {\rm as}\ a_0\rightarrow 0\, .                         \eqno{(2.20)} $$ 

\begin{figure} [tbp]
\centerline {\psfig{file=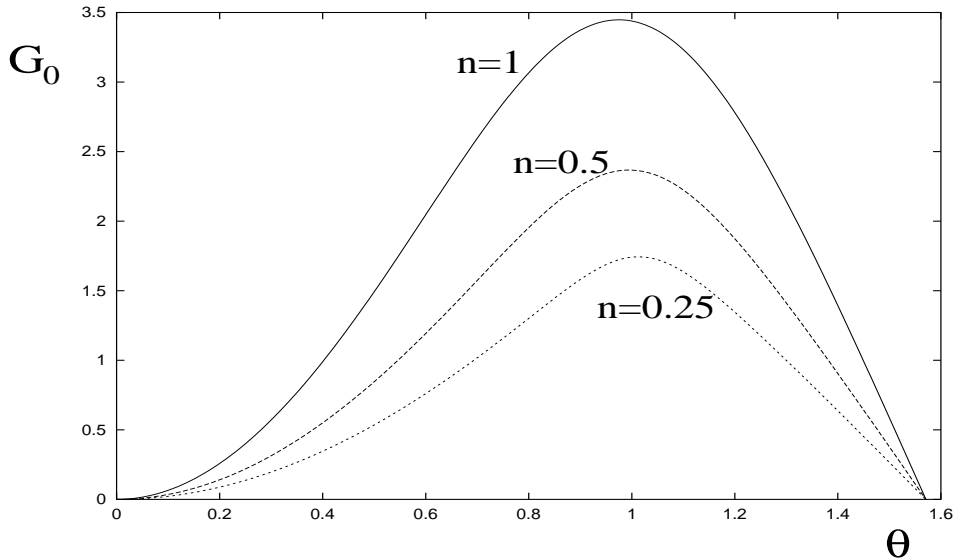,height=3 in,width=5 in}}
\caption{Plots of the function $G_0(\theta)$ for three values of~$n$.}
\label{fig-G0}
\end{figure}

Figure~\ref{fig-G0} displays $G_0(\theta)$ for several values of~$n$.
These solutions have a number of useful implications.

First, by combining equations (2.11) and (2.20), one can calculate the
critical twist angle $\Delta\Phi_{\rm c}$:
$$\Delta\Phi_{\rm c} = {1\over{n+1}} \int\limits_0^{\pi/2}
G_0^{1/n}(\theta) {{d\theta}\over{\sin{\theta}}}\, .           \eqno{(2.21)} $$

\begin{figure} [tbp]
\centerline {\psfig{file=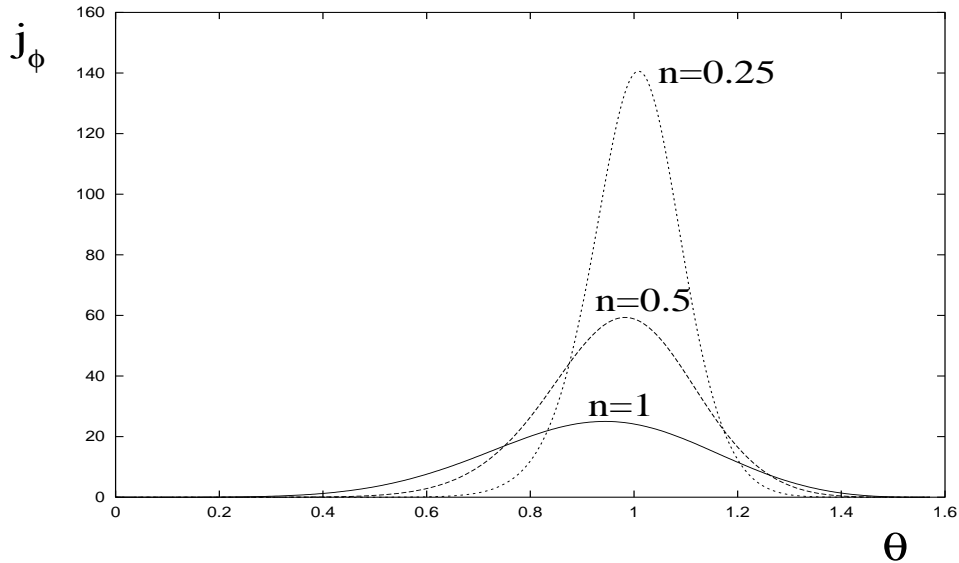,height=3 in,width=5 in}}
\caption{Plots of the azimuthal current density $j_{\phi}(\theta)$ in the
limit $\Delta\Phi \rightarrow \Delta\Phi_{\rm c}$ for three values of~$n$}
\label{fig-current}
\end{figure}

Second, it is seen from Figure~\ref{fig-G0} that the apex angle 
$\theta_{\rm ap}$, at which the function $G_0(\theta)$ reaches 
its maximum, is very close to $60^\circ$ for a wide range of 
values of~$n$. In fact, $\theta_{\rm ap} \rightarrow 60^\circ$ 
as $n\rightarrow 0$, in agreement with the finding by Lynden-Bell 
\& Boily (1994). The azimuthal current density in the magnetosphere, 
which is related to $g(\theta)$ via equations~(2.2),~(2.6),~(2.8), 
and~(2.9), is also concentrated around this angle, and in the limit 
of very small~$n$ the current distribution collapses into a narrow 
layer that extends radially along $\theta \approx \theta_{\rm ap}$ 
(see Fig.~\ref{fig-current}). As we discuss in \S~\ref{subsec-reconnection}, 
such a current layer is a natural potential site for rapid field reconnection 
in the magnetosphere.

Third, similarly to the current density, the azimuthal flux, according 
to equation~(2.9), becomes concentrated near the angle~$\theta_{\rm ap}$. 
This concentration is associated with the outward propagation of the magnetic 
field twist, as we discussed in \S~\ref{subsec-sequence}.

\section{Steady-State Configurations}
\label{sec-resistive}

In this section we consider the conditions under which a magnetically 
linked star--disk configuration can reach a steady state. We continue 
to assume, as in \S~\ref{sec-model}, that the star as well as the 
magnetosphere can be approximated as perfect conductors, but we allow 
the disk to have a finite resistivity.\footnote
{The role of resistive effects in the magnetosphere is explored in 
\S~\ref{sec-magnetosphere}.}
This picture should be adequate for capturing the basic behavior of 
real systems. For ease of presentation, we use the self-similar model 
outlined in \S~\ref{sec-model}, which corresponds to a uniform value 
of the field-line twist angle $\Delta\Phi$ at any given instant. The 
self-similarity assumption in turn imposes a condition on the radial 
dependence of the vertically integrated electrical conductivity~$\Sigma$.
However, since the conditions we derive are essentially local, we expect 
our basic conclusions to remain valid also in the more general case of a 
differentially rotating disk and a non--self-similar diffusivity. To further 
simplify the presentation, we first consider the case where the radial 
positions of the magnetic field lines in the disk remain fixed during 
the twisting process, so the diffusivity only affects the evolution of 
the azimuthal field component. We then examine the consequences of 
incorporating also radial field diffusion into the picture.

\subsection{Time Evolution of the Twist Angle}
\label{subsec-resistive-toroidal}

We start by deriving a general expression for the time evolution
of the twist angle in the presence of finite disk diffusivity. 
Focusing on a given disk radius~$r$, we assume that the disk has 
a thickness~$2H\ll r$, so that the normal field component at the 
disk surface is given approximately by $B_{{\rm d},z} \equiv B_z(z=H) 
\approx B_z(z=0)=-B_{\theta}(r,\pi/2)$ (where the approximate equality follows 
from the solenoidal condition on ${\bf B}$). The azimuthal field 
component $B_{\phi}$, which for a twisted field generally has a 
finite value $B_{{\rm d},\phi}$ at~$z=H$, is zero at $z=0$ on account 
of the reflection symmetry with respect to the midplane. Hence 
the disk carries a radial surface current density $K_r \approx 
2\, H\, j_r(z=0)$, given by
$$ K_r = - c  B_{{\rm d},\phi}/2\pi\, .                        \eqno{(3.1)} $$
Now, by Ohm's law, the radial electric field at the disk surface is
$$ E_{{\rm d},r} = K_r/\Sigma -v_{{\rm d},\phi} B_{{\rm d},z}/c \, , 
\eqno{(3.2)} $$
where $v_{{\rm d},\phi} = r \Omega_{\rm d}$ is the azimuthal speed of the 
disk matter and where $\Sigma$ is related to the magnetic diffusivity $\eta$
through $\Sigma \approx H c^2/ 2 \pi \eta$. On the other hand, the azimuthal 
speed of the field lines that thread the disk at the given radius is
$v_{{\rm B},\phi} = - c E_{{\rm d},r}/B_{{\rm d},z}$. The azimuthal speed 
of the field lines relative to the disk gas is, therefore, $v_{{\rm B},\phi} 
- v_{{\rm d},\phi} = - c K_r/\Sigma B_{{\rm d},z} \approx 
\eta B_{{\rm d},\phi}/H B_{{\rm d},z}$.

The time evolution of the twist angle is thus given by
$$ {{d\Delta\Phi}\over{dt}}=\Delta\Omega + {c^2\over{2\pi r\Sigma}}\
{B_{{\rm d},\phi}\over{B_{{\rm d},z}}}\, ,                     \eqno{(3.3)} $$
where, as in \S~\ref{sec-model}, $\Delta\Omega=\Omega_{\rm d}-\Omega_*$. 

The first term on the right-hand side of equation~(3.3)
represents the secular growth of $\Delta\Phi$ due to the
differential rotation between the disk and the star, whereas 
the second term (in which $c^2/2\pi\Sigma \approx \eta/H$) 
describes the azimuthal slippage of the field lines relative 
to the disk material brought about by the finite disk resistivity.
One can alternatively obtain this equation by generalizing the 
circuit analysis previously used to derive the steady-state 
condition (see references below) to include an inductive 
contribution to the line integral in Faraday's law (i.e., 
$c \oint{\bf E \cdot} d{\bf r} = - d\Psi/dt$). The result, 
in any case, is quite general and depends only on the disk 
being thin and on the magnetosphere being perfectly conducting; 
no other assumptions (such as mechanical equilibrium in the 
magnetosphere) need to be made.

We expect that, in general, the ratio ${B_{{\rm d},\phi}/{B_{{\rm d},z}}}$ 
in equation~(3.3) will be a function of~$\Delta\Phi$. In particular, 
if the magnetosphere is described by a force-free equilibrium, then
this function is uniquely determined, and, for a given distribution 
of $\Sigma(r)$, one obtains a closed equation for~$\Delta\Phi(r,t)$. 
For the purpose of illustration, we consider this equation in the 
context of the self-similar solutions discussed in \S~\ref{sec-model}. 
In order for these solutions to be applicable, $\Sigma(r)$ must scale 
as~$1/r$. In that case ${B_{{\rm d},\phi}/{B_{{\rm d},z}}}(\Delta\Phi)=
-h(\pi/2,\Delta\Phi)=-a_0(\Delta\Phi)/(n+1)$ [see eq.~(2.9)], so 
equation~(3.3) becomes
$$  {{d\Delta\Phi}\over{dt}}=\Delta\Omega - 
{c^2\over{2\pi r\Sigma}} {{a_0(\Delta\Phi)}\over{n+1}}\, .      \eqno{(3.4)} $$

By setting the time derivative in equation (3.3) equal to zero,
one obtains the steady-state value of ${B_{{\rm d},\phi}/{B_{{\rm d},z}}}$,
the azimuthal pitch at the disk surface. If the vertical field component in the
disk is assumed to be fixed, this expression yields
the azimuthal field component at the disk surface for which the
twisted field configuration maintains a steady state,
$$ B_{\phi,{\rm ss}} = -{{2\pi r\Sigma \Delta\Omega}\over c^2}\
B_{{\rm d},z}\, .                                           \eqno{(3.5)} $$

This result was previously obtained by Campbell (1992), LRBK,
and BH using an electric circuit formulation. The same
basic relation was used by GL to define the effective electrical
conductivity of the star--disk system in terms of the time-averaged
azimuthal pitch at the disk surface. In contrast to the GL
approach, equation (3.5) focuses on the actual conductivity of
the disk and gives the unique value of the azimuthal pitch that
corresponds to a genuine steady state. It is then possible to
inquire whether any particular system with a given distribution
of disk surface conductivity $\Sigma(r)$ and differential
rotation rate $\Delta\Omega(r)$ will be able to attain a steady
state.

To answer this question, we take the magnetospheric field to be
force free and consider the self-similar solutions described in
\S~\ref{subsec-sequence}. Typically in these solutions,
$|B_{{\rm d},\phi}(\Delta\Phi)|$ at first grows with 
increasing $\Delta\Phi$, reaches a maximum $|B_{{\rm d},\phi}^{\rm max}|$ 
at $\Delta\Phi_{\rm max}$ (with $|B_{{\rm d},\phi}^{\rm max}/B_{{\rm d},z}|\sim O(1)$ 
for~$n\simeq 1$) and then declines to zero at $\Delta\Phi_{\rm c}$.
We now show that whether or not a steady state can be reached
depends on the relative magnitude of $B_{{\rm d},\phi}^{\rm max}$
and $B_{\phi,{\rm ss}}$. 

\begin{figure} [tbp]
\centerline {\psfig{file=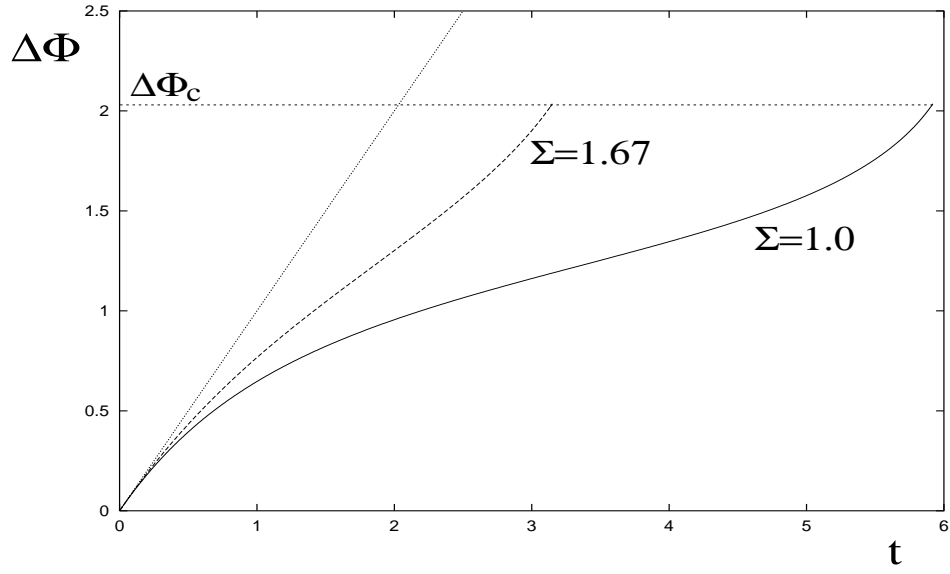,height=3 in,width=5 in}}
\caption{Time evolution of the twist angle for several values of the 
disk surface conductivity~$\Sigma$ above $\Sigma_{\rm max}$ (eq.~[3.6]) 
in the~$n=1$ self-similar solution. The time~$t$ is given in units 
of~$1/\Delta\Omega$ and the surface conductivity in units of~$c^2/
2\pi\Delta\Omega r$ ($\Sigma_{\rm max}$ in these units is equal to 
$h_{\rm d,max}=0.82$).}
\label{fig-resistive-twist-a}
\end{figure}

We define a maximum surface conductivity $\Sigma_{\rm max}$ by
$$ \Sigma_{\rm max} = \left|{c^2 \over 2\pi r \Delta\Omega}
{B_{{\rm d},\phi}^{\rm max}\over B_{{\rm d},z}} \right| = \left|{c^2
\over 2\pi r \Delta\Omega} \right | h_{{\rm d},{\rm max}}\, ,  \eqno{(3.6)} $$ 
where the second equality gives the result for our self-similar model. 
If the actual surface conductivity of the disk is {\em large}, $\Sigma>
\Sigma_{\rm max}$, then $|B_{{\rm d},\phi}^{\rm max}|<|B_{\phi,\rm ss}|$, 
and there is {\em no steady state}. In this case the azimuthal resistive 
slippage is not strong enough to balance the twist amplification of the 
azimuthal field component, and a singularity is reached in a finite time 
(corresponding to~$\Delta\Phi_{\rm c}$). The effect of the resistive diffusion 
is merely to delay the onset of the singularity but not to remove it. 
This is illustrated in Figure~\ref{fig-resistive-twist-a}, which shows 
that the time delay $\Delta t$ due to resistivity increases with 
decreasing~$\Sigma$. The slowdown rate is largest near~$\Delta\Phi_{\rm max}$ 
because, for this twist, the azimuthal field $|B_{{\rm d},\phi}|$, and hence 
the resistive slippage in the azimuthal direction, are maximized. As the 
critical twist~$\Delta\Phi_{\rm c}$ is approached, the time derivative of 
$\Delta\Phi$ reverts to its initial value $[d\Delta\Phi/dt]
(\Delta\Phi_{\rm c})=[d\Delta\Phi/dt](0)=\Delta\Omega$ because 
the azimuthal field at the disk surface vanishes at the singular point.

\begin{figure} [tbp]
\centerline {\psfig{file=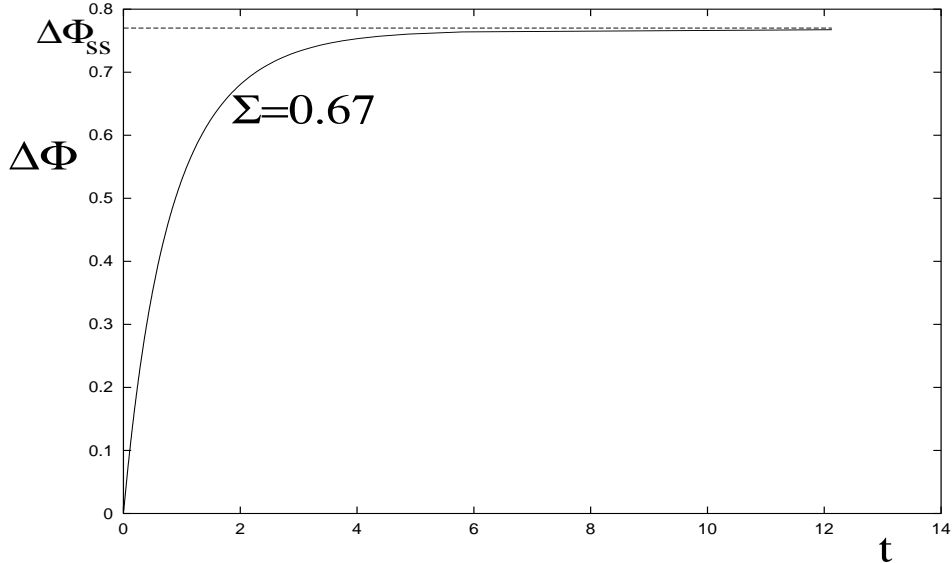,height=3 in,width=5 in}}
\caption{Time evolution of the twist angle in the~$n=1$ self-similar 
solution for the case in which the disk surface conductivity is less 
than~$\Sigma_{\rm max}$. The normalizations of $t$ and of $\Sigma$ are 
the same as in Fig.~\ref{fig-resistive-twist-a}.}
\label{fig-resistive-twist-b}
\end{figure}

Conversely, when the disk conductivity is {\em small} ($\Sigma <
\Sigma_{\rm max}$), $B_{{\rm d},\phi}^{\rm max}>B_{\phi,{\rm ss}}$ and {\em the
steady state described by equation (3.5) can be reached}. If one
starts with a purely poloidal potential field ($\Delta\Phi=0$,
$B_{{\rm d},\phi}=0$), then $\Delta\Phi$ initially grows linearly with
time but subsequently levels off, tending to a steady-state value
$\Delta\Phi_{\rm ss}$ that is given by the condition
$$ B_{{\rm d},\phi}(\Delta\Phi_{\rm ss})=B_{\phi, {\rm ss}}\, . \eqno{(3.7)} $$
This situation is illustrated in Figure~\ref{fig-resistive-twist-b}.

Note that, although in principle two solutions are possible for 
a given value of $|B_{{\rm d},\phi}| < |B_{{\rm d},\phi}^{\rm max}|$, 
the steady-state solution described here always corresponds to the left 
(ascending) branch, i.e., $\Delta\Phi_{\rm ss}<\Delta\Phi_{\rm max}$ 
(provided that one starts the evolution with the potential field 
$\Delta\Phi=0$). Also note that the steady state corresponding to 
the left branch is stable, whereas the one corresponding to the 
right (descending) branch is unstable in the following sense. Suppose 
we are at some steady state, and we decrease $\Delta\Phi$ slightly. 
Then, if the solution is on the left branch, $|B_{{\rm d},\phi}|$ decreases 
and the azimuthal slippage speed becomes smaller than $\Delta\Omega 
\, r$, causing $\Delta\Phi$ to increase back to its original value and 
thereby restoring the initial state. Solutions on the right branch exhibit 
the opposite behavior: if $\Delta\Phi$ is decreased slightly, 
$|B_{{\rm d},\phi}|$ will increase and the enhanced azimuthal resistive 
slippage will cause the twist angle~$\Delta\Phi$ to drop even further, thereby
moving away from the original state.

We now address the question of whether real magnetically linked star--disk 
systems may be expected to reach a steady state. Astrophysical plasmas are
typically highly conductive, so magnetic diffusivity is commonly
ascribed to an anomalous resistivity associated with fluid turbulence
(e.g., Parker 1979). The turbulent diffusivity can be expressed as
$\eta = \beta v_{\rm turb} H$, where $v_{\rm turb}$ is the speed of the
largest turbulent eddies and $\beta$ is a parameter $\la 1$. The eddy
speed could be determined by either thermal or magnetic processes; to
maximize the value of $\eta$, we set $v_{\rm turb} = {\rm max}\{c_{\rm s}\, ,
\, v_{\rm A}\}$. In magnetized accretion disks, the sound speed generally
exceeds the Alfv{\'e}n speed (evaluated at the midplane) except well within
the corotation radius. But this is also the region where the magnetic field 
rapidly becomes strong enough to enforce nearly complete corotation with the 
star, which, in turn, suppresses field-line twisting and inflation: this region
is therefore not of much interest in our present discussion.\footnote{We also
note in this connection that the turbulence in the disk is often attributed to
the action of the magnetic shearing instability, but that the latter ceases to
operate (neglecting for the moment resistive effects) when $v_{\rm A}$ comes
to exceed $\sim c_{\rm s}$ in an unlinked disk, or 
$\sim (c_{\rm s} v_{{\rm d},\phi})^{1/2}$ in a magnetically linked 
system (e.g., Gammie \& Balbus 1994).} Concentrating, therefore, on the case
${c_{\rm s}\gg v_{\rm A}}$, we can write 
$$\Sigma \approx {c^2 \over 2\pi\beta c_{\rm s}}\, ,        \eqno(3.8) $$
which is constant for an isothermal disk. 
Comparing equations (3.8) and (3.6), we obtain 
$${\Sigma\over \Sigma_{\rm max}} \approx \left ( {1 \over \beta \ 
h_{\rm d, max}} \right ) {r \Delta\Omega \over c_{\rm s}}\, .  \eqno(3.9) $$
This ratio is $\gg 1$ in a thin disk ($H \ll r$) threaded by a dipolar field 
($n=1$), except in the region immediately adjacent to $r_{\rm co}$.

In the case of molecular disks around YSOs, it is possible to estimate $\eta$
directly from an explicit determination of the electron--molecule collision
frequency for given temperature and density. Adopting the expressions given in
Meyer \& Meyer-Hofmeister (1999), we have $\eta = 10^{3.99}T_3^{1/2}(n_n/n_e)$,
where $T_3$ is the temperature in units of $10^3 \ {\rm K}$, and where the 
electron-to-neutral number density ratio is calculated by assuming ionization
equilibrium of alkali metals (primarily potassium) and is given
by $\log{(n_e/n_n)} = 6.48 -10.94/T_3
+0.75\log{T_3}-0.5\log{n_n}$. As an illustration, we consider T Tauri stars,
which are relatively slow rotators (mean rotation rate $\Omega_* \simeq 
10^{-5} \ {\rm s}^{-1}$), and for which we infer (assuming a $0.5\ M_{\odot}$ 
star) $r_{\rm co} \approx 8.7 \times 10^{11} \ {\rm cm}$. D'Alessio et al.
(1998) modeled accretion disks around such stars, and for a typical accretion
rate of $10^{-8} \ M_{\odot} \ {\rm yr}^{-1}$ and a disk 
``$\alpha$ parameter'' of 0.01, we infer from their results
values of $\sim 2 \times 10^3\ {\rm K}$ and $\sim 7 \times 10^{15}\ 
{\rm cm}^{-3}$ for the midplane temperature and particle density, respectively,
at $r_{\rm co}$. For these values, we get $\eta \approx 7 \times 10^{10} \
{\rm cm}\ {\rm s}^{-2}$, which is 5 orders of magnitude smaller than the 
nominal maximum turbulent diffusivity $c_{\rm s} H$. Although the estimated
diffusivity increases at larger radii, we consider the region interior to
$r_{\rm co}$ to be particularly relevant since the disk must extend to
$r \le r_{\rm co}$ if matter is to be accreted onto the star. Note also that,
as $r$ increases above $r_{\rm co}$, $|\Delta\Omega| \rightarrow \Omega_*$ and
$\Sigma_{\rm max}$ itself decreases with $r$ ($\propto 1/r$). We conclude that
the result (3.9) is likely to represent a lower bound on the ratio $\Sigma/
\Sigma_{\rm max}$ in many practical applications.

The preceding discussion indicates that it is unlikely that a disk with a
dipole-like field configuration will achieve a steady state. It can, however,
be seen from Table 1 that $h_{\rm d, max} \simeq 1/n$ for $n \le 1$.
(Note in this connection that BH demonstrate that $h_{\rm d} \rightarrow 1/n$
as $n \rightarrow 0$.) Therefore, for sufficiently small values of $n$, 
$\Sigma_{\rm max}$ could be large enough for a steady state to be attainable
even for realistic values of $\Sigma$. In \S~\ref{subsec-resistive-radial} 
we argue that small effective values of $n$ are also needed to achieve a 
steady state (at least in the time-averaged sense) if variations in the 
disk radial (and not only azimuthal) field component are taken into account.

\subsection{Radial Flux Diffusion}
\label{subsec-resistive-radial}

We now consider the role of radial resistive diffusion.
So far we have assumed that the radial distribution of 
flux in the disk does not change. However, since we include 
resistive diffusion in the azimuthal direction, we need for 
consistency to also consider diffusion in the radial direction 
and its effect on~$B_{{\rm d},z}(r)$.

If the radial motion of the disk matter is vanishingly small, 
then the radial speed of the magnetic field footpoints in their 
resistive diffusion across the disk can be written as~$v_{{\rm B},r} 
=cE_{{\rm d},\phi}/B_{{\rm d},z}=cK_{\phi}/\Sigma B_{{\rm d},z}$, 
where~$E_{{\rm d},\phi}$ is the azimuthal electric field at the disk
surface and $K_{\phi} \approx 2 H j_{\phi}(z=0)$ is the vertically 
integrated azimuthal current density (cf. eq. [3.2]). Setting $K_{\phi} 
= c B_{{\rm d},r}/2\pi$, where $B_{{\rm d},r}= B_r(r,\pi/2)$ is the 
radial magnetic field at the disk surface (cf. eq. [3.1]), one gets
$$ v_{{\rm B},r}(r,t) = {c^2 \over 2\pi\Sigma} 
{{B_{{\rm d},r}}\over{B_{{\rm d},z}}} =
{c^2 \over 2\pi\Sigma} f(\pi/2,a_0(t))\, .          \eqno{(3.10)} $$

(It is interesting to note that in the case of an effective turbulent 
conductivity described by equation~(3.8), we have $v_{{\rm B},r}(r,t)=
\beta c_s f[\pi/2,a_0(t)]$, i.e., the footpoints move radially across 
the disk with the speed of order of the sound speed!)

We have already noted in \S~\ref{subsec-singularity} that 
$B_{{\rm d},r}/B_{{\rm d},z}$ diverges as the twist angle 
approaches its critical value $\Delta\Phi_{\rm c}$. This can be 
formally demonstrated by noting that, in this limit, $a_0\rightarrow 0$ 
(see \S~\ref{subsec-sequence}), and by using equations~(2.5) and~(2.20), 
which imply $f(\pi/2)=g'(\pi/2)/n\rightarrow a_0^{-n}G_0^{\prime}(\pi/2)/n$ 
as~$a_0\rightarrow 0$. But~$G_0^{\prime}(\pi/2)$ is finite and independent 
of~$a_0$ (see Fig. 3), so it follows that 
$$f(\pi/2,a_0) \propto a_0^{-n}\, , \qquad a_0\rightarrow 0\, .\eqno{(3.11)} $$
Thus $f[\pi/2,a_0(t)]$, and hence $v_{\rm B,r}(r,t)$, diverge as $\Delta\Phi
\rightarrow \Delta\Phi_{\rm c}$.

Now, our results in \S~\ref{subsec-singularity} for the finite-time 
singularity were based on the assumption that the footpoints of the 
magnetic flux tubes are fixed in the disk throughout the field-line 
twisting process. In light of the fact that the radial diffusion 
speed diverges as the critical angle is approached, this assumption 
is not strictly valid, and in reality the footpoints will undergo 
some radial excursion $\Delta r$ during the twisting time~$t_{\rm c}$ 
from $\Delta\Phi=0$ to~$\Delta\Phi=\Delta\Phi_{\rm c}$. If~$\Delta r$ 
remains much smaller than~$r$ then our results will not change much. 
However, if $\Delta r/r \ga 1$, then our conclusions about the approach 
to a singularity will need to be reexamined.

We can obtain a conservative estimate of $\Delta r$ by working in the 
high-conductivity limit ($\Sigma \gg c^2/\Delta\Omega r$, corresponding to a 
diffusivity $\eta \ll \Delta\Omega H r$). In this limit we neglect 
resistive slippage in the azimuthal direction and
use equation~(2.3); this approximation becomes even better justified near 
the critical point, where $B_{{\rm d},\phi}\rightarrow 0$ (see eq. [3.3]).
Assuming that the surface conductivity $\Sigma$ does not change with time, the
total radial footpoint displacement over the time interval $[0,t_{\rm c}]$ is
$$ \Delta r={c^2 \over 2\pi\Sigma} \int_0^{t_{\rm c}} f(\pi/2,a_0(t)) dt=  $$
$$ {c^2 \over 2\pi\Sigma \Delta\Omega} \int_0^{\Delta \Phi_{\rm c}} 
f(\pi/2,a_0(\Delta \Phi)) d\Delta\Phi =                                    $$
$$ {c^2 \over 2\pi\Sigma\Delta\Omega}\int_{\Delta\Phi=0}^{\Delta\Phi_{\rm c}} 
\left({{d\Delta\Phi}\over{d a_0}}\right) f(\pi/2,a_0) da_0\, .  \eqno{(3.12)}$$

In the special case when $\Sigma \propto 1/r$, resistive slippage 
does not break the self-similarity assumption. Indeed, in this case
the radial displacement at any given radius scales as~$r$, with the 
coefficient of proportionality increasing with twist angle but having 
the same value at all locations. Therefore, only the overall magnitude 
of the flux distribution is affected by the footpoint migration, but 
its self-similar scaling (and the value of the power-law index~$n$) 
remain unchanged.

We have already obtained the asymptotic behavior of~$f(\pi/2)$ 
in the limit of small~$a_0$ as $\Delta\Phi \rightarrow \Delta\Phi_{\rm c}$ 
(see eq.~[3.11]). The calculation of $(d\Delta\Phi/da_0)$ in this limit is 
more cumbersome and is given in Appendix~A. On the basis of equation~(A11) 
from that Appendix, we can write 
$$\Delta \Phi \simeq \Delta\Phi_{\rm c} + \xi a_0^n\, , 
\qquad a_0\rightarrow 0\, ,                                   \eqno{(3.13)} $$
which implies
$$ {{d\Delta\Phi}\over{d a_0}} \propto a_0^{n-1}, \qquad 
a_0 \rightarrow 0\, .                                          \eqno{(3.14)} $$
Using equations (3.11) and (3.14) in equation (3.12), we obtain
$$\int^{\Delta\Phi_{\rm c}} f(\pi/2,a_0(\Delta \Phi)) d\Delta\Phi \sim
\int^{a_0=0} \left({{d\Delta\Phi}\over{da_0}}\right) f(\pi/2,a_0) da_0 \sim $$
$$ \int^{a_0=0} a_0^{-n} a_0^{n-1} da_0 \sim \log{a_0}\, ,     \eqno{(3.15)} $$
which diverges logarithmically as $a_0\rightarrow 0$. 

The result~(3.15) demonstrates that, even if the conductivity is 
relatively large, one cannot neglect the radial field diffusivity 
in the disk as the critical point is approached. As $\Delta \Phi 
\rightarrow \Delta\Phi_{\rm c}$, the radial displacement of the magnetic 
footpoints gets progressively larger and eventually reaches a level 
where the approximation $B_{{\rm d},z}(r) \approx {\rm const}$ that 
was used in the derivation of equation (3.3) becomes inadequate. The 
outward motion of the footpoints would have the effect of decreasing 
$B_{{\rm d},r}$ and preventing the surface radial field component from 
blowing up. It can be argued, however, that this does {\em not} mean 
that the finite-time singularity discussed in \S~\ref{subsec-singularity} 
would be avoided, although its onset would be delayed somewhat. The reason 
is that, as $t \rightarrow t_{\rm c}$, $\Delta r$ only scales logarithmically 
with $a_0(t)$, but the radius $r(r_0,\theta_{\rm ap})$ of the apex 
point of a field line that remains anchored in the disk at some given 
value of $r_0$ increases as $1/a_0(t)$ (see eq. [4.28] below). [In that 
limit, $a_0(t)$ decreases with time as $(t_{\rm c}-t)^{1/n}$; see eq.~(A11) in 
Appendix~A.] This implies that the twisted field lines will expand much 
more rapidly in the magnetosphere than their footpoints will migrate 
inside the disk, so the conclusion that (barring inertial effects) they 
open in a finite time should continue to hold. Although inertial effects 
will intervene to reduce the field-line expansion speed in the magnetosphere 
to roughly the local Alfv\'en speed (see \S~\ref{subsec-inertial}), the 
latter will still be much higher than the field diffusion speed in 
the disk (which, for a turbulent diffusivity, is of the order of 
$v_{\rm turb}={\rm max}\{c_{\rm s}\, , \, v_{\rm A}\}$; see \S~\ref
{subsec-resistive-toroidal}). The effect of radial field diffusivity 
will be even less of an issue if reconnection effects in the magnetosphere 
terminate the field-line expansion before the critical twist angle 
is attained (see \S~\ref{subsec-reconnection}), since the value of 
$B_{{\rm d},r}$ in this case will remain bounded.

In one proposed scenario (see \S~\ref{sec-intro}), the twisted magnetic 
field lines reconnect as $\Delta\Phi$ approaches the critical value, and 
the process of twisting, expansion, and reconnection recurs in a periodic 
fashion. If this indeed is what happens, then it is interesting to inquire 
whether the system can maintain a {\em time-averaged} steady state despite 
the expected radial diffusion of the field lines. VB, who first addressed 
this question, proposed that this could happen if $n<1$, since, in that 
case, the radial magnetic field component at the disk surface (and hence 
$\Delta r$) changes sign in the course of the field-line evolution (see 
Fig.~\ref{fig-contour}). Therefore, depending on the value of the twist 
angle at which reconnection occurs, there will be a value of $n\in (0,1)$ 
for which $B_{{\rm d},r}$ averages to zero between the start of the twisting 
cycle (when the field is potential) and the instant of reconnection. 
An alternative possibility, which also requires $n<1$, was proposed 
by BH for an {\em exact} steady state. It is based on the fact that 
a self-similar field configuration corresponding to $0<n<1$ attains 
$B_{{\rm d},r}=0$ at some twist angle $\Delta\Phi_0$ that is generally 
less than $\Delta\Phi_{\rm max}$ (see Table 1). The possibility then 
arises that $\Delta\Phi_0$ is equal to $\Delta\Phi_{\rm ss}$ (which, 
according to the discussion in \S~\ref{subsec-resistive-toroidal}, is 
also generally less than $\Delta\Phi_{\rm max}$), so that a genuine 
steady state with $B_{{\rm d},r}=0$ (and with the twisting due to the 
differential rotation balanced by the toroidal resistive slippage) is 
established. However, for a given~$n$, this is only possible for some 
special value of~$\Sigma$, since $\Delta\Phi_{ss}$ is directly related 
to~$\Sigma$. For $n \la 1$, the required value of $\Sigma$
is unrealistically large (of order~$c^2/r\Delta\Omega$; see 
\S~\ref{subsec-resistive-toroidal}).

Both the VB and the BH proposals could probably be realized only in 
systems in which $n\ll 1$. In the case of the VB picture, this follows 
from the expectation that the twist angle at which reconnection occurs 
is quite close to $\Delta\Phi_{\rm c}$ (see \S~\ref{subsec-reconnection}), 
whereas in the case of the BH suggestion it is a consequence of the 
fact that, for characteristic values of $\Sigma$, a steady state will 
typically not be established unless $n$ is very small. We defer
until \S\ref{sec-discussion} a discussion of the plausibility of
real astrophysical systems being represented by such low-$n$ configurations.

\section{Effects of Plasma Inertia and Reconnection in the Magnetosphere}
\label{sec-magnetosphere}

As we noted in \S~\ref{sec-intro}, magnetic field reconnection across 
the current concentration region that forms when a twisted flux tube 
begins to expand is a possible alternative to the opening of the field 
lines. Whether reconnection can compete with the opening process depends 
in large measure on the magnitude of the plasma resistivity in the region 
of current concentration. Classical collisional resistivity is generally 
too small to play a role, and one appeals to anomalous resistivity induced 
by current-driven microinstabilities. The instability criterion is that the 
current density exceed a certain threshold, which is typically proportional 
to the electron density~$n_e$. 
This suggests that the relevant quantity to 
examine is the ratio of the current density $j_{\phi}$ to~$n_e$ (or~$\rho$). 
Anomalous resistivity could thus be triggered not only by the current 
concentration near $\theta_{\rm ap}$ but also by the drop in density 
brought about by the rapid expansion of the toroidal field as the critical 
twist angle is approached (see \S~\ref{subsec-singularity}).%
\footnote 
{It is very tempting to suggest that a similar mechanism (i.e., 
the triggering of anomalous resistivity through a rapid density 
decrease due to the expansion of magnetic field lines) may also 
work for coronal mass ejections in the sun. However, this does 
not seem to be likely, since the density in the solar corona is 
relatively large.}

In this section we explore this possibility by calculating the velocity 
and density evolution outside the disk using the sequence of force-free
equilibria presented in \S~\ref{sec-model}. As part of this analysis we 
also evaluate the inertial effects in the expanding magnetosphere, which 
must remain small for the force-free equilibrium approximation to be 
applicable. In order to isolate the resistive effects in the magnetosphere 
and to distinguish them from the corresponding processes in the disk (which 
were considered in \S~\ref{sec-resistive}), we assume (as in 
\S~\ref{sec-model}) that the magnetic field remains frozen into 
the disk material at all times.

\subsection{Magnetospheric Velocity and Density Evolution}
\label{subsec-velocity}

We assume that the magnetosphere remains force-free throughout its 
evolution, with the plasma on the whole remaining a good conductor 
and responding passively to the motion of the field lines. Subsequently, 
in \S~\ref{subsec-inertial}, we address the question of when inertial 
effects become important. Then, in \S~\ref{subsec-reconnection}, we
address the issue of reconnection.

We determine the rate of change of the density from the mass continuity 
relation, which requires us to first obtain the velocity field. Writing the 
total velocity as the sum of the parallel and the perpendicular components,
$$ {\bf v} = {\bf v_{\parallel}} + {\bf v_{\perp}} = 
v_{\parallel}{\bf{\hat b}} + {\bf v_{\perp}}                  \eqno{(4.1)} $$
(where ${\bf{\hat b}}$ is the unit vector along the magnetic field), we first
calculate the perpendicular velocity~${\bf v_{\perp}}$. We do this by applying
the ideal MHD flux-freezing condition to the given magnetic field configuration
${\bf B}({\bf r},t)$. [We do not use the equation of motion for 
${\bf v_{\perp}}$ since the latter has already been employed (in lowest order,
neglecting inertial and pressure terms) to obtain the underlying sequence of
force-free equilibria.]

Consider a surface of constant $\theta (=\theta_0)$ and a field line 
that at some moment of time $t$ intersects this surface at a point
with coordinates ($r[\theta_0,t],\theta_0,\phi[\theta_0,t]$). After
some infinitesimal time interval $dt$ the same field line intersects
the surface $\theta=\theta_0$ at a new point ($r(\theta_0,t+dt),\theta_0,
\phi(\theta_0,t+dt)$). We can thus define the vector ${\bf u}(\theta_0,t)$
as representing the rate of motion of the point of intersection of the 
given field line with the surface $\theta={\rm const}$:
$$ {\bf u} = (\dot{r}(\theta,t), 0, \dot{\phi}r\sin\theta)\, ,  \eqno{(4.2)} $$
where the dot symbol denotes an explicit time derivative ($\partial/
\partial t$). By the flux freezing condition,
${\bf v_\perp}={\bf u_\perp}$. Using equation~(2.7) to express 
$\dot{r}(\theta,t)$ in terms of $\dot{g}(\theta,t)$, one thus gets
$$ v_{\perp r}={{r\dot{g}}\over{ng}} \left(1-{{f^2\sin^2\theta}\over{P^2}}
\right) -r\dot{\phi} {{fh\sin^3\theta}\over{P^2}}\, ,        \eqno{(4.3)} $$
$$ v_{\perp\theta} = - {{r\dot{g}f\sin\theta}\over{nP^2}}-
r\dot{\phi} {{gh\sin^2\theta}\over{P^2}}\, ,                    \eqno{(4.4)} $$
$$ v_{\perp\phi} = -{{r\dot{g}fh\sin^2{\theta}}\over{ngP^2}} + r\dot{\phi}
\sin\theta\left(1-{{h^2\sin^2\theta}\over{P^2}}\right)\, ,      \eqno{(4.5)} $$
where 
$$ P^2(\theta) \equiv \left({{\vert{\bf B}\vert r^{n+2}}\over C}\right)^2
\sin^2{\theta}=f^2\sin^2{\theta}+g^2(\theta)+h^2\sin^2{\theta}\, . 
\eqno{(4.6)} $$

We now proceed to calculate the parallel component of the 
velocity,~${\bf v_{\parallel}}$. Unlike the perpendicular 
component, $v_{\parallel}$ is determined from the equation 
of motion, whose component along ${\bf{\hat b}}$ is given by 
$$ \left({{d{\bf v}}\over{dt}}\right)_{\parallel}=
{d\over{dt}} v_{\parallel}-{\bf v}_{\perp}\cdot{{d{\bf{\hat b}}}\over{dt}}=
F_{\parallel}\, ,                                               \eqno{(4.7)} $$
where $F_{\parallel}$ is the acceleration due to the sum of the 
parallel projections of all forces (including, in general, the 
centrifugal, Coriolis, gravitational, and pressure forces). 
Equation~(4.7) implies that
$$ \dot{v}_\parallel=-({\bf v}\cdot\nabla)v_\parallel+
{\bf v}_{\perp}\cdot{{d{\bf{\hat b}}}\over{dt}}+F_\parallel\, . \eqno{(4.8)} $$

We first calculate the parallel projections of the centrifugal 
and Coriolis forces. (These two inertial forces appear because we 
are working in a frame  of reference that rotates with the angular velocity
$\Omega_*$ of the star.) We have
$$ F_{\rm cent \parallel} = {\bf F}_{\rm cent}\cdot {\bf \hat{b}} =
{{\Omega_*^2 r}\over P}\sin\theta [f(\theta)\sin^2\theta+
g(\theta)\cos\theta]                                        \eqno{(4.9)} $$
and
$$ F_{\rm cor \parallel}=2[{\bf v\times\Omega_*}]\cdot\hat{\bf b}=
2[{\bf v_\perp \times\Omega_*}]\cdot\hat{\bf b}=                            $$
$$ {{2\Omega_*}\over P} \left[v_{\perp\phi}(f\sin^2\theta+g\cos\theta)
-h\sin\theta(v_{\perp r}\sin\theta + v_{\perp\theta}\cos\theta) \right]\, .
                                                               \eqno{(4.10)} $$

Next, we note that the gravitational force scales with distance as~$r^{-2}$, 
whereas the centrifugal and Coriolis forces are proportional to~$r$. In 
this analysis we limit ourselves to large distances ($r\gg r_{\rm co}$) 
in order for our self-similar model to be valid. Then we can neglect the 
gravitational force. We also neglect any pressure forces. Thus we get
$$ F_\parallel=F_{\rm cent\parallel} + F_{\rm cor \parallel}.\eqno{(4.11)} $$

Now we are ready to calculate~$v_\parallel$.
Using the fact that, due to self-similarity, $v_\parallel\propto r$, 
so that $\partial v_\parallel/\partial r = v_\parallel/r$, we 
can rewrite equation~(4.8) as 
$$ \dot{v_\parallel} = - {{v_r v_\parallel}\over r}- 
v_\theta{{\partial v_\parallel}\over{r \partial \theta}}+
{\bf v}_\perp \cdot {{\partial{\bf\hat{b}}}\over{\partial t}}+ 
{\bf v_\perp\cdot (v\cdot\nabla)\hat{b}}+F_\parallel\, .       \eqno{(4.12)} $$

We need to derive the expressions for the two terms involving
the change in the unit vector~${\bf\hat{b}}$. First, using ${\bf 
v_\perp}\cdot {\bf\hat{b}}=0$, we rewrite ${\bf v}_\perp \cdot 
\partial{\bf\hat{b}}/\partial t$ in terms of the known functions 
$f(\theta,t)$, $g(\theta,t)$, and~$h(\theta,t)$ as
$$ {\bf v_\perp} \cdot {{\partial{\bf\hat{b}}}\over{\partial t}} =
{1\over P}\left(v_{\perp r} \dot{f} \sin{\theta}+
v_{\perp\theta}\dot{g}+v_{\perp\phi}\dot{h}\sin{\theta}\right).\eqno{(4.13)} $$

Similarly, after some algebra one obtains
$$ {\bf v_\perp \cdot (v\cdot\nabla)\hat{b}} =
{v_\theta\over{Pr}} \left[ v_{\perp r} (g''/n-g)+
\left(v_{\perp\theta}(n+1)+v_{\perp\phi}a_0 g^{1/n}\right)f\sin\theta\right]+$$
$${v_\phi\over{Pr}}\left[-v_{\perp r}h\sin{\theta}-v_{\perp\theta}h\cos\theta+
v_{\perp\phi}\left(f\sin\theta+g\cot\theta\right)\right].      \eqno{(4.14)} $$

Equation~(4.12), supplemented by equations~(4.9)--(4.11), (4.13), and (4.14),
is the partial differential equation that determines~$v_\parallel(\theta,t)$. 
We solve it numerically by the finite differences method, assuming 
the following boundary conditions at $\theta=0$ and~$\theta=\pi/2$:
$$ {{\partial v_\parallel}\over{\partial\theta}}(0,t) = 0, \qquad  
{{\partial v_\parallel}\over{\partial\theta}}(\pi/2,t) = 0,    \eqno{(4.15)} $$
and the initial condition at $t=0$ (corresponding to 
$\Delta\Phi=0$):
$$ v_\parallel (\theta, 0) =0\, .                              \eqno{(4.16)} $$

Once the velocity field is known for all times, 
the density evolution can be determined 
from the continuity equation,
$$ \dot{\rho} = -\nabla\cdot(\rho{\bf v}) = 
-{1\over{r^2}} {\partial\over{\partial r}} (r^2 \rho v_r) -
{1\over{r\sin\theta}} {\partial\over{\partial\theta}}
(\sin\theta \rho v_\theta)\, .                                \eqno{(4.17)} $$
It is easy to see that, on account of
the self-similarity of the velocity field, the 
evolution equation~(4.17) preserves any power-law 
dependence of~$\rho$ on~$r$. Thus, if we assume 
that $\rho(r,t=0) \propto r^{-p}$, we can write
$$ \rho(r,\theta,t)=r^{-p} \tilde{\rho}(\theta,t)\, .          \eqno{(4.18)} $$
The continuity equation~(4.17) can then be expressed as
$$ \dot{\tilde{\rho}} = - (3-p) \tilde\rho {v_r\over r} -
{1\over{r\sin\theta}} {\partial\over{\partial\theta}}
(\sin\theta \tilde{\rho} v_\theta)\, .                         \eqno{(4.19)} $$
We solve this partial differential equation numerically
with the initial condition $\tilde\rho(\theta,0)=1$, and 
the boundary conditions
$$ {{\partial\tilde{\rho}}\over{\partial\theta}}(0,t)=0\, , \qquad
\tilde{\rho}(\pi/2,t) = 1\, .                                  \eqno{(4.20)} $$

\begin{figure} [tbp]
\centerline{\psfig{file=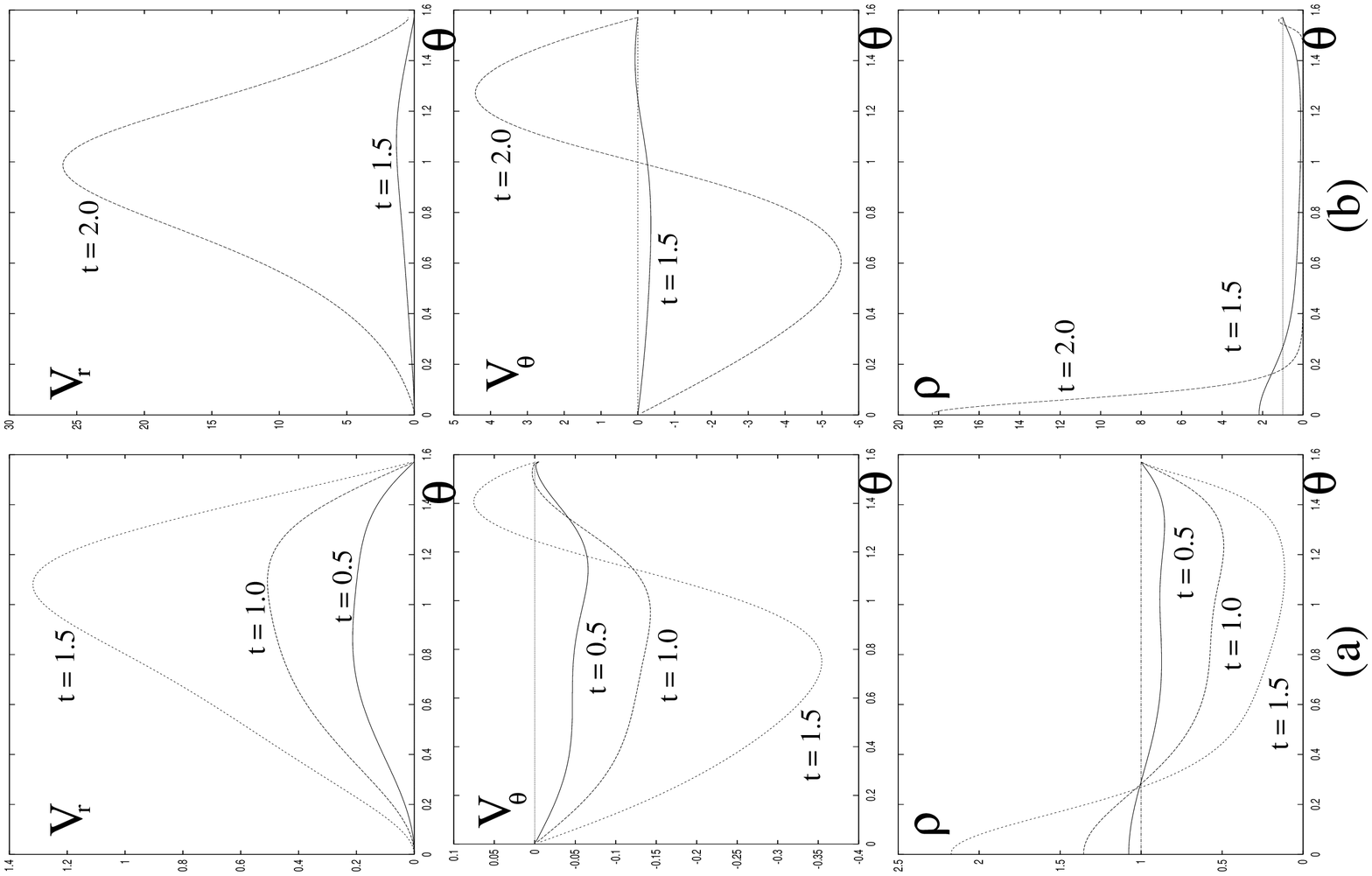,width=6in,height=6.5in,angle=-90}}
\caption {Radial and $\theta$ velocity components and the plasma
density as functions of $\theta$ at fixed radius at: 
{\bf (a)} $\Delta\Phi=0.5$, 1.0, and 1.5, 
and {\bf (b)} $\Delta\Phi=1.5$, 2.0. 
These plots are for the $n=1$ self-similar field configuration with
a uniform initial density distribution ($p=0$). Time is measured in 
units of $1/\Delta\Omega$: in these units the critical twist angle 
$\Delta\Phi_{\rm c}$ is attained at $t_{\rm c} = 2.036$.
\label{fig-velocity}}
\end{figure}

Figure~\ref{fig-velocity} presents the results of our numerical 
computations for the case $n=1$, $p=0$, and~$\Omega_*=-1.0$ (we 
choose the units of time so that $\Delta\Omega = 1$; then our
choice of~$\Omega_*$ corresponds to a star rotating with angular 
velocity equal to~1, while the disk is not rotating). It is seen 
that both the~$r$ and the $\theta$~components of the velocity 
vary smoothly as functions of~$\theta$, and that they grow rapidly
as~$\Delta\Phi_{\rm c}$ is approached. The radial velocity component has 
a maximum at the apex point $\theta_{\rm ap}\simeq 0.98$, whereas 
$v_{\theta}$ changes sign there.%
\footnote {Note that $\theta_{\rm ap}$ changes slowly with 
time, starting at $\theta_{\rm ap}=\pi/2$ at $t=0$ and then
gradually approaching its asymptotic value~$\theta_{\rm ap}(t_{\rm c})$,
which depends only weakly on~$n$. Generally, $\theta_{\rm ap}(t_{\rm c})$ 
is very close to one radian. Since in this section we are particularly
interested in the behavior near~$t=t_{\rm c}$, we consistently evaluate 
$\theta_{\rm ap}$ at that point.}
The plasma density around $\theta_{\rm ap}$ decreases rapidly 
with~$\theta$, which corresponds to a significant amount of 
plasma being moved toward the symmetry axis, where it is 
concentrated in a narrow sector ($\Delta \theta$ of a few 
degrees). As we discuss in \S~\ref{sec-discussion}, this 
mass concentration may be relevant to the formation of 
stellar jets.

\begin{figure} [tbp]
\centerline{\psfig{file=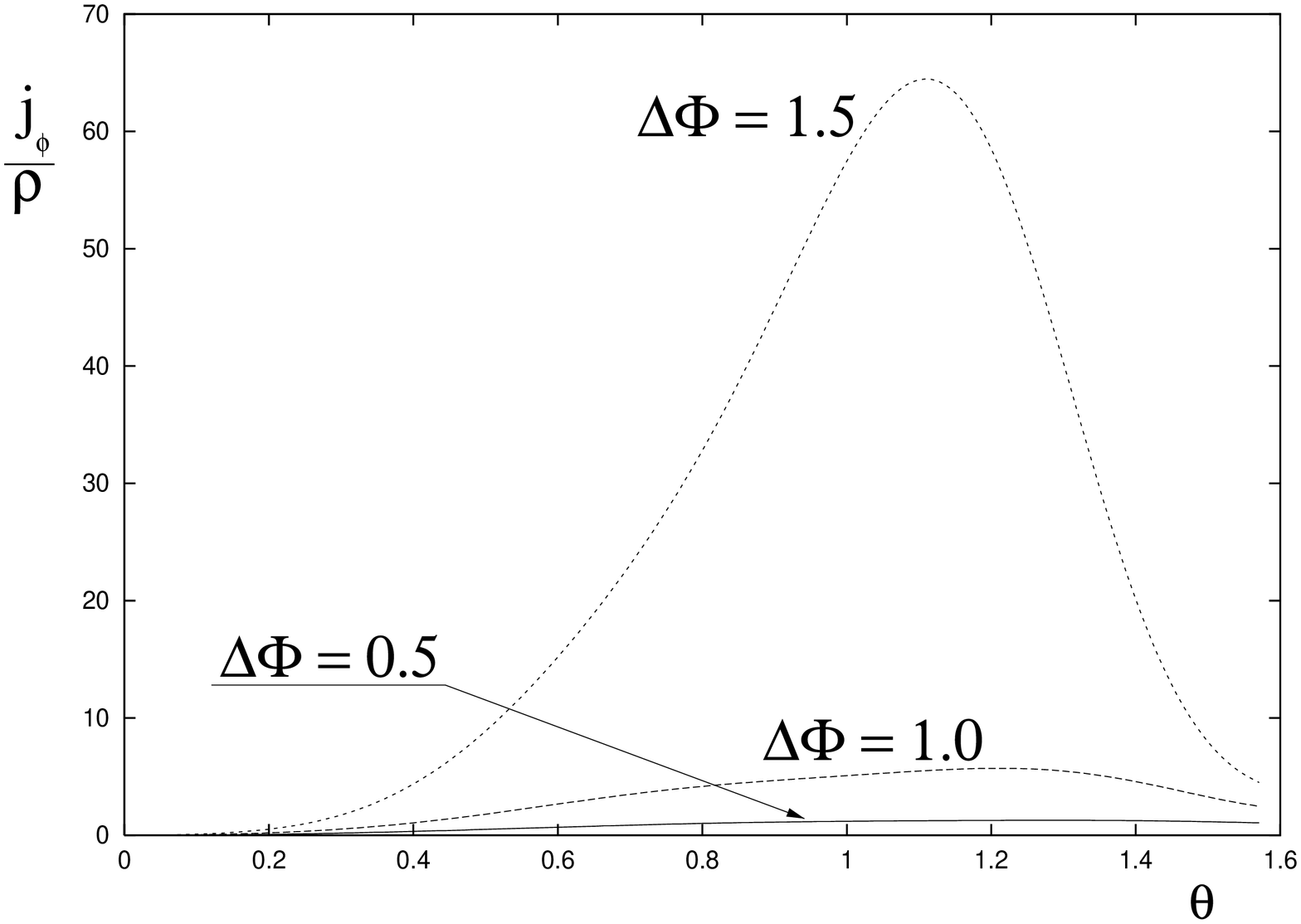,height=3in,width=5in}}
\caption {The ratio of the current density~$j_{\phi}$ to the plasma 
density~$\rho$ (in arbitrary units) as a function of $\theta$ for 
$\Delta\Phi=0.5$, 1.0, and 1.5 in the $n=1$ self-similar model.
\label{fig-j-over-rho}}
\end{figure}

Figure~\ref{fig-j-over-rho} shows the ratio of the current
density to the mass density, which we consider to be a field
reconnection diagnostic, as a function of~$\theta$. It is seen 
that, as the field lines expand, this ratio increases rapidly 
near~$\theta=\theta_{\rm ap}$, which suggests that the threshold 
for triggering anomalous resistivity (and, thereby, reconnection)
could be exceeded. However, in order to reach a definitive conclusion, 
one needs to find out whether the threshold is reached before inertial 
effects become important. The answer to this question requires a more 
detailed knowledge of the behavior of the velocity and density at the 
apex in the limit~$t\rightarrow t_{\rm c}$. We therefore perform an 
asymptotic analysis of the magnetospheric velocity field in this limit.

The details of this analysis are given in Appendix B.
Here we only summarize the main results by giving the 
asymptotic expressions for the velocity:
$$ v_{\perp r}(r,\theta_{\rm ap},t) \simeq
-{r\over n}\ {1\over{t-t_{\rm c}}}\, ,                        \eqno{(4.21)} $$
$$ v_{\perp\theta}(r,\theta\simeq\theta_{\rm ap},t) \simeq -r 
{{n+1}\over n}{{\theta-\theta_{\rm ap}}\over{t-t_{\rm c}}}\, , \eqno{(4.22)} $$
$$ v_{\perp\phi}(r,\theta\simeq\theta_{\rm ap},t) \simeq 
-r {{G_0^{1/n}}\over n} {{\theta-\theta_{\rm ap}}\over{t-t_{\rm c}}}\, ,
                                                             \eqno{(4.23)} $$
and
$$ v_\parallel(r,\theta,t) \simeq
r (\theta-\theta_{\rm ap}) {A\over{t-t_{\rm c}}}\, ,          \eqno{(4.24)} $$
where 
$$ A={\kappa\over{2n}}\left(3n+4\pm\sqrt{5n^2+12n+8}\right)   \eqno{(4.25)} $$
and $\kappa\equiv\sqrt{1+G_0^{2/n}(\theta_{\rm ap})/(n+1)^2}$.
For example, for $n=1$ we have $A_1=\kappa\simeq 2$ and 
$A_2=6\kappa\simeq 12$. 

We can now proceed to calculate the behavior
of the density at $\theta_{\rm ap}$ as $t\rightarrow t_{\rm c}$.
Substituting the expressions for ${\bf v}_\perp$ and $v_\parallel$
into equation~(4.19), we readily obtain
$$ \rho \propto r^{-p} (t_{\rm c}-t)^q\, ,                  \eqno{(4.26)} $$
where 
$$ q(n) \equiv {{3-p}\over n}+{{n+1}\over n}-{A\over\kappa}\, .\eqno{(4.27)} $$

Table~\ref{table-apex} lists the values 
of~$\theta_{\rm ap}$, $G_0(\theta_{\rm ap})$, $\kappa$, 
$A$, and $q$ for two representative values of~$n$. Note 
that because we have a quadratic equation for~$A$, two 
roots ($A_1$ and $A_2$) are possible for each~$n$, giving 
rise to two different values ($q_1$ and $q_2$) of the density 
power exponent~$q$.
It is not a priori clear how to choose the physically relevant
value for~$A$, and hence for~$q$. However, based on the good agreement 
between the value of $q_1$ and the result of the full numerical solution 
for the case $n=1$ (see Fig. 9), we conjecture that the appropriate root 
is generally the lower one, $A_1$.

\begin{table}[tbp]
\caption{ASYMPTOTIC VELOCITY AND DENSITY PARAMETERS AT THE APEX ANGLE}
\vskip 10 pt
\begin{tabular}{|@{\hspace{1cm}}c@{\hspace{1cm}}| 
*{7}{@{\hspace{0.3cm}}c@{\hspace{0.3cm}}}|}
\hline \hline
$n$  & $\theta_{\rm ap}$ & $G_0(\theta_{\rm ap})$ & 
$\kappa$ & $A_1$ & $A_2$ & $q_1$ & $q_2$ \\ 
\hline
1.0 & 0.976 & 3.45 & 1.99 & 1.99 & 11.96 &  4-p &-1-p \\  
0.5 & 0.995 & 2.37 & 3.87 & 6.17 & 36.4 & 7.4-2p &-0.4-2p\\
\hline \hline
\end{tabular}
\label{table-apex}
\end{table}

\begin{figure} [tbp]
\centerline {\psfig{file=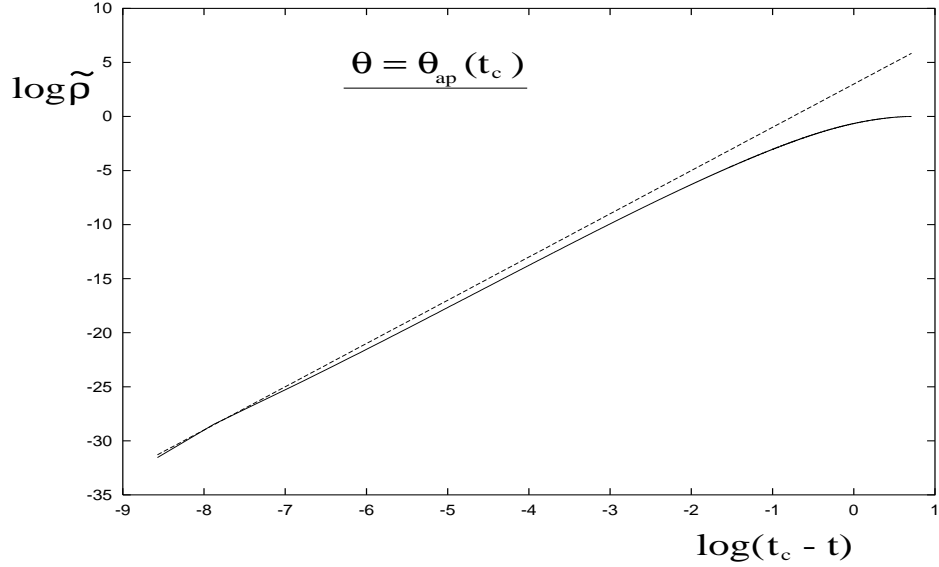,height=3 in,width=5 in}}
\caption {Log-log plot of $\tilde{\rho}(\theta_{\rm ap}(t_{\rm c}),t)$ 
near~$t=t_{\rm c}$ for the~$n=1$, $p=0$ self-similar solution. The density 
is plotted in arbitrary units and the time in units of~$1/\Delta\Omega$. 
The dashed line represents the asymptotic solution $\tilde{\rho}
(\theta_{\rm ap},t) \propto (t_c-t)^4$ derived in the text.
\label{fig-logrho-logtime}}
\end{figure}

\begin{figure} [tbp]
\centerline {\psfig{file=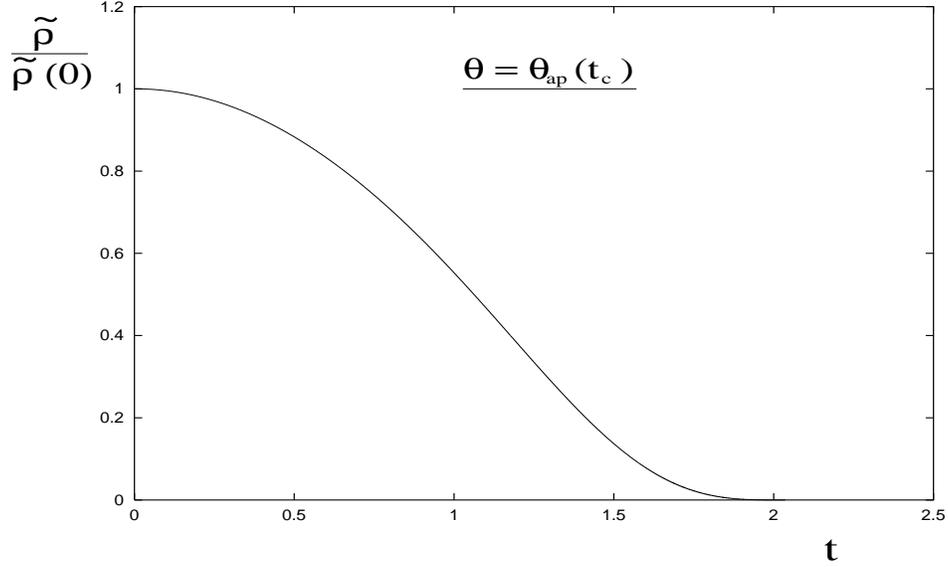,height=3in,width=5in}}
\caption {Normalized plasma density at the final apex point 
$\theta_{\rm ap}(t_{\rm c})$ as a function of time (in units 
of $1/\Delta\Omega$) for~$n=1$, $p=0$.
\label{fig-rho-time}}
\end{figure}

Figure~\ref{fig-logrho-logtime} details the comparison between
the predictions of the asymptotic analysis and the results of 
the direct solution of equation~(4.19) for the case~$p=0$. It 
is seen that a reasonably good agreement is achieved when $t$ 
approaches $t_{\rm c}$ to within $\sim 0.1/\Delta\Omega$ or so. 
Furthermore, by examining the plot in Figure~\ref{fig-rho-time} 
of the entire time evolution of~$\tilde{\rho}(\theta_{\rm ap}
(t_{\rm c}),t)$, one sees that there is a gradual transition 
in the interval $t \simeq 1.6-2.0\, \Delta\Omega^{-1}$ from the
early linear dependence on~$t$ to the behavior [$\propto (t_c-t)^4$] 
predicted by equation~(4.26). We note that a high value of~$q$ (as
chosen in this figure) results in $\tilde{\rho}(\theta_{\rm ap})$ 
becoming quite small well before the singularity. It is also interesting 
to note that the drop in density near $t=t_{\rm c}$ is, in fact, faster 
than that corresponding to uniform expansion [$q_{\rm uniform}=(3-p)/n$]. 
This is due to the $\theta$-divergence of the flow near $\theta=
\theta_{\rm ap}$.

If, instead of fixing~$r$, a given field line~$\Psi$ is considered, 
one can use equations~(2.7) and~(B4) to obtain
$$ r(\Psi,\theta_{\rm ap},t)=r_0(\Psi) a_0^{-1} G_0^{1/n}(\theta_{\rm ap}) 
\propto r_0(\Psi) \Delta\Omega^{-1/n} (t_{\rm c}-t)^{-1/n}\, , \eqno{(4.28)} $$
and therefore, by equation (4.26),
$$ \rho(\Psi,\theta_{\rm ap},t) \propto \rho_0(\Psi)\, \Delta\Omega^{q+p/n}
(t_{\rm c}-t)^{q+p/n}\propto (t_{\rm c}-t)^{1+4/n-A/\kappa}\, , \eqno{(4.29)}$$
where $\rho_0$ is some typical initial density (at $\Delta\Phi=0$) on this 
field line. 

Having found the asymptotic behavior of the plasma density, we 
can now address the issues of inertial effects and reconnection 
in the magnetosphere. We start with the role of inertial effects.

\subsection{Inertial Effects in the Magnetosphere}
\label{subsec-inertial}

As the field lines expand near the singular point and the plasma
velocity at the apex becomes larger and larger, there is a concern
that it will become large enough to be comparable with the local 
Alfv{\'e}n speed, thus invalidating the equilibrium assumption. 

In order to investigate this question, we first need to find the 
behavior of the Alfv{\'e}n speed on an expanding field line at 
$\theta=\theta_{\rm ap}$ as~$t \rightarrow t_{\rm c}$. Consider 
a field line~$\Psi$ intersecting the disk at the radius~$r_0(\Psi)$. 
Initially, at zero twist angle, the Alfv{\'e}n speed is large,
$$ v_{A,0} \equiv v_A(r_0,\theta_{\rm ap},\Delta\Phi=0) \sim
{C\over{r_0^{n+2}\sqrt{\rho_0}}} \gg \Delta\Omega \, r_0 \, ,  \eqno{(4.30)} $$
The inequality~(4.30) justifies our underlying assumption that the 
magnetic  field in the magnetosphere is in equilibrium. In fact, 
the ratio $\Delta\Omega r_0/v_{A,0}\ll 1$ plays the role of a small 
parameter in the asymptotic analysis of the inertial effects in the 
magnetosphere.

We now use equations~(B7) and (4.28)--(4.29) to find the behavior of 
the Alfv{\'e}n speed near the critical point,
$$ v_A(\Psi,\theta_{\rm ap},t) \sim 
v_{A,0} [\Delta\Omega(t_{\rm c}-t)]^{2/n-q/2-p/2n}  \sim 
v_{A,0} [\Delta\Omega(t_{\rm c}-t)]^{A/2\kappa -1/2}.       \eqno{(4.31)} $$
(Interestingly, for~$n=1$ one gets $A=\kappa$,
so~$v_A \rightarrow {\rm const} \sim v_{A,0}$ in this case.)

Then, using equations~(4.21) and~(4.28),
$$ {{v_r(\theta_{\rm ap},t)}\over{v_A(\theta_{\rm ap},t)}} |_\Psi \sim
{{r_0\Delta\Omega}\over{v_{A,0}}} [\Delta\Omega(t_{\rm c}-t)]^
{-({1\over 2}+{1\over n}+{A\over{2\kappa}})}.                  \eqno{(4.32)}$$

It is seen that in this case the power-law exponent is independent 
of~$p$ and is always negative (for example it is equal to --2 for 
$n=1$). This means that, sooner or later, the inertial effects on 
a given field line will become important. We can even estimate the 
time when this happens as
$$ \Delta\Omega(t_{\rm c}-t)_{\rm in} \sim 
\left({{r_0\Delta\Omega}\over{v_{A,0}}} \right)^
{{2n\over{n+2+nA/\kappa}}}.                                 \eqno{(4.33)} $$

Note that inertial effects may, in fact, become important
even earlier, but not near~$\theta_{\rm ap}$. Indeed, on 
a significant part of a field line, $\rho$ does not decrease 
as much as near~$\theta_{\rm ap}$, whereas the typical velocities
there are comparable to~$v_r(\theta_{\rm ap})$. However, 
we shall use the estimate~(4.33) anyway, because we are mostly 
interested in the neighborhood of~$\theta_{\rm ap}$.

The onset of inertial effects could in principle remove the
finite-time singularity induced by the twisting of the field
lines (see \S~\ref{subsec-singularity}). However, we expect that
in practical applications the effective opening of the field
lines will only be delayed by these effects, rather than
eliminated altogether (see \S~\ref{sec-discussion} for a fuller
discussion of this point). Furthermore, we anticipate that, even 
in the absence of inertial effects, the field-line expansion in real
systems will be terminated by reconnection --- although, as we
show in the next subsection, inertial effects could also delay 
the onset of the latter process.

\subsection{Reconnection in the Magnetosphere}
\label{subsec-reconnection}

In this subsection we address the prospects for reconnection 
in the disk magnetosphere. We first derive the relevant 
equations in the self-similar model framework, and then use them to
draw some quantitative conclusions using representative
parameters for YSO systems.

Reconnection may occur when the critical twist angle is
approached. It is important to recognize, however, that 
the twisted field configuration does {\em not} give rise 
to a fully developed current sheet. Formally, a current
sheet is characterized by an infinitesimal thickness in 
the limit $\eta \rightarrow 0$, whereas in our self-similar 
solution the current is concentrated in a region of finite 
angular extent about $\theta=\theta_{\rm ap}$ (see 
Fig.~\ref{fig-current}). The angular half-width of this 
region becomes very small only in the limit~$n\rightarrow 0$, 
but for $n \simeq 1$ it is of order $\sim 10^\circ$. Hence one 
can say that current is concentrated in a layer with an aspect 
ratio of order 10 or so; therefore, in order for reconnection 
to be of any significance, one needs a very low Lundquist number.
In the framework of the Sweet--Parker (Sweet 1958, Parker 1963) 
reconnection model, one would need
$S \equiv v_{\rm A} L/\eta \approx 100$ (where $L$ is the 
characteristic length of the current layer). 
As in almost any other astrophysical context,
the required value of~$S$ is too small to be 
accounted for by the classical Spitzer resistivity.

It can, however, be expected that near the critical point the
ideal MHD approximation will cease to be valid in the region 
of high current concentration. One interesting possibility is 
the development of an effective anomalous resistivity that is 
triggered by current-driven instabilities. For definiteness, 
we focus on the anomalous resistivity associated with the 
{\em ion-acoustic/Buneman instabilities} --- the mechanism 
most commonly considered in the literature (e.g., Coroniti 
\& Eviatar 1977; Galeev \& Sagdeev 1984; Parker 1994). Another 
possible mechanism for anomalous resistivity is lower-hybrid drift
turbulence initiated by a perpendicular 
current (see Krall \& Liewer 1971 and Drake et al. 1984, and 
references therein). However, lower-hybrid instability requires 
perpendicular electron drift, which is absent in the force-free 
configurations employed in our model.

A current-driven anomalous resistivity is triggered when the 
current density exceeds a certain critical value,
$$ j = e n_{\rm e} v_{\rm d} > j_{\rm c} = e n_{\rm e} v_{\rm c}\, ,  $$
where $v_{\rm d}$ is the drift speed of the current-carrying 
electrons, and the critical speed $v_{\rm c}$ is of the order 
of the thermal speed of either electrons or ions (depending on 
the nature of the instability) and is thus a function of the
temperature~$T$. The criterion for triggering reconnection can 
therefore be written as
$$ {j\over\rho} > \left({j\over\rho}\right)_{\rm c} (T) = 
{ev_{\rm c}(T)\over{\mu_{\rm e} m_{\rm p}}}\, ,                \eqno{(4.34)} $$
where $\mu_{\rm e}$ is the molecular weight per electron in
units of the proton mass.

Assuming first, for simplicity, that the magnetosphere is
isothermal ($T = \rm const$), we infer from equation (4.34) 
that reconnection is most likely to start at the point where 
the ratio $j/\rho$ is maximized. From our analysis we know that, 
at a fixed~$r$, the current density is maximal, and the density 
is minimal, at the apex angle $\theta=\theta_{\rm ap}$ (see 
Fig.~\ref{fig-j-over-rho}). The radial position of the possible 
reconnection site cannot be determined within the framework of 
the self-similar model adopted here. In a realistic situation 
involving a Keplerian disk, however, we can argue that the ratio 
$j/\rho$ will be greatest at the apex point of the field line that 
underwent the largest expansion, i.e., the field line with the 
largest twist angle.

We now examine how the ratio $j/\rho$ at the apex angle
changes with time as $t\rightarrow t_{\rm c}$. According 
to equations~(2.6), (2.8), and~(B7), 
$$ j(r,\theta_{\rm ap},t)= {c\over{4\pi}} \alpha B = 
{c\over{4\pi}} {C\over{r^{n+3}}} G_0^{1+1/n}(\theta_{\rm ap}) 
{\kappa \xi\over{\Delta\Omega\sin\theta_{\rm ap}}}
{1\over{t_{\rm c}-t}} \sim c{C\over{r^{n+3}}}\ 
{1\over{\Delta\Omega(t_{\rm c}-t)}}\, ,                      \eqno{(4.35)} $$
where we reintroduced the speed of light~$c$. If we fix the field 
line instead of the radius, we get
$$ j(\Psi,\theta_{\rm ap},t) \sim {cC\over{r_0^{n+3}(\Psi)}}\
[\Delta\Omega(t_{\rm c}-t)]^{3/n}\, .                      \eqno{(4.36)} $$
Using equation~(4.29) for $\rho(\Psi,\theta_{\rm ap},t)$, we
finally obtain the ratio~$j/\rho$:
$$ {j\over\rho}(\Psi,\theta_{\rm ap},t) \sim 
\left({cC\over{r_0^{n+3}\rho_0}}\right)\
[\Delta\Omega(t_{\rm c}-t)]^{A/\kappa-1-1/n}\, .            \eqno{(4.37)} $$

It is seen that the power-law exponent characterizing the asymptotic
time evolution is independent of~$p$, and, interestingly, is exactly 
equal to the exponent in the asymptotic expression for~$v_\theta$ that 
follows from equations~(B9) and~(B12). This implies that the time 
evolution of $j_\phi/\rho$ {\em on a given field line} is governed 
by the divergence of~$v_\theta$ at~$\theta=\theta_{\rm ap}$. It is 
now also possible to estimate when the condition~(4.34) for triggering 
anomalous resistivity will be satisfied, enabling reconnection to take 
place. The result is
$$ \Delta \Omega(t_{\rm c}-t)_{\rm onset,iso} \sim 
\left({v_{\rm c}\over{v_{\rm d}}}\right)^{n\over{nA/\kappa-n-1}} \sim
\left[ {ev_{\rm c}\over \mu_{\rm e} m_{\rm p}}\
{{r_0^{n+3}\rho_0}\over{cC}} \right]^ {n\over{nA/\kappa-n-1}}\, . 
\eqno{(4.38)} $$

Finally, by comparing the estimates (4.33) and (4.38),
we can derive the condition for the onset of
reconnection to occur before inertial effects 
become important in the magnetosphere. For example, 
for~$n=1$, we find
$$ {ev_{\rm c}\over \mu_{\rm e} m_{\rm p}}\ {{r_0^4 \rho_0}\over{cC}}\
\sqrt{r_0 \Delta\Omega\over{v_{A,0}}} < 1\, .                 \eqno{(4.39)} $$

So far in this subsection we have assumed that the temperature stays fixed 
during the evolution. However, because the field-line expansion is very
rapid near the critical point, it might be more realistic to assume 
that the evolution of the plasma is {\em adiabatic} rather than isothermal.
Indeed, let us consider the time scale for temperature equalization due 
to electron thermal conduction along the magnetic field lines. For the
adopted fiducial parameter values (see below), the electron collisional 
mean free path is $\lambda_{\rm e} \simeq 10^8\, {\rm cm} \ll r_0 \simeq 
10^{12}\, {\rm cm}$. The electron thermal conductivity is then $\chi_{\rm e} 
\simeq \lambda_{\rm e} v_{\rm th,e}\simeq 10^{16}{\rm cm^2\ s^{-1}}$, and 
the temperature equalization time due to electron thermal diffusion can be 
estimated as $\tau_{\rm e} \simeq r_0^2/\chi_{\rm e}\simeq 10^8 \, {\rm s} 
\gg (\Delta\Omega)^{-1}$. This motivates an examination of the adiabatic 
expansion limit.\footnote
{The thermal evolution of the magnetospheric plasma could also be affected by
wave dissipation, heating by the stellar and disk radiation
fields, recombination,
etc., which complicate its precise determination. These effects 
may restrict the range of applicability of the adiabatic approximation to 
temperatures above some minimum value, which may be of
order~$10^3 \, {\rm K}$.}

Under the adiabatic assumption, the density drop near~$\theta_{\rm ap}$ 
is accompanied by a drop in temperature~$T \propto
\rho^{(\gamma-1)}$ (where~$\gamma$ is the adiabatic index), and 
by a corresponding decrease in the critical drift speed $v_{\rm c} \propto
\sqrt{T}\propto \rho^{(\gamma-1)/2}$. One can thus write $v_{\rm
c}\approx v_{\rm c,0}(\rho/\rho_0)^{(\gamma-1)/2}$, 
where $v_{\rm c,0}$ is the critical drift speed for the 
initial temperature~$T_0$.

The criterion for triggering anomalous resistivity can then be written as
$$ {j\over{\rho}} \left({\rho_0\over\rho}\right)^{{\gamma-1}\over 2} >
{{ev_{\rm c,0}}\over \mu_{\rm e} m_{\rm p}}\, .         \eqno{(4.40)} $$
Correspondingly, we get
$$ \Delta\Omega (t_{\rm c}-t)_{\rm onset,ad} \sim 
\left[ {ev_{c,0}\over \mu_{\rm e} m_{\rm p}}\
{{r_0^{n+3}\rho_0}\over{cC}} \right]^
{{2n}\over{(\gamma+1)n(A/\kappa-1)-4\gamma+2}}\, .            \eqno{(4.41)} $$
By comparing the expressions (4.41) and (4.33), we derive the
condition for reconnection to occur before inertial effects
become important. For $n=1$, it is given by
$$ \left[ {ev_{c,0}\over \mu_{\rm e} m_{\rm p}}\ 
{{r_0^4\rho_0}\over{cC}} \right]^{1 \over {2\gamma-1}}\, 
\sqrt{r_0 \Delta\Omega\over{v_{A,0}}} < 1\, ,          \eqno{(4.42)} $$
which generalizes equation (4.39).

To obtain quantitative estimates, we choose
the following set of fiducial parameters, which may apply to
accreting, magnetized YSOs:\\
$R_*=1.5\times 10^{11} \ {\rm cm}$; \\
$r_0=5R_*=7.5\times 10^{11} \ {\rm cm}$; \\
$B(R_*,\theta=0)=1\, {\rm kG}\ 
\Rightarrow C=1.7\times 10^{36}\,{\rm G\, cm^3}$ (for $n=1$);\\
$\Delta\Omega=10^{-5}{\rm s^{-1}}$;
$n_{e,0} = 10^6 {\rm cm^{-3}}$;
$T_0 = 10^5\ {\rm K}$;
$\mu_{\rm e} = 1$.

Using these values, we get:\\
$ \Delta\Omega\, r_0= 7.5 \times 10^6 \ {\rm cm \ s^{-1}}$;\\
$ v_{A,0} = C/r_0^3 \sqrt{4\pi\rho_0} = 9 \cdot 10^8  \ {\rm cm \ s^{-1}}$;\\
$\Delta\Omega (t_{\rm c}-t)_{\rm in} \sim 0.09 \ll 1$;\\
$v_{c,0} \sim v_{{\rm th},e} = 1.2\times 10^8  \ {\rm cm \ s^{-1}}$;\\
$j_0\equiv cC/4\pi r_0^4 \sim 0.013\ {\rm G \ s^{-1}}$;\\
$v_{\rm d,0} = j_0/en_{\rm e,0} \sim 27  \ {\rm cm \ s^{-1}}$;\\
$\Delta\Omega (t_{\rm c}-t)_{\rm onset, iso} \sim 
3 \times 10^{-6} \ll \Delta\Omega (t_{\rm c}-t)_{\rm in}$;\\
$\Delta\Omega (t_{\rm c}-t)_{\rm onset, ad} \sim 4 \times 10^{-3} \ll 
\Delta\Omega (t_{\rm c}-t)_{\rm in}$ --- for~$\gamma=5/3$.

On the basis of these estimates we conclude that inertial effects
should become important much earlier than the time when
reconnection is first triggered. We note, however, that this
conclusion depends on the rather uncertain value of the initial
plasma density $n_{e,0}$: if that density were lower, then
reconnection could occur while the equilibrium assumption was
still valid. For the isothermal case with $n=1$, this requires
(using eq. [4.39] and the adopted fiducial values of $r_0$, $T_0$, 
$B_0$ and~$\Delta\Omega$) 
$$ n_{\rm e,0}^{\rm iso} < 300\, {\rm cm}^{-3}\, ,            \eqno{(4.43)} $$
corresponding to $\Delta\Omega(t_{\rm c}-t)_{\rm onset,iso} \ga
0.01$, whereas for the adiabatic case (using eq. [4.42] and
setting $\gamma=5/3$), the upper limit on the density is
$$ n_{\rm e,0}^{\rm ad} < 1 \times 10^4\, {\rm cm}^{-3}\, ,    \eqno{(4.44)} $$
which implies $\Delta\Omega(t_{\rm c}-t)_{\rm onset,ad} \ga 0.03$.

Besides addressing the issue of the {\em onset} of anomalous resistivity, 
one has to consider whether the current-driven microinstabilities,
once triggered, will in fact lead to an anomalous resistivity that
is large enough to provide the required low value of the
effective Lundquist number ($S_{\rm eff} \la 100$). Using the
Sagdeev (1967) estimate of the anomalous collision frequency
associated with the ion-acoustic instability,
$$ \nu_{eff}=0.01\, \omega_{\rm pi}(v_{\rm d}/c_{\rm s}) (T_{\rm
e}/T_{\rm i})    \eqno{(4.45)} $$
(where $c_{\rm s}$ is the speed of sound, $\omega_{\rm p}$ is
the plasma frequency, and the subscripts $e$ and $i$ refer to
the electrons and ions, respectively),
we obtain the corresponding anomalous resistivity from $\eta_{\rm eff} 
= c^2 m_{\rm e} \nu_{\rm eff}/4\pi n_{\rm e} e^2$.
Assuming that the electron and ion temperatures are equal,
we then find the effective Lundquist number to be
$$ S_{eff} = 2\times 10^5\, (\omega_{\rm ci}L/c)
(c_{\rm s}/v_{\rm d})\, ,                                     \eqno{(4.46)} $$
where $\omega_{\rm c}$ denotes the cyclotron frequency.
We shall use $v_{\rm d}=v_{\rm th,e}$ even though the ion-acoustic 
instability can arise at lower values of the drift speed, 
since the effective resistivity is not significantly enhanced
unless $v_{\rm d} \simeq v_{\rm th,e}$ (e.g., Kadomtsev 1965). 
We then get an expression that, interestingly enough, is
independent of both density and temperature,
$$ S_{eff} \simeq 4 \times 10^3\, (\omega_{\rm ci}L/c).       \eqno{(4.47)} $$

The expression (4.47) is applicable once the condition~(4.34)
(or its adiabatic counterpart [4.40]) is satisfied. After the
ion-acoustic microturbulence is initiated, the value of $S_{\rm eff}$ 
at the apex of the expanding field lines decreases on account of the
time evolution of $\omega_{\rm ci}\propto B \sim (C/r_0^3)\, 
[\Delta\Omega(t_{\rm c}-t)]^2$ and $L\sim r\sim r_0/[\Delta\Omega
(t_{\rm c}-t)]$. In fact, 
$$ S_{eff} \simeq S_0\, \Delta\Omega(t_{\rm c}-t) \rightarrow 0,
\qquad t\rightarrow t_{\rm c}\, ,                            \eqno{(4.48)} $$
where $S_0 = 4\times 10^3\, \omega_{\rm ci}(t=0) r_0/c \simeq 
4 \times 10^9$. This demonstrates that, even when excited, 
the ion-acoustic anomalous resistivity will saturate at a level 
that is too low to provide an efficient route for reconnection.

It is, however, worth noting that, after being triggered,
anomalous resistivity will be strongly localized. Over the 
past two decades, numerical simulations (Ugai \& Tsuda 1977;
Sato \& Hayashi 1979; Scholer 1989; Erkaev et al. 2000) have 
shown that a strong local enhancement of resistivity may lead
to a transition to Petschek's (1964) regime of fast reconnection.
Nevertheless, it appears that, even if one assumes the maximum
Petschek reconnection inflow velocity, $u_{\rm rec} = v_{\rm A}/\ln S$, 
the length of time available for reconnection (from the moment when 
anomalous resistivity is triggered until the singular point~$t_{\rm c}$) 
will be too short for any significant amount of magnetic flux to be
reconnected. Indeed, the condition that all magnetic flux 
within the area of angular width $\Delta \theta \sim 0.1$
around the apex is reconnected over the time interval
$(t_{\rm c}-t)_{\rm rec}$ can be written roughly as
$$ u_{\rm rec} (t_{\rm c}-t)_{\rm rec} > \Delta \theta \, r = 
\Delta\theta \, r_0 \, {1\over{\Delta\Omega(t_{\rm c}-t)_{\rm
rec}}}\, . \eqno{(4.49)}   $$
Using $u_{\rm rec} = v_{\rm A}/\ln S$ together with the expression (4.31)
for $v_{\rm A}$, we obtain
$$ [\Delta\Omega (t_{\rm c}-t)_{\rm rec}]^2 = \Delta \theta
\left({{\Delta\Omega r_0}\over{v_{\rm A,0}}}\right)
[\ln S_0 + \ln (t_{\rm c}-t)_{\rm rec}]\, .     \eqno{(4.50)}              $$
Neglecting the term  $\ln (t_{\rm c}-t)$ compared with $\ln S_0$
in equation (4.50), and taking $\Delta\theta = 0.1$, we get
$(t_{\rm c}-t)_{\rm rec} > 0.13/\Delta\Omega$, which
greatly exceeds even our highest estimate [$(t_{\rm c} - t)_{\rm
res,ad}$] for the available time.

Our expressions for the onset time of ion-acoustic
microturbulence (eqs. [4.38] and [4.41]) have been obtained on
the assumption that the equilibrium model remains applicable,
which, as we have noted, requires the initial densities to be
sufficiently low ($n_{\rm e,0} \la 10^3-10^4\, {\rm cm}^{-3}$; 
see eqs. [4.43] and [4.44]). We emphasize, however, that
even if the initial densities are higher and inertial effects
set in early enough to render our onset-time estimates inaccurate, 
we do not expect the evolution of the field lines to be affected 
in a qualitative way. Rather, we anticipate that inertial effects 
will merely {\em delay} the onset of reconnection. The actual
triggering time of the microsinstability could well exceed 
the nominal critical time $t_{\rm c}$, but it may be expected to
remain smaller than the effective opening time of the field
lines, which will also be increased by the inertial effects
(see \S~\ref{subsec-inertial}). In a similar vein, we expect the
relative ordering of $(t-t_{\rm c})_{\rm onset}$ and $(t-t_{\rm c})_{\rm
rec}$ (eq. [4.50]), and therefore our conclusions about the expected
efficiency of reconnection, to remain unchanged by the time
dilation induced by inertial effects.

Finally, we note that a strongly developed {\em MHD tearing turbulence} 
could, in principle, provide an alternative route to reconnection 
(Strauss 1988). For the estimated aspect ratio of the current layer 
that arises in our model, the required fluctuation level that needs 
to be sustained in order for reconnection to be efficient is $|\delta B/B| 
\sim 0.1$. It is unclear whether such a comparatively high level of 
turbulence saturation could be attained. Additional uncertainty is 
created by the expected suppression of the onset and growth of the 
tearing modes by line-tying effects (see Mok \& van~Hoven 1982); 
this issue has not yet been addressed in the literature and may 
provide an interesting direction of future research. Notwithstanding 
these caveats, we consider hyperresistivity produced by tearing-mode 
turbulence to be a promising mechanism for fast reconnection of the 
twisted field lines.

\section{Keplerian Disk}
\label{sec-keplerian}

The analysis presented in the previous sections employed the
self-similar solutions presented in \S~\ref{sec-model}. In this
section we examine the more realistic case of a Keplerian disk
and compare its behavior to that of the self-similar,
uniform-rotation model. Since the Keplerian disk has a
characteristic radial scale, namely, the corotation
radius $r_{\rm co}$, the self-similar semianalytic approach 
is clearly inapplicable. The problem becomes fully
two-dimensional and requires numerical tools. 

To tackle this problem we developed a numerical code that enables
us to find sequences of equilibria once the following two functions 
are specified on the disk surface. The first is the rotation law, i.e., 
the rate of relative twist as a function of radius along the disk 
surface, $\Delta\Omega(r)$. The second is the magnetic flux distribution on 
the disk surface, $\Psi_{\rm d}(r)$. Unlike in the self-similar case, 
these two functions do not have to be power laws. We first describe 
our numerical procedure and then present the results of our simulations.

\begin{figure} [tbp]
\centerline {\psfig{file=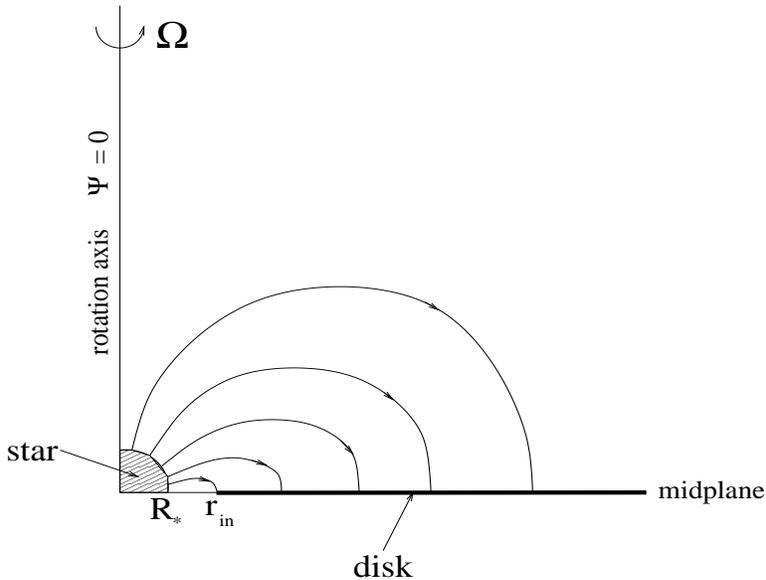,height=3 in,width=4 in}}
\caption{Geometry of the problem.
\label{fig-geometry}}
\end{figure}

The computational domain consists of the outside of a sphere of 
radius~$R_*$ (see Fig.~\ref{fig-geometry}). In the remainder of
this section, we normalize the radius $r$ by~$R_*$. Using the 
symmetry with respect the disk plane, we only consider the upper 
halfspace.

To find the force-free magnetostatic equilibria, 
we solve the Grad-Shafranov equation 
$$ {{\partial^2 \Psi}\over{\partial r^2}}+
{\sin{\theta}\over{r^2}} {\partial\over{\partial\theta}}
\left( {1\over{\sin{\theta}}} {{\partial\Psi}\over{\partial\theta}}\right)=
-F(\Psi) F'(\Psi)\, .                                           \eqno{(5.1)} $$
The main difficulty in this equation is that the nonlinear term 
on the right-hand side [$F(\Psi) F'(\Psi)$] is not given explicitly. 
Rather, it is determined implicitly by the rotation law via the condition
$$ \Delta\Phi(\Psi) =\Delta\Omega(\Psi)t = F(\Psi) I(\Psi)\, , \eqno{(5.2)} $$
where 
$$ I(\Psi) \equiv \int\limits_\Psi {1\over{B_\theta r}} 
{d\theta\over{\sin{\theta}}}                                 \eqno{(5.3)} $$
is an integral along the magnetic field line~$\Psi$.

The time~$t$ in equation~(5.1) is the parameter controlling the sequence 
of equilibria, and $\Delta\Omega(\Psi_{\rm d}(r))$ is a prescribed function 
representing the rotation law. For example, for a Keplerian disk, one has 
$\Delta\Omega=-\Omega_* [1 - (r_{\rm co}/r)^{3/2}]$. In the remainder of 
this section, we normalize $t$ by~$\vert \Omega_*\vert^{-1}$.

Our goal is to find the time sequence of equilibria for a
given rotation law. We start at $t=0$ with the potential 
field, which corresponds to $F \equiv 0$. We then give a
small increment to $t$ (corresponding to an increment in
$\Delta\Phi$ of the order of a small fraction of a radian), 
find the solution using the procedure described below, then 
proceed to the next moment of time~$t$, etc.

For each moment of time~$t$ we solve the system (5.1)--(5.3)
iteratively: at the $k$-th iteration we use the result $F^{(k)}(\Psi)$ 
of the previous iteration to approximate the function [taking $F(\Psi)$ 
for the initial guess ($k=0$) to be the solution for the previous moment 
of time, or zero for $t=0$], solve the elliptic equation~(5.1), then use 
the solution $\Psi^{(k)}(r,\theta)$ to calculate the integral $I^{(k)}(\Psi)$ 
along field lines, and then we update the function~$F(\Psi)$ 
according to
$$ F^{(k+1)}(\Psi) = \Delta\Omega(\Psi) t /I^{(k)}(\Psi)\, .    \eqno{(5.4)} $$
We repeat this procedure until the process converges.

When solving the elliptic equation~(5.1) for given~$t$ and~$k$, 
we use the relaxation method. We introduce a fictitious time 
variable~$\tau$ and then evolve $\Psi(\tau,r,\theta)$ according 
to
$$ {{\partial \Psi}\over{\partial \tau}} = 
{{\partial^2 \Psi}\over{\partial r^2}}+
{\sin{\theta}\over{r^2}} {\partial\over{\partial\theta}}
\left( {1\over{\sin{\theta}}} {{\partial\Psi}\over{\partial\theta}}\right)
+F(\Psi) F'(\Psi)\, .                                        \eqno{(5.5)} $$

We first tried to solve this system using a uniform grid in 
spherical coordinates~$(r,\theta)$. However, in this case it is 
necessary to introduce an outer boundary at some large radius 
$r=r_{\rm max}$, where one encounters several serious problems,
such as the choice of boundary conditions and the treatment of 
the integral~(5.3) for field lines that cross this boundary. 
To bypass these issues, we effectively place the outer boundary 
at infinity and use the transformation
$$ x = {1\over{\sqrt{r}}}\, ,                                \eqno{(5.6)} $$
which maps $r=\infty$ to $x=0$, while keeping the 
inner boundary (the surface of the star $R_*=1$) 
at~$x=1$. Correspondingly, we replace the uniform 
$(r,\theta)$ grid with a uniform $(x,\theta)$ one.\footnote
{We also tried the mapping $x=1/r$ but have found that 
$x=1/\sqrt{r}$ works somewhat better because it allows 
one to pack more gridpoints at larger radii.} 

After applying the transformation~(5.6), equation (5.5) becomes
$$ {{\partial \Psi}\over{\partial \tau}} = 
{1\over 4} x^6 {{\partial^2 \Psi}\over{\partial x^2}}+
{3\over 4} x^5 {{\partial \Psi}\over{\partial x}}+
x^4 \sin{\theta} {\partial\over{\partial\theta}}
\left( {1\over{\sin{\theta}}} {{\partial\Psi}\over{\partial\theta}}\right)
+F(\Psi) F'(\Psi)\, .                                        \eqno{(5.7)} $$
This equation is integrated on a rectangular domain in the 
($x,\theta$) plane with~$x$ running from~0 to~1 and~$\theta$ 
running from~0 to~$\pi/2$. There are four boundaries: the 
surface of the star $x=1$, the axis $\theta=0$, the outer 
boundary $x=0$, and the surface of the disk $\theta=\pi/2$. 
On three of these the boundary conditions are particularly 
simple:
$$ \Psi(x,\theta=0) = \Psi(x=0,\theta) =0\, ,                  \eqno{(5.8)} $$
and
$$ \Psi(x=1,\theta) = \Psi_*(\theta)\, ,                       \eqno{(5.9)} $$
where the latter condition represents a prescribed magnetic flux
distribution on the surface of the star, which is assumed to be infinitely
conducting. The results we show correspond to a dipole field,
$$ \Psi_*(\theta) = \sin^2{\theta}\, ,                        \eqno{(5.10)} $$
normalized so that the total amount of flux through the stellar surface is~1.

The boundary conditions on the fourth boundary (the equatorial 
plane~$\theta=\pi/2$) are somewhat more complicated because our 
model incorporates an inner gap between the disk and the star, 
which breaks this boundary into two pieces (see Fig.~\ref{fig-geometry}). 
Typically we place the inner edge of the disk at $r_{\rm in}=1.5$ 
(corresponding to~$x_{\rm in}=2/3$, $\Psi_{\rm in}= 2/3$). The space 
inside the gap, $1<r<r_{\rm in}$, is filled with very tenuous plasma, 
just like the magnetosphere above the disk. Hence, the field lines 
crossing the equatorial plane inside the gap must be potential, and, 
because of the symmetry with respect to the midplane, the magnetic 
field inside~$r_{\rm in}$ has to be perpendicular to this plane,
$$ {{\partial\Psi}\over{\partial\theta}}
(x>x_{\rm in},\theta={\pi\over 2}) =0\, .                      \eqno{(5.11)} $$

In the region $r>r_{\rm in}$ the magnetic field lines are frozen 
into the disk surface; thus the flux distribution in this region 
is fixed, similar to the situation at the stellar surface:
$$ \Psi(x<x_{\rm in},\theta={\pi\over 2}) = \Psi_{\rm d}(x)\, ,\eqno{(5.12)} $$
where $\Psi_{\rm d}(x)$ is a prescribed function. In our
illustrative examples we again employ a dipole representation,
$$ \Psi_{\rm d}(x) = x^2 = {1\over r}\, .                     \eqno{(5.13)} $$ 

\begin{figure} [tbp]
\centerline{\psfig{file=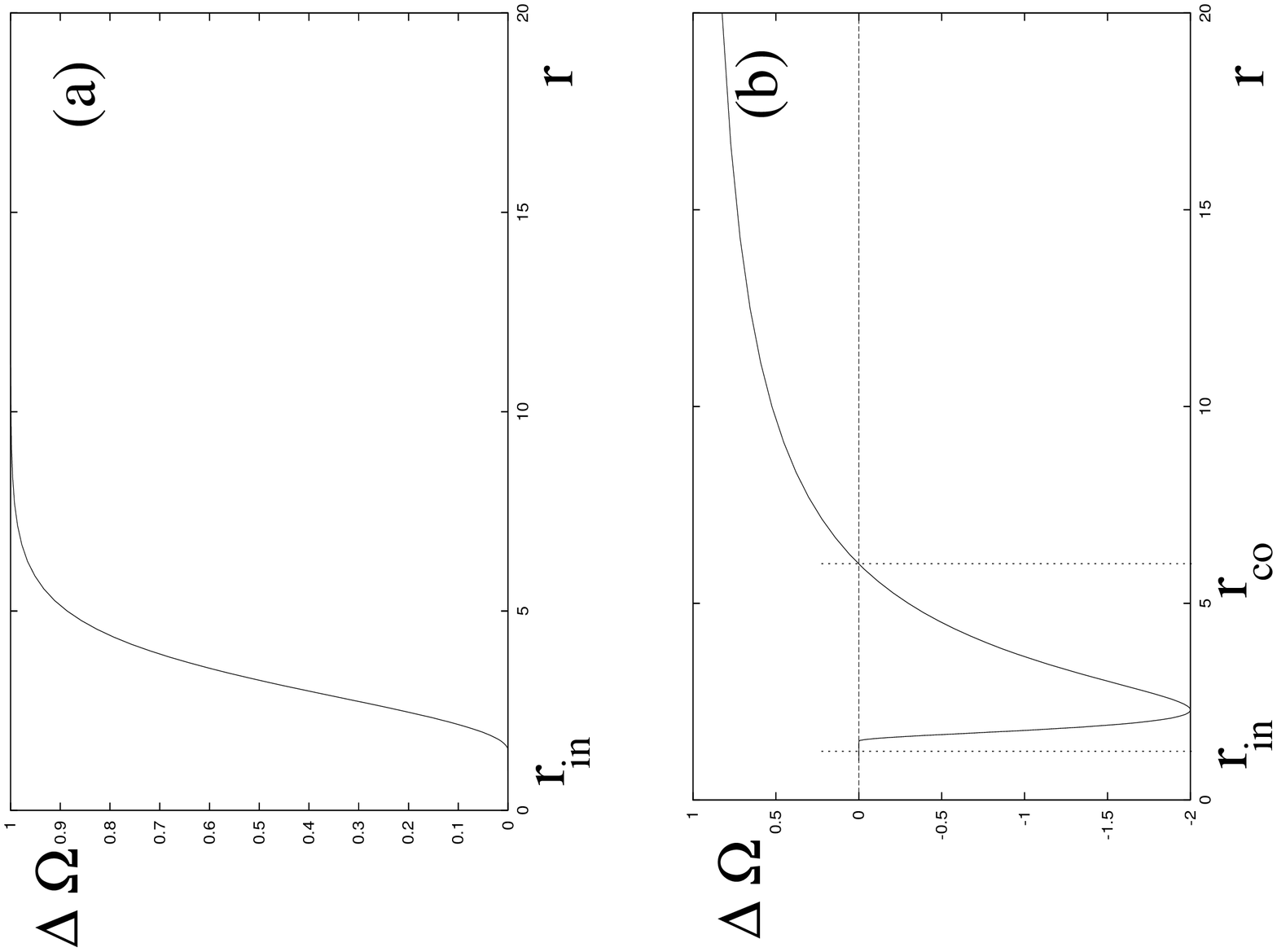,height=6in,width=4in,angle=-90}}
\caption{Angular velocity profile $\Delta\Omega(r)$ for: 
(a) uniformly rotating disk; (b) Keplerian disk.
\label{fig-twist}}
\end{figure}

We start with the potential dipole field at~$t=0$. We then proceed 
through the sequence of equilibria by gradually increasing~$t$ (and 
therefore the twist angle), and using the solution for the previous 
value of~$t$ as the initial guess for the next value of~$t$.
Although our procedure can be used with any choice 
of~$\Delta\Omega(r)$, our choice of the twist function 
was guided by the need for $\Delta\Omega(r)$ to vanish
in the inner gap. For numerical convenience, we want 
this function to remain smooth as it approaches zero 
at~$r_{\rm in}$ (which also makes physical sense, since 
we expect that near the inner gap the accreting gas will
undergo a gradual transition from a Keplerian rotation 
law to corotation with the star). Along the rest of the 
disk surface, however, this function can be arbitrary. 
We investigated two particular cases: 
{\em uniform rotation} (Fig.~\ref{fig-twist}a), wherein 
$\Delta\Omega(r)$ approaches a constant value as one moves 
away from the inner edge; and {\em Keplerian rotation} 
(Fig.~\ref{fig-twist}b), where, for $r>r_{\rm in}$, the 
rotation law becomes Keplerian with~$r_{\rm co}=6$.

\begin{figure} [tbp]
\centerline{\psfig{file=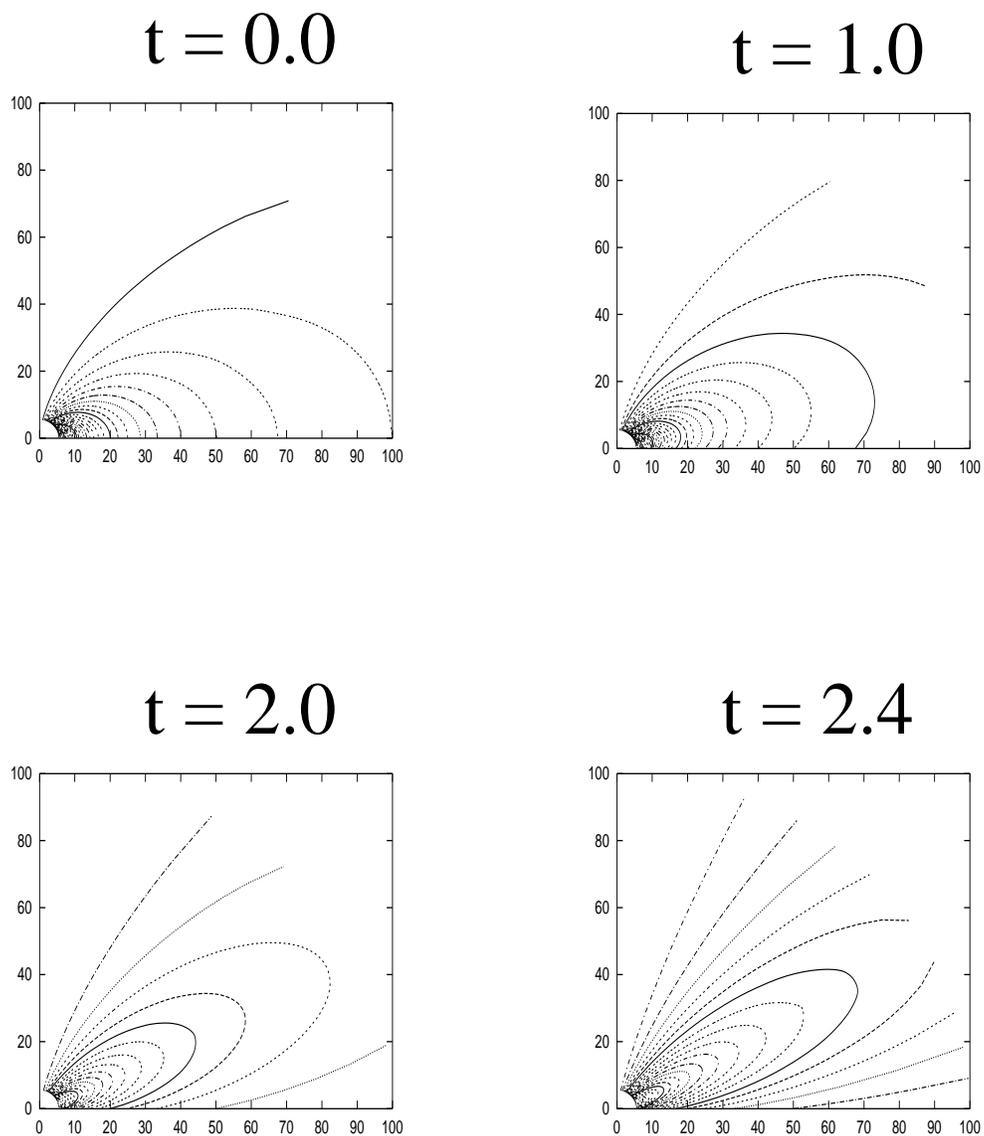,height=7in,width=6.5in,angle=90}}
\caption{Sequence of magnetic field contours for the uniformly
rotating disk model.
\label{fig-flux-uniform}}
\end{figure}

\begin{figure} [tbp]
\centerline{\psfig{file=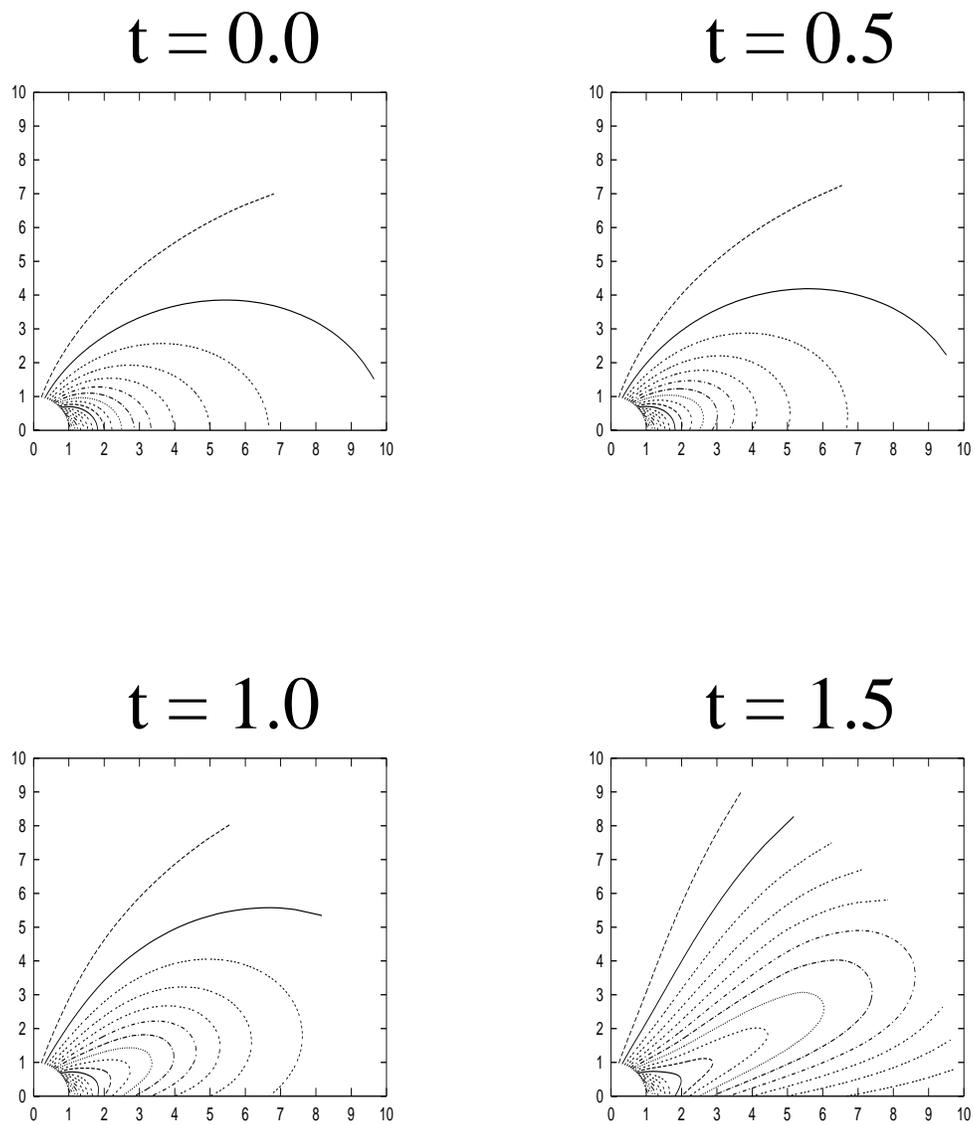,height=7in,width=6.5in,angle=90}}
\caption{Sequence of magnetic field contours for the Keplerian
disk model.
\label{fig-flux-Keplerian}}
\end{figure}

\begin{figure} [tbp]
\centerline{\psfig{file=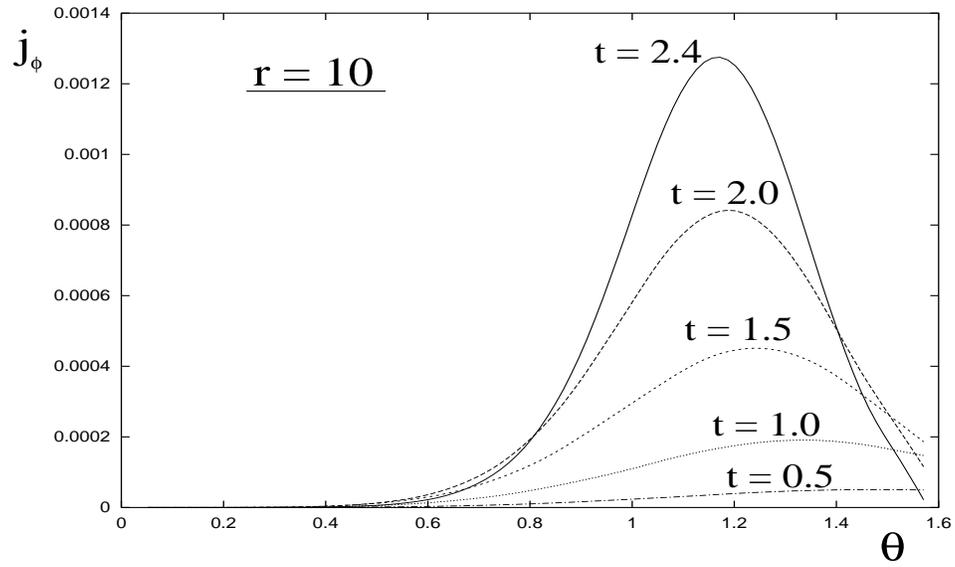,height=3in,width=5in}}
\caption{Toroidal current density as a function of~$\theta$ at a 
given radius ($r=10$) at different moments of time for the uniformly 
rotating disk model.
\label{fig-current-uniform}}
\end{figure}

\begin{figure} [tbp]
\centerline{\psfig{file=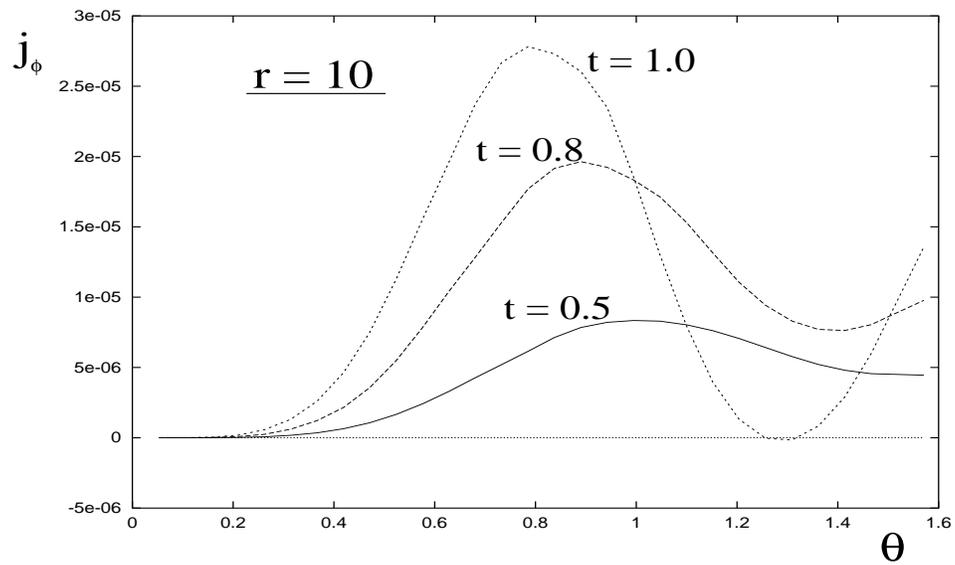,height=3in,width=5in}} 
\caption{Toroidal current density as a function of~$\theta$ at a given 
radius ($r=10$) at different moments of time for the Keplerian disk model.
\label{fig-current-Keplerian}}
\end{figure}

\begin{figure} [tbp]
\centerline{\psfig{file=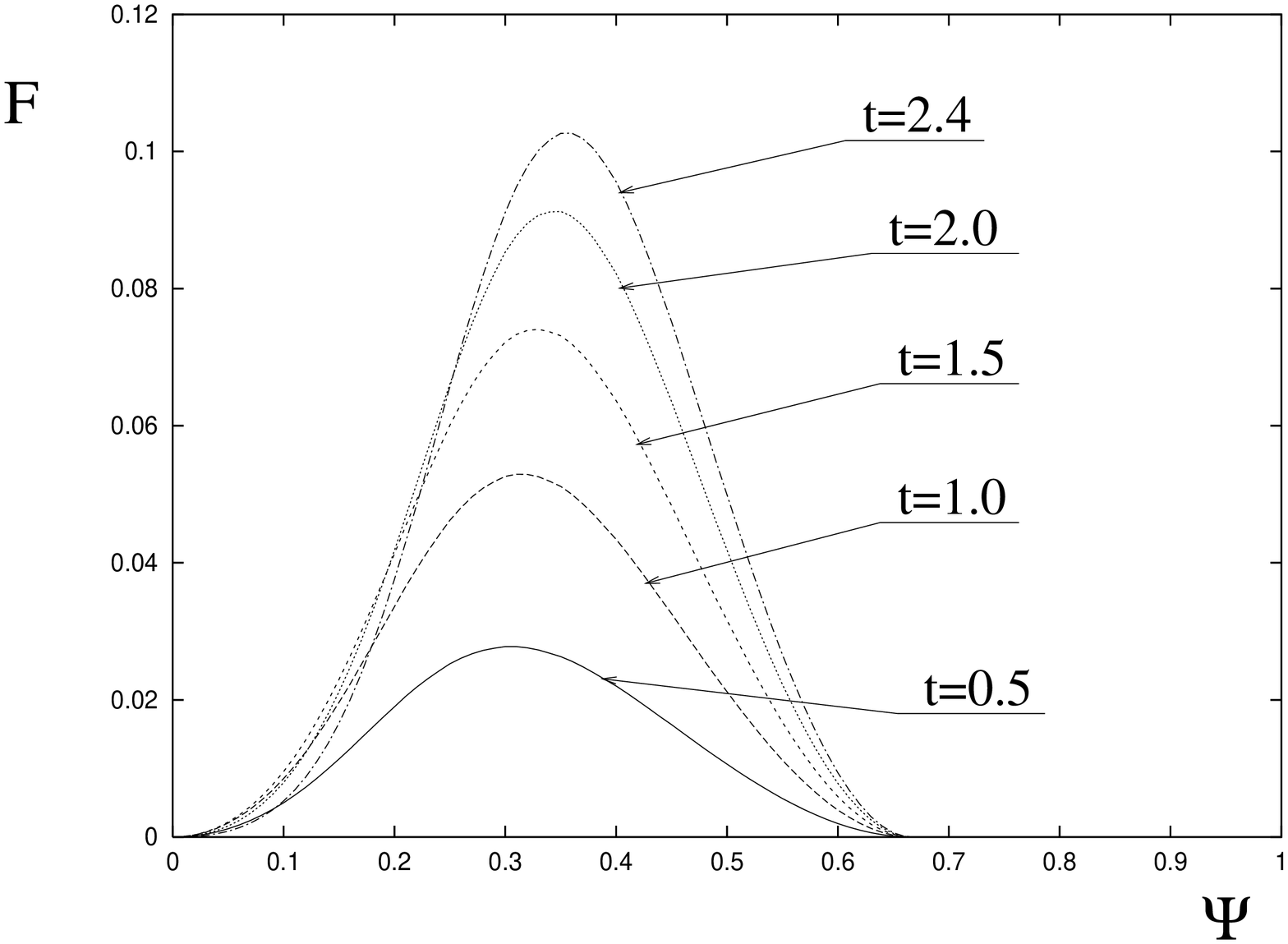,height=3 in,width=5 in}}
\caption{$F(\Psi,t)$ for the uniformly rotating disk model.
\label{fig-F-uniform}}
\end{figure}

\begin{figure} [tbp]
\centerline{\psfig{file=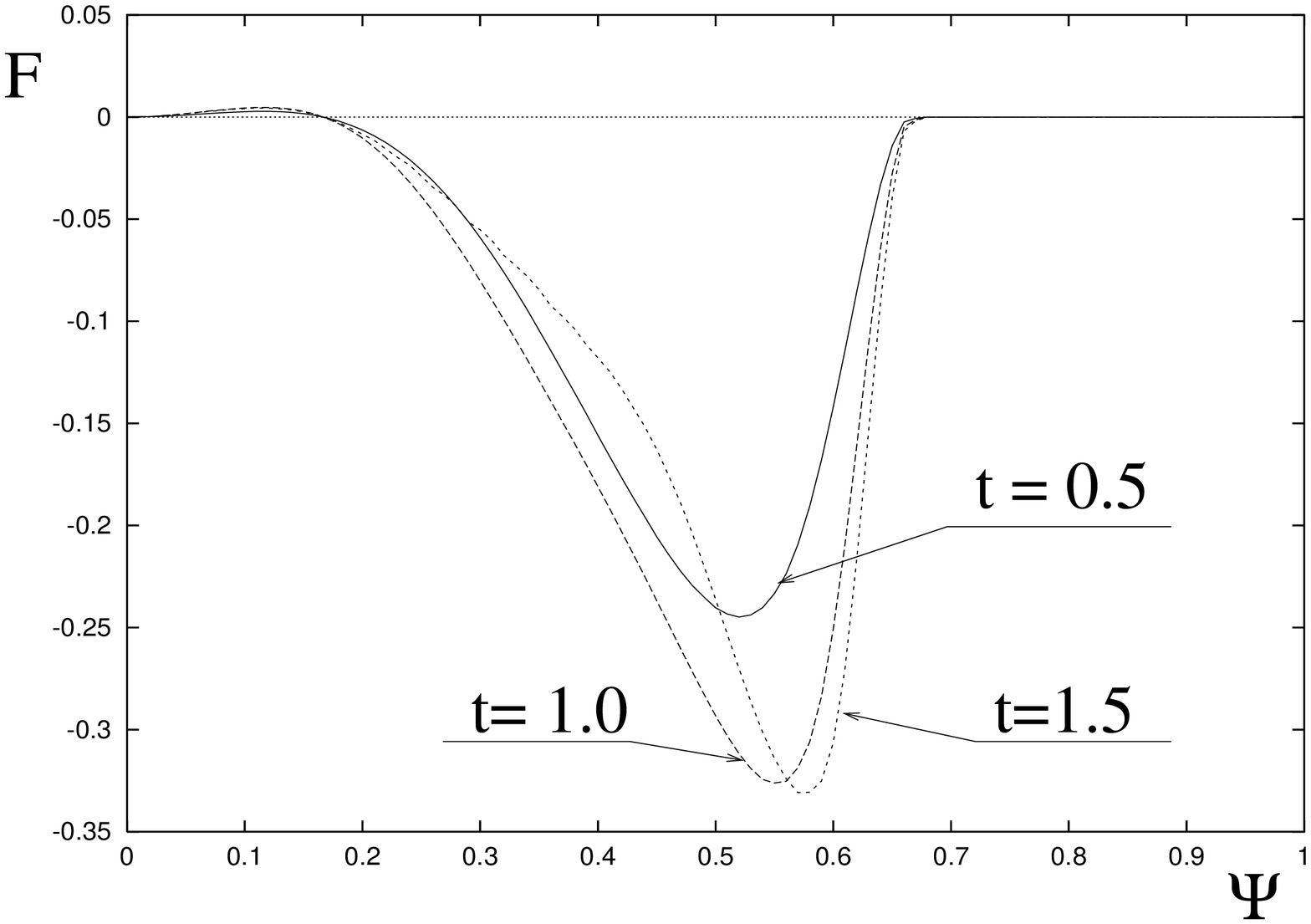,height=3 in,width=5 in}}
\caption{$F(\Psi,t)$ for the Keplerian disk model.
\label{fig-F-Keplerian}}
\end{figure}

We now turn to a description of our results. Figures~\ref
{fig-flux-uniform} and \ref{fig-flux-Keplerian} present a
series of contour plots of the magnetic flux function for
several instances of time for the uniformly rotating and
Keplerian disk models, respectively. It is seen that the 
basic behavior is very similar in both cases, the most 
important qualitative feature being the rapid expansion 
of the field lines near~$\theta\simeq 60-70^{\circ}$. 
Just as in the self-similar model, this expansion is 
accompanied by a rapid rise of the toroidal current 
density~$j_\phi$ in this region, as shown in Figures~\ref
{fig-current-uniform} and~\ref{fig-current-Keplerian}.
Figures~\ref{fig-F-uniform} and~\ref{fig-F-Keplerian} describe
the evolution of the function~$F(\Psi,t)$ for the two models.

\begin{figure} [tbp]
\centerline {\psfig{file=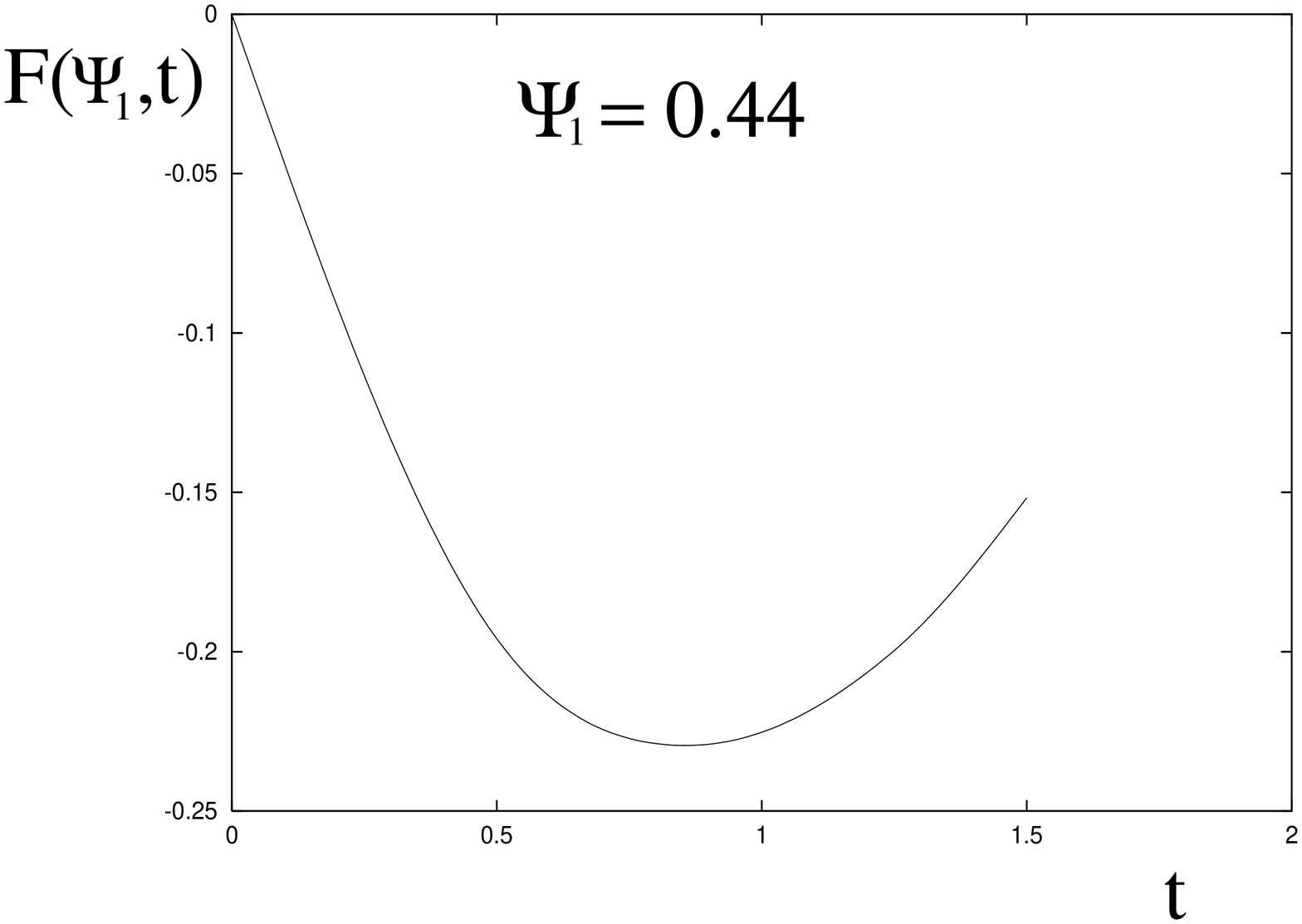,height=3 in,width=5 in}}
\caption{Evolution of the function $F(\Psi_1,t)$ for the
Keplerian disk model; $\Psi_1=0.44$ is the field line with 
the largest twist.
\label{fig-F1-of-t-Keplerian}}
\end{figure}

In all the cases we studied, the evolution can be divided into
two stages, distinguished by the time behavior of $F(\Psi,t)$ on
the field lines that have undergone the largest twist. For the
Keplerian disk model, this field line is given by~$\Psi_1=0.44$, 
on which the absolute value of $\Delta\Phi$ is twice the asymptotic 
value at infinity (see Fig.~\ref{fig-twist}b). Note that 
the function~$F(\Psi=\Psi_1,t)$ serves as an indirect analog of 
$a_0(\Delta\Phi)$ in the self-similar model. The plot of 
$F(\Psi_1,t)$ is shown in Figure~\ref{fig-F1-of-t-Keplerian}. 

In the uniform-rotation disk model the analogy can be made more direct.
A convenient way to describe the behavior in this case is to look
at the evolution of the second derivative of $F(\Psi)$ at $\Psi=0$. 
Indeed, at large distances, $r \gg r_{\rm in}$, the twist angle 
$\Delta \Phi$ approaches a constant (see Fig.~\ref{fig-twist}a), 
so one can expect a self-similar power-law asymptotic behavior of 
$F(\Psi)$ in the limit~$\Psi\rightarrow 0$. Using $\alpha(\Psi)=F'(\Psi)$ 
and $F(0)=0$, one can express $F(\Psi,t)$ using the notation of
\S~\ref{subsec-van-Ballegooijen} as
$$ F(\Psi,t) = \int\limits_0^\Psi \alpha d\Psi = 
\int\limits_0^\Psi {{a_0(t) d\Psi}\over{r_0(\Psi)}}\, .    \eqno{(5.14)} $$
In the case $n=1$ considered in our numerical calculations
one has $r_0(\Psi) = 1/\Psi$, so
$$ F(\Psi,t)={{a_0(t)\Psi^2}\over 2},\qquad\Psi\rightarrow 0. \eqno{(5.15)} $$
This quadratic behavior is indeed exhibited by our calculated
solution. 

\begin{figure} [tbp]
\centerline {\psfig{file=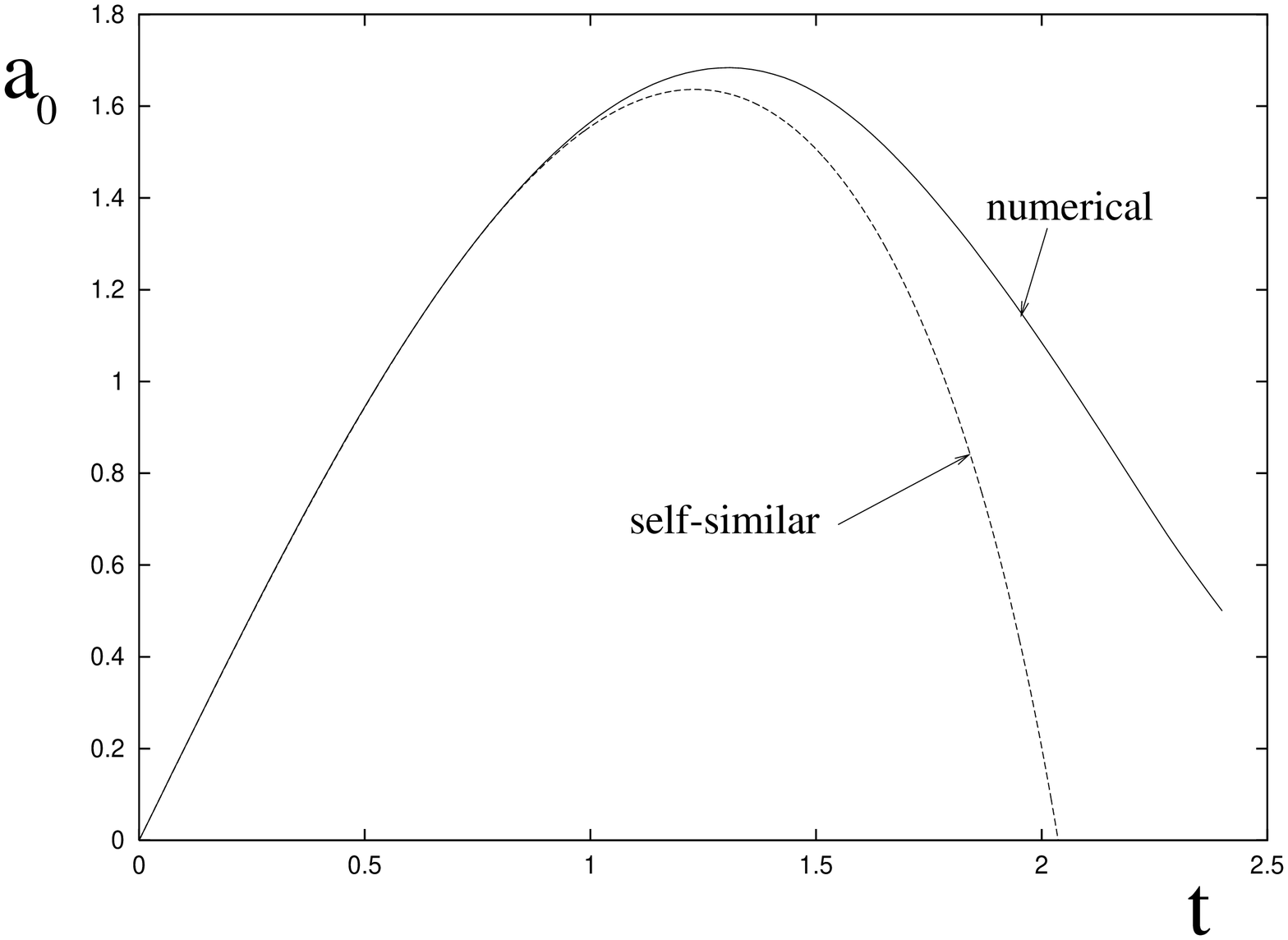,height=3 in,width=5 in}}
\caption {Evolution of the function $a_0(t)$ [eq.~5.16)] for 
the uniformly rotating disk model. The solid line represents 
the result of the numerical calculations discussed in this
section, whereas the dashed line shows the behavior of the
corresponding function in the self-similar model discussed 
in~\S~\ref{sec-model}.
\label{fig-a0-of-t-uniform}}
\end{figure}

The foregoing considerations have motivated us to select the time evolution of 
$$ a_0(t) \equiv {{d^2 F}\over{d\Psi^2}}|_{\Psi=0}             \eqno{(5.16)} $$
for a direct comparison with the corresponding function in the
self-similar model. Figure~\ref{fig-a0-of-t-uniform} demonstrates 
that on the ascending branch of the solution the agreement is very 
good, but that on the descending branch there is a significant 
deviation that is reflected in different values of the critical 
twist angle~$\Delta\Phi_{\rm c}$. This discrepancy can be attributed 
to the fact that the innermost field lines have smaller twist angle 
and therefore are not inflated as much as in the self-similar case. 
Consequently, the magnetic stresses driving the expansion are somewhat
weaker and the opening of the field lines is delayed. Still, we see 
from Figure~\ref{fig-a0-of-t-uniform} (as well as 
Fig.~\ref{fig-F1-of-t-Keplerian}) that the basic 
behavior is the same. As $t$ is increased, $a_0(t)$ (and $F(\Psi_1,t)$ 
in the Keplerian case) at first rises until it reaches a maximum at some 
instant~$t_{\rm max}$, and subsequently starts to decrease, just as in 
the self-similar case. During the first stage (the ascending 
branch, $t<t_{\rm max}$) the shape of the field lines does not 
change significantly, but during the second stage (the descending 
branch, $t>t_{\rm max}$) there is a rapidly accelerating
expansion of the field lines, which
approach an open state (see Figs.~\ref{fig-flux-uniform} and 
\ref{fig-flux-Keplerian}). This qualitative behavior appears to be 
universal and is independent of the details, such as the particular 
rotation law of the disk. However, certain quantitative features,
such as the values of $t_{\rm max}$ and $a_{0,\rm max}$ [or $F(\Psi_1,
t_{\rm max}$)] do depend on the particular parameters (e.g., $r_{\rm in}$, 
$r_{\rm co}$, $\Psi_1$, etc.).

\begin{figure} [tbp]
\centerline{\psfig{file=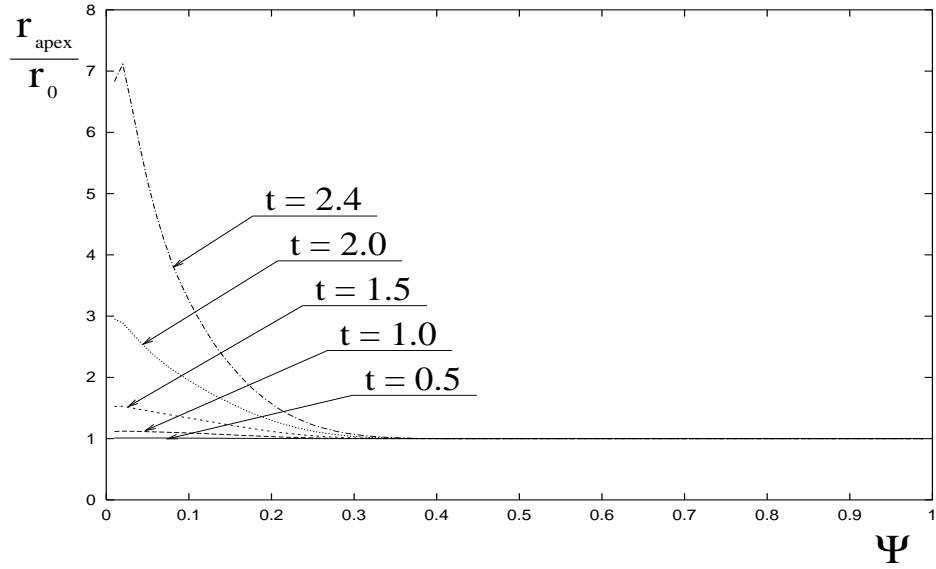,height=3in,width=5 in}}
\caption{The expansion factor as a function of magnetic flux 
in the uniformly rotating disk model.
\label{fig-expansion-uniform}}
\end{figure}

\begin{figure} [tbp]
\centerline{\psfig{file=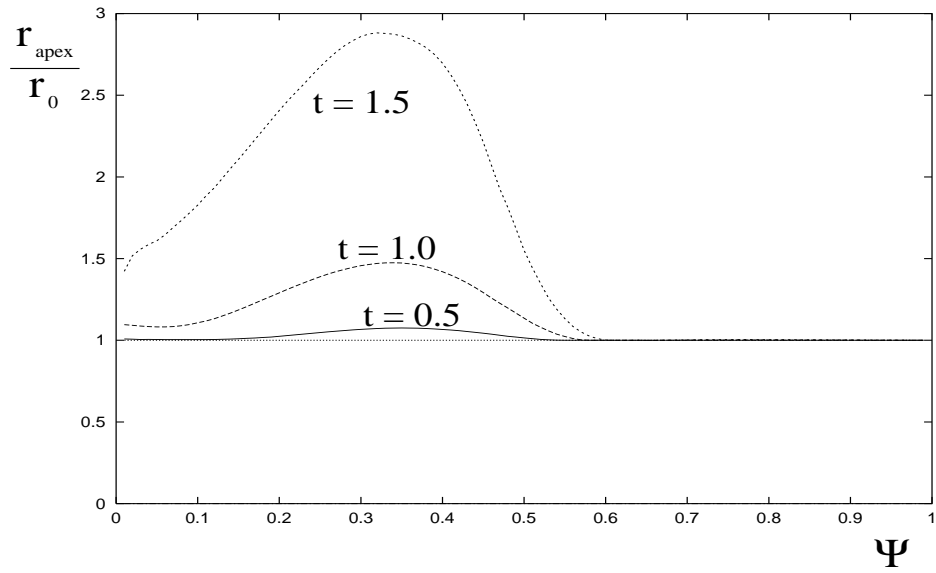,height=3in,width=5 in}}
\caption {The expansion factor as a function of magnetic flux 
in the Keplerian disk model.
\label{fig-expansion-Keplerian}}
\end{figure}

The two evolutionary stages are also characterized by distinct
convergence properties of the numerical solutions. During the
first stage the convergence is rapid and robust, but during 
the second stage it progressively slows down and one has to update
the function~$F(\Psi)$ more frequently, and finally we reach a
point where we need to terminate the computation. The reason for
this is that the field lines have by that time expanded very 
strongly, and our iteration process no longer converges at a 
given resolution. Instead, we get an unphysical reconnection 
in the middle of the domain, which renders the numerical 
procedure inconsistent. Although increasing the resolution 
helps, the computation time necessary for convergence grows 
dramatically.\footnote
{This behavior can be understood from the fact that our 
relaxation procedure represents a diffusion process with 
a diffusion coefficient~$D_x \propto x^6$, and the diffusion 
time over a distance $\delta x\sim x$ is proportional to 
$x^2/D_x\sim x^{-4}=r^2$, which diverges as the field lines 
expand.}
Fortunately, by the time we are forced to stop the computation 
the field lines have already expanded by a large factor (see 
Figs.~\ref {fig-expansion-uniform} and~\ref{fig-expansion-Keplerian}), 
so an extrapolation of the functions~$a_0(t)$ and~$F(\Psi_1,t)$
becomes possible. We deduce that these functions reach zero at 
a finite twist angle (about~2.7~rad for the uniform-rotation
case and~2.0~rad for the Keplerian case), which corresponds 
to opening of the field lines in a finite time, similar to 
the behavior of the self-similar solutions.

We thus see that a good case can be made for the finite-time
singularity occurring not only in the self-similar solution
(see \S~\ref{subsec-singularity}) but also in a broader class
of models. To strengthen this point, we present a simple physical
argument showing why this should be the case.\footnote
{For a different, more mathematical argument in favor 
of a finite time-singularity in sheared, axisymmetric, 
force-free magnetic fields, see, e.g., Aly 1995.}

Consider the small-$F$ limit of equation (5.1), in which
the nonlinear term $FF'(\Psi)$ for some chosen field line
$\Psi$ is smaller than, say, the first linear term on the 
left-hand side at typical distances of order the footpoint 
radius $r_0(\Psi)$. By a dimensional analysis, this implies
$$ F(\Psi) \ll {\Psi\over{r_0(\Psi)}}\, .                    \eqno{(5.17)} $$
Small values of $F$ may correspond to two qualitatively different solutions.
The first solution is close to the potential field, with the nonlinear
term being unimportant everywhere on the field line. The other
solution corresponds to greatly expanded field lines.
In the latter solution, a given field line stretches out to distances so large
that the linear terms (dimensionally proportional to $\Psi/r^2$) become small
and the nonlinear term gives an important contribution to the
equilibrium structure of the distant portion of the expanded field. For
a given value of~$F$, we can define a characteristic radial
scale~$r_F$ where this state of affairs will prevail,
$$ r_F(\Psi,t) \equiv {\Psi\over{F(\Psi,t)}} \gg r_0(\Psi)\, . \eqno{(5.18)} $$
The expression (5.18) also gives an order-of-magnitude estimate
of the position $r_{\rm ap}(\Psi,t)$ of the apex of the field line 
$\Psi$ at the time when $F(\Psi)$ has the value $F(\Psi,t)$. 

Our next step is to use this estimate to calculate the twist angle 
corresponding to the field line $\Psi$ at time~$t$. This can be 
easily accomplished by using equations~(5.2) and (5.3) and {\em 
assuming} that there are no thin scales (i.e., current sheets) in 
the $\theta$~direction. This means that the part of the field line 
that corresponds to $r\sim r_F$ is finite in~$\theta$. Then one can 
write:
$$ \Delta\Phi(\Psi,t) = F(\Psi,t) \int\limits_{\Psi}
{{d\theta}\over{B_{\theta} r(\Psi,\theta) \sin^2(\theta)}} \sim
F(\Psi,t) {1\over{(B_{\theta}r)_{\rm min}}}.                \eqno{(5.19)} $$

Now, $(B_{\theta} r)_{\rm min} = 
-[(\partial\Psi/\partial r)/\sin{\theta}]_{\rm min}$ can be estimated 
simply as $(\Psi/r)_{\rm min} = \Psi/r_{\rm ap}(\Psi,t) \sim \Psi/r_F$.
Thus, 
$$ \Delta\Phi(\Psi,t)\sim F(\Psi,t){r_F\over\Psi} \sim O(1).  \eqno{(5.20)} $$

Therefore, to lowest order in $F$, the twist angle
becomes independent of~$F$ as $F\rightarrow 0$ for a given field 
line. That is, as~$F$ goes to zero and the field lines open, $\Delta\Phi$ 
approaches a finite value and we have a finite-time singularity.

\section{Discussion}
\label{sec-discussion}

In this section we consider the implications of our work to the various 
modeling issues discussed in \S~\ref{sec-intro}. We start with one of 
our key results, namely, the finding (\S~\ref{sec-resistive}) that real 
astrophysical systems are unlikely to reach an exact steady state unless 
they have significant {\em negative} radial field components at the disk 
surface in their untwisted (potential) state (corresponding to an $n\ll 1$ 
magnetic field configuration in the self-similar representation). We now 
argue that such low-effective-$n$ configurations are not likely to be 
realized in axisymmetric, magnetically linked star--disk systems.

Although models in which the stellar magnetic field is pulled into the 
disk have been considered in the literature (e.g., Ostriker \& Shu 1995), 
and $B_{{\rm d},r}$ is $<0$ even for a strictly dipolar field if the disk 
has a finite height, it is difficult to imagine that strong negative radial 
field components would naturally develop in the course of disk accretion onto 
a magnetized star. In fact, simple global disk models (e.g., Spruit \& Taam
1990) indicate that, so long as the disk is at least partially conductive, it 
will tend to exclude the (dipolar) stellar magnetic field and cause some of 
the magnetic flux to remain trapped inside its inner radius, thereby giving
rise to a {\em positive} radial field component at the disk surface. 
Furthermore, there has been no indication from any of the numerical 
simulations carried out to date that the magnetic field in a disk 
interacting with a central dipole can evolve into an $n\ll 1$ configuration. 
Although the published simulations have all been strictly 2-D, 
such systems are even less likely in 3-D. This is because the 
steady-state values of $|B_{{\rm d},\phi}/B_{{\rm d},z}|$ are 
expected to be well in excess of 1 for $n \ll 1$ (see Table 1), 
implying that the field configuration might be susceptible to a 
disruptive, nonaxisymmetric instability (e.g., Kadomtsev 1966).

It is conceivable, however, that negative values of $B_{{\rm d},r}$ would
occur in the case of a {\em nonaxisymmetric} field configuration.
Nonaxisymmetry could be introduced by the misalignment of a stellar dipolar
field axis with the disk axis, which is, in fact, the common interpretation 
of many of the observed periodicities in accreting magnetic stars (e.g., 
Lamb 1989; Patterson 1994; Bertout et al. 1988). Alternatively, it could 
arise from the presence of isolated stellar magnetic loops that thread the 
disk (e.g., Bertout et al. 1996; Safier 1998). The latter possibility is 
consistent with the inferred (by radio observations) existence of extended 
($\ga 10 \, R_*$), organized magnetic structures in certain YSOs (e.g., 
Andr\'e  et al. 1992). The axisymmetric self-similar model with $n\, < \, 1$
could provide a qualitative representation of the field evolution in this case,
but some important aspects, such as the possibility that only a fraction of the
field lines open first (e.g., Amari et al. 1996b), or the development of kink 
and/or buckling instabilities (e.g., Parker 1979), might be captured only in 
a fully 3-D calculation. In any case, although such systems might attain a
steady time-averaged configuration, they could never be in a strictly steady
state.

Another new result of our work has been the demonstration 
(\S~\ref{subsec-resistive-radial}) that, even in disks with 
a comparatively low diffusivity, in which the field lines are
twisted until they open up, one cannot consistently neglect the 
{\em radial} migration of the magnetic footpoints in the disk as 
the critical twist angle $\Delta\Phi_{\rm c}$ is approached. We have argued 
that it is unlikely that a state with $B_{{\rm d},r} = 0$ (even just in 
a time-averaged sense) can be attained unless, again, the magnetic field 
configuration corresponds to that of an $n\ll 1$ self-similar model. Under 
these circumstances, a disk that undergoes cycles of successive opening and 
closing (by reconnection) of twisted field lines (and, to a lesser extent, 
even a disk in which a steady-state twist is maintained) should manifest 
a secular evolution of the radial flux distribution. We have explicitly 
shown, however, that the rate of outward flux diffusion in the disk is 
significantly smaller than the rate of expansion of the twisted field 
lines in the overlying magnetosphere. This provides a strong argument 
{\em against} the suggestion made by BH that the field lines in 
low-diffusivity disks remain perpendicular to the disk surface at 
all times. BH's conclusion that the twisting process will tend to 
expel to the outer edge of the disk almost all the stellar field 
lines that initially threaded it can be questioned on similar grounds. 
Although an outward flux migration might take place, we consider it 
unlikely that it would lead to a flux rearrangement in the disk
that is as drastic as envisioned by BH. In fact, in certain cases a 
steady time-averaged flux distribution may be established. This is 
evidently the case in the time-dependent model presented by Goodson 
et al. (1999) and Goodson \& Winglee (1999), in which inward radial
mass motions that counter the outward radial flux diffusion take place 
in the disk.

We have also investigated (\S~\ref{sec-magnetosphere}) the density 
redistribution in the magnetosphere brought about by the motion of 
the twisted field lines. We showed that, as the critical twist angle 
is approached, the density in the apex region of the elongated flux 
ropes drops precipitously, with some of the depleted mass being pushed 
toward the rotation axis, where it becomes concentrated in a narrow 
angular sector about $\theta=0$. A similar axial density enhancement 
has been found in the numerical simulations of Goodson et al. (1997, 
1999; see also Hayashi et al. 1996), where it is directly associated 
with the formation of an overdense, ``knotty'' jet. As already noted 
by Goodson et al. (1999), this mechanism of producing axially outflowing
condensations requires a good alignment between the rotation and magnetic 
dipole axes. Although our sequence-of-equilibria representation cannot 
capture the complex dynamical evolution of the simulated jets, it is 
gratifying that this simple model nonetheless allows a quantitative 
description of the basic process that generates their density 
structure.\footnote
{The simulations by Hayashi et al. (1996) and Goodson et al. (1997, 1999) 
also show an equatorial density enhancement induced by mass being pushed 
away from the apex regions of the expanding flux loops toward lower 
latitudes. This feature of the density evolution is not reproduced 
by our calculations on account of the fixed-density boundary condition 
that we imposed at the disk surface (see~eq.~[4.20]).}

Although inertial effects slow down the field-line expansion, one can 
expect that the twisted field lines will still {\em effectively open} 
in a finite time. By this we mean that for any region of a given size 
threaded by a finite amount of flux, the apexes of the majority
of the field lines will leave the region under condsideration
after being twisted by a finite angle. This is indeed the
behavior found in numerical 
simulations. For example, in the model of Goodson et al. (1999), 
even though inertial effects are enhanced by a relatively high 
initial mass loading of the magnetosphere and by subsequent episodes 
of post-reconnective mass ejection from the disk, the field lines 
typically open in less than one differential star--disk rotation 
period (as stated in Goodson \& Winglee 1999),
which is to be compared with the time ($\la 1/2$ period) predicted by 
our model in the absence of inertial effects (see \S~\ref{sec-model} 
and \S~\ref{sec-keplerian}). The basic question is then whether the field 
lines eventually reconnect, resulting in a repetitive cycle of
inflation and reconnection, or whether the opening process is
essentially a one-time affair.

The 2-D numerical simulations carried out to date do not provide
a clear answer to this question. Although reconnection was
found to take place in the work of Hayashi et al. (1996), this was probably a
consequence of their adoption of
an unrealistically low threshold value of $j/\rho$ for the onset of 
anomalous resistivity and of a rather strong functional dependence 
[$\eta \propto (j/\rho)^2$] above the threshold. Numerical
diffusivity was evidently also responsible for the reconnection
events observed in the simulations by Romanova et al. (1998)
of twisted, {\em disk-anchored} magnetic loops. In the case of 
the numerical model of Goodson et al. (1999), the nominal 
value of the numerical resistivity-based Lundquist number in the innermost 
zone was $\sim 100$ at the start of their run. As
noted in \S~\ref
{subsec-reconnection}, our self-similar model results indicate that a
value in this range is compatible with current-driven Sweet-Parker
reconnection. However, the reconnection that actually took
place in this simulation was triggered by inward mass 
motions at the inner radius of the disk that strongly pinched the 
magnetosphere (see also Miller \& Stone 1997). 

Our analysis of the 2-D self-similar model has led us to the conclusion 
that ion-acoustic microinstability will be triggered too late to result 
in significant reconnection before the field lines effectively open. The 
crux of the problem is that the current layer that forms in this geometry 
along the field-line elongation direction is comparatively wide (aspect 
ratio $\sim 10$ for an initially dipolar field), so the effective Lundquist
number that is required for efficient reconnection is rather low and is 
attained only relatively late in the field-line opening process. 
Tearing-mode turbulence could, in principle, lead to reconnection on
the required scale if the tearing fluctuation level is sufficiently
high. As discussed by Strauss (1988), such turbulence can occur in 3-D,
giving rise to hyperresistivity that may result in fast reconnection. We can, 
in fact, argue more generally that, in contrast with the idealized 2-D 
configurations that we employed in our model, real 3-D systems may exhibit
nonaxisymmetric behavior that could significantly broaden the range of 
possible scenarios. For example, it seems quite probable that, as the 
field lines are twisted, the 2-D equilibrium will become unstable to 
nonaxisymmetric MHD perturbations, such as the kink mode (a possibility 
already mentioned by LRBK). The detailed MHD stability analysis that is 
required for a closer examination of this suggestion is beyond the scope 
of this paper, but it seems likely that such instabilities could induce 
a transition to a new, nonaxisymmetric configuration that might have real 
(i.e., infinitesimally thin) current sheets. Such current sheets could 
form sites of rapid reconnection that would enable the system to relax 
to a less stressed state, and it is quite probable that the field-line 
expansion would then terminate much sooner than in the 2-D scenario 
considered above. This possibility could be checked by means of future 
3-D numerical simulations.

We thus regard it as highly plausible that reconnection, leading to periodic 
(or quasi-periodic) evolution, would be the realistic outcome of the field-line
twisting process.  This inherently nonsteady behavior enhances the
attractiveness of this scenario as a modeling framework. In particular, in the
case of YSOs, it could be relevant to the interpretation of such phenomena as 
X-ray flares, the formation of a ``knotty'' axial jet and of a less collimated 
and slower (but more massive) disk outflow, and the heating of the permitted 
line-emitting gas in the magnetosphere (e.g., Hayashi et al. 1996; Hirose et 
al. 1997; Miller \& Stone 1997; Hartmann 1997; Shu et al. 1997; Goodson  et 
al. 1997, 1999; Goodson \& Winglee 1999). Furthermore, since the successive 
openings and reconnections of the field lines have been inferred to modulate 
the gas inflow rate onto the star and the net torque exerted on the star by 
the accreting matter, they would also have a direct impact on the mass and 
spin evolution of the protostar. It may nevertheless be possible to construct 
stationary, global models of accretion disks in such systems by obtaining a 
suitable time average of the magnetic field configuration. This is, in fact, 
the approach originally undertaken by GL and subsequently elaborated upon by 
Zylstra (1988) and Daumerie (1996). In particular, Daumerie developed a model 
in which the flux distribution along the disk is calculated by approximating 
the magnetic field in the magnetosphere as being current-free and by using an 
effective conductivity (based on the value of $B_{{\rm d},\phi}/B_{{\rm d},z}$ 
through a relation analogous to our eq. [3.5]) to evaluate the
poloidal field components in the disk. We note, however, that
inasmuch as the azimuthal currents (which determine the disk
contribution to the poloidal magnetic field) flow entirely inside
the disk, one should employ the actual disk conductivity in this
calculation instead of the above ansatz, as the latter could significantly 
overestimate the correct value in regions where the effective surface 
conductivity exceeds the limit (3.6). It may also be possible to refine 
the treatment of the magnetospheric field by using the approach outlined 
in \S~\ref{sec-keplerian} to calculate the magnetic field configuration in 
the fully force-free formulation (i.e., without neglecting current flows in 
the magnetosphere). After identifying the time at which reconnection is most 
likely to occur, one could proceed to derive the time-averaged values of 
$B_{{\rm d},r}(r)/B_{{\rm d},z}(r)$ and $B_{{\rm d},\phi}(r)/B_{{\rm d},z}(r)$ 
and use them as boundary conditions for the disk model.

Past numerical simulations have proved invaluable in clarifying some apparent
difficulties and in pointing the way to further theoretical
progress. For example,
Safier (1998) has raised concerns about the ability of the magnetic field 
lines to thread the disk rather than be swept out in a thermally driven 
stellar wind. However, the numerical simulations (which typically do include
thermal effects) have demonstrated that a field with an initially dipolar
topology can maintain a configuration in which it continuously threads the
disk even as some field lines open and drive an outflow, and that various 
time-dependent effects (such as field-line twisting and mass motions in and out
of the disk plane) combine to increase its strength above naive, steady-state
estimates. Nevertheless, more refined simulations (affording
better control on the field diffusivity and incorporating 3-D
effects) are needed to fully resolve this and other still-open issues.

\section{Summary}
\label{sec-summary}

Disk accretion onto a magnetic star occurs in a variety of astrophysical
contexts, ranging from T Tauri stars to X-ray pulsars. The magnetohydrodynamic
interaction between the stellar field and the accreting matter can have a
strong effect on the disk structure, the transfer of mass and angular momentum
between the disk and the star, and the production of bipolar
outflows. In this paper we concentrated on a key element of this
interaction --- the time evolution of the magnetic field lines
that thread the disk resulting from the relative rotation between the
disk and the star --- and studied it using simplified models of
axisymmetric, force-free fields. In particular, we employed the
semianalytic similarity solution first derived by van
Ballegooijen (1994) and Lynden-Bell \& Boily (1994) to construct
a sequence of magnetospheric equilibria parametrized by the relative
twist angle $\Delta \Phi$, but we tested the generality of the
basic conclusions of this model (which strictly applies only to a uniform
relative angular speed) by constructing numerical solutions of the
Grad-Shafranov equation for systems with rotation laws that
approximate a Keplerian disk. Our main results can be summarized
as follows.

\begin{itemize}
\item On the assumption that both the star and the
magnetosphere can be approximated as being perfectly conducting,
the behavior of the twisted field lines depends on the magnitude
of the surface conductivity $\Sigma$ of the disk. A steady-state
configuration can be established only if, at a radius $r$ in the
disk, $\Sigma(r) < \Sigma_{\rm max}(r) \sim c^2/r|\Delta \Omega(r)|$
(eq. [3.6]). Weakly ionized protostellar disks as well as disks
characterized by a turbulent diffusivity typically do not
satisfy this condition except very close to the corotation
radius, implying that the field lines undergo secular inflation
and will open up when a critical twist angle $\Delta\Phi_c$ ($
\sim 2\ {\rm rad}$ for an initially dipolar field configuration)
is reached. 

\item If inertial effects in the magnetosphere could be
neglected, then the field lines would open in a fraction of
the differential star--disk rotation period. However, inertial
effects intervene, slowing down the expansion speed at the apex
of the twisted field line to the local Alfv\'en speed (which,
for an initial dipolar field and a power-law density distribution,
remains constant throughout the evolution in the self-similar
model). This will delay the effective opening of the field
lines, but will not by itself prevent it from occurring.

\item If $\Sigma(r) > \Sigma_{\rm max}(r)$ and the field lines
undergo a strong expansion, then the radial
magnetic stresses at the disk surface will cause the field lines
to start migrating outward. A similar conclusion was reached by Bardou \&
Heyvaerts (1996), who went on to suggest that the field lines will
be rapidly expelled from the disk.  We argued, however, that
this is unlikely to happen in view of the fact that the radial
diffusion in the disk is much slower than the field-line
expansion in the magnetosphere. If the field lines expand and
reconnect in a periodic manner then a time-averaged steady state in
which the mean surface radial field is zero is in principle
possible (van Ballegooijen 1994), but this situation will not
arise if the initial field configuration is dipolar (although it
might conceivably pertain to isolated stellar flux tubes
that thread the disk).

\item The opening of the field lines induces a redistribution of
the magnetospheric density $\rho$. In particular, it leads to a
strong depletion of
mass near the apex of the expanding field lines (which typically elongate in a
direction making an angle $\ga 60^{\circ}$ to the rotation
axis) and to a pronounced density enhancement near the rotation
axis. The decrease in density along the direction of elongation
(which also marks the location of a $j_\phi$ current layer)
is conducive to the triggering of microinstabilities (such as
ion-acoustic), since the instability onset condition corresponds to
a threshold value of $j/\rho$. The density enhancement near
the axis (which is largely a byproduct of the axisymmetry
assumption) represents a possible mechanism for the formation of
an axially outflowing condensation (generalizing to a ``knotty'' jet
if the process is repetitive), as suggested previously (e.g.,
Goodson et al. 1999) on the basis of numerical simulations.

\item The question of whether the evolution of the twisted field
culminates in the effective opening of the field lines (e.g.,
Lovelace et al. 1995), or whether it is a repetitive process
(e.g., Aly \& Kuijpers 1990) of field-line expansion and
reconnection (not necessarily back to exactly the same field
configuration that existed before the most recent expansion, in
which case it will not be strictly periodic), is still open. The
reconnection events observed in the numerical simulations carried 
out to date are by and large due to numerical diffusivity and 
therefore do not provide conclusive answers. Based on an analysis 
of the 2-D self-similar model, we concluded that anomalous resistivity
derived from the ion-acoustic/Buneman instability will be
triggered too late for significant reconnection to take place 
across the comparatively wide (aspect ratio $\sim 10$ for an 
initially dipolar field) current layer that forms in this case 
before the field lines effectively open.
We suggested, however, that hyperresistivity associated with
tearing-mode turbulence (Strauss 1988) might lead to efficient
reconnection, although whether such turbulence could arise in 
the configuration under consideration is not clear. 
In addition, it is conceivable 
that even faster reconnection could take place as a result of 3-D 
kinking of the twisted field lines that would create genuine (i.e., 
infinitesimally thin) current sheets. This issue might be resolved 
by MHD stability calculations and 3-D numerical simulations.

\end{itemize}

\vskip 20pt

We thank Bob Rosner, Oded Regev, Aad van Ballegooijen, 
Fred Lamb, and B.C. Low for interesting and fruitful discussions.
This research was supported in part by NASA grants NAG 5-3687 and 
NAG 5-1485 and by the ASCI/Alliances Center for Astrophysical 
Thermonuclear Flashes at the University of Chicago under DOE 
subcontract B341495.

\appendix
\section{CALCULATION OF $({d\Delta\Phi}/{da_0})$
NEAR THE CRITICAL POINT}

In this Appendix we present a more detailed analysis of the behavior 
of the magnetic field near the critical point. Our main goal is 
to find the asymptotic behavior of $({d\Delta\Phi}/{da_0})$ 
near~$\Delta\Phi_{\rm c}$. We use the results of this analysis in 
\S~\ref{subsec-resistive-radial} to determine 
the effects of resistive diffusion near the critical point, 
and also in \S~\ref{sec-magnetosphere} in studying 
the role of inertial effects and reconnection in the magnetosphere.

First, we note that calculating $({d\Delta\Phi}/{da_0})$ is not
a straightforward task, because it requires the knowledge of the
deviation of $g(\theta)$ 
from the asymptotic solution~$G_0(\theta)a_0^{-n}$. Indeed, if one simply 
uses the lowest order expression $g(\theta)\simeq G_0(\theta)a_0^{-n}$,
then, upon substituting it into equation~(2.13), one gets equation~(2.21). 
We see that in order to find the behavior of $(d\Delta\Phi/da_0)$ near the 
critical point, one needs to know the next-order (in $a_0$) correction to 
$g(\theta$) as~$a_0\rightarrow 0$.

Let us write
$$ g(\theta)=G(\theta)a_0^{-n} = G_0(\theta) a_0^{-n} +
\delta G(\theta) a_0^{-n}\, ,                                   \eqno{(A1)} $$
where $G(\theta)$ is the solution of equation~(2.19) with boundary conditions
$G(0)=0$, $G(\pi/2)=a_0^n \ll 1$.

We want to find the correction $\delta G(\theta)$. In order to do this, 
let us introduce a small parameter~$\varepsilon$. We choose $\varepsilon$ 
so that $\delta G(\pi/2-\varepsilon) \ll G_0(\pi/2-\varepsilon)$ and at 
the same time $\varepsilon \ll 1$. As can be seen from Figure~\ref{fig-G0}, 
the derivative $dG_0/d\theta$ at $\theta=\pi/2$ is in general a finite 
(meaning that it does not scale with~$a_0$) number that depends only on~$n$.
Thus, $G_0(\pi/2-\varepsilon)\simeq \varepsilon dG_0/d\theta(\pi/2)$.
Next, $\delta G(\pi/2-\varepsilon)$ is of order $\delta G(\pi/2)=a_0^n$,
so we must choose $\varepsilon$ in the range
$$ a_0^n \ll \varepsilon \ll 1\, .                              \eqno{(A2)} $$

Inside the $\varepsilon$-vicinity of $\theta=\pi/2$ we can 
approximate $\sin{\theta}\simeq 1={\rm const}$, so we get
$$ G'' (\theta)+n(n+1)G +{n\over{n+1}} G^{1+2/n}  = 0\,  .       \eqno{(A3)} $$
with $G(\pi/2)=a_0^n$.

In this region $G(\theta) \ll 1$, so we can neglect the nonlinear 
term compared with the other terms, and hence the solution can be 
written as
$$ G_{\rm inner}(\theta) \simeq a_0^n + (\theta-\pi/2) 
\cdot{\rm const}\, .                                          \eqno{(A4)} $$

In the region $\theta< \pi/2-\varepsilon$, the correction $\delta G$ is 
small compared with~$G_0(\theta)$, and we can linearize equation~(2.19). 
We then obtain the following linear equation for~$\delta G$:
$$ \sin(\theta){d\over{d\theta}}\left({1\over{\sin{\theta}}}
{{d\delta G}\over{d\theta}}\right) +n(n+1)\delta G(\theta)+
{{n+2}\over{n+1}} \delta G(\theta) [G_0(\theta)]^{2/n} = 0\, .  \eqno{(A5)} $$

Actually, near $\theta=\pi/2$ the last term of equation (A5) becomes
negligible compared with the other terms (because
$G_0(\theta)\ll 1$ there). Since this is the only term that distinguishes this
equation from the one describing
the inner region $\theta>\pi/2-\varepsilon$, we can ignore the 
difference and extend the range of applicability of equation~(A5) all the 
way up to $\theta=\pi/2$, where we set the boundary condition 
$\delta G(\pi/2)=a_0^n$.

Thus, the solution of the linear equation~(A5) with the boundary 
conditions
$$\delta G(0)=0,  \qquad \delta G(\pi/2)=a_0^n                 \eqno{(A6)} $$
provides accurate (to the lowest order in~$a_0$) correction to 
the function~$g(\theta)$.

Using the function $\delta G(\theta)$ we can now determine
the asymptotic behavior of $\Delta\Phi(a_0)$ in the limit
$a_0\rightarrow 0$. We rite
$$ \Delta\Phi =\Delta \Phi_{\rm c} +\delta \Delta\Phi\, ,     \eqno{(A7)} $$
where
$$ \delta \Delta \Phi = {1\over{n+1}} \int_0^{\pi/2}
\delta [G^{1/n}(\theta)] {d\theta\over{\sin{\theta}}}\, .       \eqno{(A8)} $$

In order to estimate this integral, let us choose some 
$\epsilon_1\sim a_0^{n-x}\ll 1$, $0<x<n$. Then, within
the~$\epsilon_1$-vicinity of $\theta = \pi/2$ we have 
$G(\theta) \simeq a_0^n + G_0(\theta)$ and $G_0(\theta) 
\simeq [dG_0/d\theta]|_{\theta=\pi/2}(\theta-\pi/2) \sim 
C_1 \epsilon_1$. We can thus estimate that in this region
$|\delta [G^{1/n}(\theta)]| < C_2 \epsilon_1^{1/n}$,
where~$C_1$ and~$C_2$ are positive constants of order 1.

Correspondingly, we estimate the contribution 
to the integral~(A8) from this region as
$$ \left | {1\over{n+1}} \int_{\pi/2-\epsilon_1}^{\pi/2}
\delta [G^{1/n}(\theta)] {d\theta\over{\sin{\theta}}}\right | <
C_3 \epsilon_1^{1+1/n}\, ,                             \eqno{(A9a)} $$
where~$C_3$ is a positive constant of order 1.

In the rest of the integration domain,
we can write $\delta [G^{1/n}(\theta)] \simeq 
(1/n) (\delta G) G_0^{1/n-1}(\theta)$, which yields
$$ {1\over{n+1}} \int_0^{\pi/2-\epsilon_1}
\delta [G^{1/n}(\theta)] {d\theta\over{\sin{\theta}}} \simeq
{1\over{n+1}} \int_0^{\pi/2-\epsilon_1} {1\over n} 
\delta G G_0^{1/n-1}(\theta) {d\theta\over{\sin{\theta}}}\, .  \eqno{(A9b)} $$
Since $\delta G \sim a_0^n$, the contribution 
from this region is of order $a_0^n$. We see that, under 
the additional condition that 
$$ \epsilon_1^{1+1/n} \ll a_0^n \Rightarrow  
x < {n\over{n+1}}\, ,                                         \eqno{(A10)} $$
this contribution is much larger than the contribution from 
the~$\epsilon_1$-vicinity of $\theta=\pi/2$, which we therefore 
can neglect. Thus, we immediately see that the correction to the 
twist angle $\Delta\Phi$ is of the order~$a_0^n$. Now let us 
determine the coefficient.

Notice that, as one takes the upper limit of the integral~(A9b)
to $\theta=\pi/2$, i.e., as $\epsilon_1\rightarrow 0$, the integral 
converges as a function of $\epsilon_1$. This is because, as 
$\theta\rightarrow\pi/2$, $\delta G \rightarrow a_0^n=\rm const$ 
and $G_0^{1/n-1}(\theta) \sim (\pi/2-\theta)^{1/n-1}$, so
$\int^{\pi/2-\epsilon_1}\delta G G_0^{1/n-1}(\theta) 
d\theta/\sin{\theta} \sim \int^{\pi/2-\epsilon_1}
(\pi/2-\theta)^{1/n-1} d\theta \sim \epsilon_1^{1/n} \rightarrow 0$ 
as $\epsilon_1\rightarrow 0$.

Therefore, to lowest order in~$a_0$, we can write
$$ \delta \Delta \Phi \simeq {1\over{n(n+1)}} \int_0^{\pi/2} \delta G 
G_0^{1/n-1}(\theta){d\theta\over{\sin{\theta}}} =\xi(n) 
a_0^n\, .                                             \eqno{(A11)} $$ 
Using the functions $\delta G$ and $G_0(\theta)$ obtained above, 
we get, for example,
$$ \delta \Delta \Phi (n=1, a_0 \rightarrow 0) \approx  -0.17\
a_0    $$
and 
$$ \delta \Delta \Phi (n=0.5, a_0\rightarrow 0) = -0.22\
a_0^{1/2}\, .    $$

\section{ASYMPTOTIC ANALYSIS OF THE VELOCITY FIELD NEAR THE
APEX ANGLE}

In this Appendix we present an asymptotic analysis of 
the velocity field near the apex angle~$\theta_{\rm ap}$ 
in the limit $t\rightarrow t_{\rm c}$.

First we do some preparatory work. Near the critical point $t=t_{\rm c}$ 
the behavior of $g(\theta)$ in the vicinity of the apex angle 
$\theta_{\rm ap}$ can be described by
$$ g(\theta,a_0)\simeq a_0^{-n} G_0(\theta) = 
a_0^{-n} G_0(\theta_{\rm ap}) 
\ [1-\mu{{(\theta-\theta_{\rm ap})^2}\over 2}]\, ,            \eqno{(B1)} $$
where we used equation~(2.20) and defined $\mu= 
\mu(n) \equiv -G_0''(\theta_{\rm ap})/
G_0(\theta_{\rm ap})$. Using equation~(2.19), 
$$ \mu =n(n+1)\ [1+{{G_0^{2/n}}\over{(1+n)^2}}]\, .             \eqno{(B2)} $$

Next, in order to get the time dependence, we need to 
relate $a_0$ to time. Using equation~(A11) we can write:
$$ a_0^{-n}={\xi\over{\delta\Delta\Phi}}=
{\xi\over{\Delta\Omega}}\ {1\over{t-t_{\rm c}}}\, .              \eqno{(B3)} $$

Therefore,
$$ g(\theta,t)={\xi\over{\Delta\Omega}}\ {{G_0(\theta_{\rm
ap})}\over{t-t_{\rm c}}}
\ [1-\mu{{(\theta-\theta_{\rm ap})^2}\over 2}]\, .             \eqno{(B4)} $$
Furthermore, on account of equations (2.5) and (2.9), 
$$f(\theta\simeq\theta_{\rm ap})=
-{{a_0^{-n}G_0(\theta_{\rm ap})}\over{n\sin\theta_{\rm ap}}}
\mu (\theta-\theta_{\rm ap})=
-{1\over{n\sin\theta_{\rm ap}}} {\xi\over{\Delta\Omega}}
{{G_0(\theta_{\rm ap})}\over{t-t_{\rm c}}}\mu(\theta-\theta_{\rm
ap}) \eqno{(B5)} $$
and
$$ h(\theta_{\rm ap})=
{{a_0^{-n}G_0^{1+1/n}(\theta_{\rm ap})}\over{(n+1)\sin\theta_{\rm ap}}}=
{{G_0^{1+1/n}(\theta_{\rm ap})}\over{(n+1)\sin\theta_{\rm ap}}}
{\xi\over{\Delta\Omega}} {1\over{t-t_{\rm c}}}\, .              \eqno{(B6)} $$

In addition, similarly to $\dot{\Delta\Phi}=\Delta\Omega$, we 
can write $\dot{\phi}(\theta_{\rm ap},t)=\Delta\Omega\zeta/\xi$,
where $\zeta$ is some constant of order 1.

Finally, we derive the expression for~$P(\theta_{\rm ap},t)$
(eq. [4.6]):
$$ P(\theta_{\rm ap},t) =\kappa g(\theta_{\rm ap})= \kappa 
G_0(\theta_{\rm ap}){\xi\over{\Delta\Omega}}{1\over{t-t_{\rm
c}}}\, ,    \eqno{(B7)} $$
where $\kappa\equiv\sqrt{1+G_0^{2/n}(\theta_{\rm ap})/(n+1)^2}\, .$

Substituting these expressions into equations~(4.3)--(4.5) 
for the components of ${\bf v_\perp}$, we get
$$ v_{\perp r}(\theta_{\rm ap})=
r{{\dot{g}(\theta_{\rm ap})}\over{ng(\theta_{\rm ap})}}=
-{r\over n}\ {1\over{t-t_{\rm c}}}\, ,
\eqno{(B8)} $$ 
$$ v_{\perp\theta}(\theta\simeq\theta_{\rm ap}) =
-r {{\dot{g}f\sin\theta_{\rm ap}}\over{nP^2}} \simeq
-r {{n+1}\over n} {{\theta-\theta_{\rm ap}}\over{t-t_{\rm c}}}\, , 
\eqno{(B9)} $$
and
$$ v_{\perp\phi}(\theta\simeq\theta_{\rm ap}) =
-r {{\dot{g}fh\sin^2\theta_{\rm ap}}\over{ngP^2}} \simeq
-r {{G_0^{1/n}}\over n} {{\theta-\theta_{\rm ap}}\over{t-t_{\rm
c}}}\, . \eqno{(B10)} $$

We can now calculate the parallel velocity using the equation of
motion~(4.12). The centrifugal and Coriolis terms in this
equation can be neglected in the vicinity of the critical point~$t=t_{\rm c}$ 
because their contributions are, respectively, of zeroth and first order in 
$(t-t_{\rm c})^{-1}$, whereas the contributions of most of the
other terms are of order $(t-t_{\rm c})^{-2}$ (and therefore
dominate in the limit $t\rightarrow t_{\rm c}$).

Substituting the expressions~(B4)--(B10) for~$g$, $f$, $h$, $P$,
and~${\bf v}_\perp$ into equations~(4.12)--(4.14), we obtain an equation 
for $v_\parallel(\theta\simeq\theta_{\rm ap},t)$:
$$ \dot{v}_\parallel=
-{v_\parallel\over{\kappa r}}{{\partial v_\parallel}\over{\partial\theta}}+
{{n+1}\over n} {{\theta-\theta_{\rm ap}}\over{t-t_{\rm c}}}
{{\partial v_\parallel}\over{\partial\theta}}+
{v_\parallel\over n} {{n+3}\over{t-t_{\rm c}}} -
r \kappa {{\theta-\theta_{\rm ap}}\over{(t-t_{\rm c})^2}}
{{(n+1)(n+2)}\over{n^2}}\, .                                   \eqno{(B11)} $$ 

Equation (B11) implies that $v_\parallel$ can
be written as 
$$ v_\parallel(r,\theta,t)=
r (\theta-\theta_{\rm ap}) {A\over{t-t_{\rm c}}}\, ,           \eqno{(B12)} $$ 
where $A=A(n)$ is a numerical constant. After substituting this 
expression into equation~(B11), we get a quadratic equation for~$A$:
$$ A^2-A\kappa(3+4/n)+\kappa^2{{(n+1)(n+2)}\over{n^2}} = 0\, . \eqno{(B13)} $$
  
Equation (B13) has two solutions for each value of~$n$:
$$ A_{1,2}={\kappa\over{2n}} \left(3n + 4 \pm
\sqrt{5n^2+12n+8} \right)\, .                                  \eqno{(B14)} $$
For example, for $n=1$, we find $A_1=\kappa\simeq 2$ and 
$A_2=6\kappa\simeq 12\, .$

\section*{REFERENCES}
\parindent 0 pt

Aly, J.~J. 1984, ApJ, 283, 349

Aly, J.~J. 1985, A\&A, 143, 19

Aly, J.~J., \& Kuijpers, J. 1990, A\&A, 227, 473

Aly, J.~J. 1995, ApJ, 439, L63

Amari, T., Luciani, J.~F., Aly, J.~J., \& Tagger, M. 1996a,
A\&A, 306, 913

Amari, T., Luciani, J.~F., Aly, J.~J., \& Tagger, M. 1996b,
ApJ, 466, L39

Andr\'e, P., Deeney, B. D., Phillips, R. B., \& Lestrade, J.-P. 1992, ApJ,
401, 667

Bardou, A., \& Heyvaerts, J. 1996, A\&A, 307, 1009 (BH)

Bertout, C., Basri, G., \& Bouvier, J. 1988, ApJ,
330, 350

Bertout, C., Harder, S., Malbet, F., Mennessier, C.,
\& Regev, O. 1996, AJ, 112, 2159

Bouvier, J., Carbit, S., Fernandez, M., Martin, E.~L.,
\& Matthews, J.~M. 1993, A\&A, 272, 176

Campbell, C.~G. 1992, Geophys. Astrophys. Fluid Dyn., 
63, 179

Collier Cameron, A., \& Campbell, C.~G. 1993, A\&A, 274, 309

Collier Cameron, A., Campbell, C.~G., \& Quaintrell, H.
1995, A\&A, 298, 133

Coroniti,~F.~V., \& Eviatar,~A. 1977, ApJS, 33, 189

D'Alessio, P., Cant\'o, J., Calvet, N., \& Lizano, S. 1998, ApJ, 500, 411

Daumerie, P. R. 1996, Ph.D. Thesis, University of Illinois at Urbana-Champaign

Drake~J.~F., Guzdar~P.~N., Hassam~A.~B., \& Huba,~J.~D. 1984,
Phys. Fluids,  27, 1148.

Edwards, S., Hartigan, P., Ghandour, L., \& Andrulis, C.
1994, AJ, 108, 1056

Edwards, S., et al. 1993, AJ, 106, 372

Erkaev, N.V., Semenov, V.S., \& Jamitzky, F. 2000, PRL, 84, 1455

Galeev,~A.~A., \& Sagdeev,~R.~Z. 1984, in Handbook of Plasma Physics,
Vol.~II, ed. A.~A.~Galeev \& R.~N.~Sudan 
(Amsterdam: North-Holland), 271

Gammie, C. F., \& Balbus, S. A. 1994, MNRAS, 270, 138

Ghosh, P., \& Lamb, F.~K. 1978, ApJ, 223, L83 (GL)

Ghosh, P., \& Lamb, F.~K. 1979a, ApJ, 232, 259 (GL)

Ghosh, P., \& Lamb, F.~K. 1979b, ApJ, 234, 296 (GL)

Ghosh, P. 1995, MNRAS, 272, 763

Goodson, A.~P., Winglee, R.~M., \& B{\"o}hm, K.-H. 1997,
ApJ, 489, 199
 
Goodson, A.~P., B{\"o}hm, K.-H., \& Winglee, R.~M. 1999,
ApJ, 524, 142
 
Goodson, A.~P., \& Winglee, R.~M. 1999, ApJ, 524, 159

Hartmann, L. 1997, in Herbig-Haro Flows and the the Birth of Low Mass Stars,
ed. B. Reipurth \& C. Bertout (Dordrecht: Kluwer), 391

Hartmann, L., Hewett, R., \& Calvet, N. 1994, ApJ, 426, 669

Hayashi, M.~R., Shibata, K., \& Matsumoto, R. 1996, ApJ,
468, L37

Herbst, W., Rhode, K.~L., Hillenbrand, L.~A., \& Curran, G.
2000, AJ, 119, 261


Johns-Krull, C.~M., \& Basri, G. 1997, ApJ, 474, 433

Johns-Krull, C.~M., \& Hatzes, A.~P. 1997, ApJ, 487, 896

Kadomtsev, B.~B. 1965, Plasma Turbulence (London: Academic Press), 77

Kadomtsev, B. B. 1966, Rev. Plasma Phys., 2, 153

K{\"o}nigl, A. 1991, ApJ, 370, L39

Krall,~N.~A., \& ~Liewer,~P.~C. 1971, Phys. Rev. A, 4, 2094

Lamb, F. K. 1989, in Timing Neutron Stars, ed. H. {\"O}gelman \& E. P. J. van 
den Heuvel (Dordrecht: Kluwer), 649

Lamb, F. K., Hamilton, R. J., \& Miller, M. C. 1993, in Isolated Pulsars, ed. 
K. van Riper, R. Epstein, \& C. Ho (Cambridge: Cambridge Univ. Press), 364

Lamzin, S.~A. 1995, A\&A, 295, L20

Lovelace, R.~V.~E., Romanova, M.~M., \& Bisnovatyi-Kogan, G.~S.
1995, MNRAS, 275, 244 (LRBK)

Low, B.~C. 1986, ApJ, 307, 205

Lynden-Bell, D., \& Boily, C. 1994, MNRAS, 267, 146

Martin, E.~L. 1997, A\&A, 321, 492

Meyer, F., \& Meyer-Hofmeister, E. 1999, A\&A, 341, L23

Miki{\'c}, Z., \& Linker, J.~A. 1994, ApJ, 430, 898

Miller, K.~A., \& Stone, J.~M. 1997, ApJ, 489, 890

Mok, Y., \& van~Hoven,~G. 1982, Phys. Fluids, 25, 636

Ostriker, E. C., \& Shu, F. H. 1995, ApJ, 447, 813

Parker, E.N. 1963, ApJS, 8, 177

Parker, E.~N.  1979, Cosmical Magnetic Fields
(Oxford: Oxford Univ. Press)

Parker, E.~N.  1994, Spontaneous Current Sheets in Magnetic Fields
(Oxford: Oxford Univ. Press)

Patterson, J. 1994, PASP, 106, 697

Petchek, H. E. 1964, in AAS-NASA Symposium on the Physics of Solar 
Flares, NASA-SP 50, ed. W. N. Hess (Washington: NASA), 425

Romanova, M. M., Ustyugova, G. V., Koldoba, A., V., Chechetkin,
V. M., \& Lovelace, R. V. E. 1998, ApJ, 500, 703

Safier, P.~N. 1998, ApJ, 494, 336

Sagdeev~R.~Z. 1967, Proc. Symp. Appl. Math., 18, 281

Sato, T., \& Hayashi, T. 1979, Phys. Fluids, 22, 1189

Scholer, M. 1989, J. Geophys. Res., 94, 8805

Shu, F. H., Shang, H., Glassgold, A. E., \& Lee, T. 1997, Science, 277, 1475

Spruit, H. C., \& Taam, R. E. 1990, A\&A, 229, 475

Strauss,~H. 1988, ApJ, 326, 412

Sweet, P. A. 1958, in Electromagnetic Phenomena in Cosmic Physics,
ed. B. Lehnert (London: Cambridge Univ. Press), 123

Ugai, M., \& Tsuda, T. 1977, J. Plasma Phys., 17, 337

van~Ballegooijen, A.~A. 1994, Space Sci. Rev., 68, 299 (VB)

Wang, Y.-M. 1987, A\&A, 183, 257

Yi, I., Wheeler, J.~C., \& Vishniac, E.~T. 1997, ApJ, 491, L93

Yi, I. 1994, ApJ, 428, 760 

Yi, I. 1995, ApJ, 442, 768

Zylstra, G. J. 1988, Ph.D. Thesis, University of Illinois at Urbana-Champaign


\end{document}